\documentclass[format=acmsmall, review=false]{acmart}
\usepackage{acm-ec-26}
\usepackage{booktabs} 
\usepackage[ruled]{algorithm2e} 

\SetAlFnt{\small}
\SetAlCapFnt{\small}
\SetAlCapNameFnt{\small}
\SetAlCapHSkip{0pt}
\IncMargin{-\parindent}

\setcitestyle{authoryear}

\usepackage{amssymb,amsmath,amsthm,amsfonts}
\usepackage{graphicx}
\usepackage{color,xcolor}
\usepackage{balance}
\usepackage{multirow,makecell,tabu}
\usepackage[english]{babel}
\usepackage{cleveref}
\usepackage{booktabs}
\setcitestyle{nosort}

\makeatletter
\let\old@section\section
\usepackage[bibliography=common]{apxproof}
\let\section\old@section
\makeatother

\newcommand{\anonymize}[1]{\textcolor{gray}{\footnotesize [Redacted for anonymity]}}

\newcommand{\peak}{\mathbf{p}}
\newcommand{\other}{\mathbf{q}}
\newcommand{\extreme}{\mathbf{v}}
\newcommand{\diff}{\mathbf{d}}

\newtheoremrep{theorem}{Theorem}[section]
\newtheoremrep{proposition}[theorem]{Proposition}
\newtheoremrep{lemma}[theorem]{Lemma}

\theoremstyle{definition}

\newcommand{\er}[1]{\textcolor{blue}{#1}}
\newcommand{\erel}[1]{\er{(Erel says: #1)}}

\newcommand{\ay}[1]{\textcolor{red}{#1}}
\newcommand{\ayelet}[1]{\ay{(Ayelet says: #1)}}

\newcommand{\fullversion}[1]{}

\DeclareMathOperator*{\argmax}{arg\,max}

\makeatletter

\title{How Do People Evaluate Non-Ideal Budgets? Evidence from Pairwise Comparisons}

\author{Ayelet Amster}
\affiliation{%
  \institution{The Open University}
  \city{Ra'anana}
  \country{Israel}}
\email{ayelet.amster@gmail.com}

\author{Lioz Akirav}
\affiliation{%
  \institution{Ariel University}
  \city{Ariel}
  \country{Israel}}
\email{lioz.akirav@gmail.com}

\author{Rica Gonen}
\affiliation{%
  \institution{The Open University}
  \city{Ra'anana}
  \country{Israel}}
\email{ricagonen@gmail.com}

\author{Erel Segal-Halevi}
\affiliation{%
  \institution{Ariel University}
  \city{Ariel}
  \country{Israel}}
\email{erelsgl@gmail.com}

\begin{abstract}

Budget aggregation is a process in which citizens declare their individual ideal budget allocations, and a predetermined rule aggregates these declarations into a single collective allocation. Recent theoretical work has proposed various budget-aggregation rules and established impossibility results concerning the simultaneous satisfaction of desirable axioms. The analysis of both aggregation rules and impossibility results typically relies on assumptions about how voters evaluate allocations that differ from their ideals; different assumptions may lead to substantially different theoretical conclusions. Despite their importance, these assumptions have received little direct empirical validation.

We present a framework for empirically testing hypotheses about voters' utility functions using simple pairwise comparisons. We introduce a modular, open-source polling system that first elicits a participant's ideal budget allocation and then presents several carefully generated pairs of non-ideal alternatives. Different pair-generation algorithms enable researchers to test specific utility models as well as more general structural properties of preferences. To illustrate the applicability of the framework, we conducted 82 polls, yielding 3,408 participant-poll observations, of which 1,778 passed the attention checks and were included in the analyses.

Our results show that standard utility models, including $\ell_1$, $\ell_2$, and Leontief utilities, did not consistently predict the choices of most participants. Allowing relative-distance noise of up to $20\%$ substantially increased consistency for some models, particularly $\ell_2$, but did not explain the majority of remaining inconsistencies.

We further find that most screened participants' responses are incompatible with the sign and issue symmetry imposed by standard unweighted $\ell_p$ metrics: participants often evaluated increases and decreases differently and assigned different importance to deviations in different issues. Overall, our findings suggest that simple symmetric distance-based models provide an incomplete description of observed budget-allocation choices. They motivate the development and further empirical evaluation of more flexible utility models for practically applicable budget-aggregation mechanisms.

\end{abstract}

\ccsdesc[500]{Computing methodologies~Algorithmic game theory and mechanism design}
\keywords{participatory budgeting, budget aggregation, utility models}

\begin{document}

\begin{titlepage}

\maketitle
\makeatletter \gdef\@ACM@checkaffil{} \makeatother

\end{titlepage}

\section{Introduction}
This research is motivated by the growing interest in \emph{participatory budgeting} --- a process by which the citizens can participate in deciding how to divide the budget of their city or state. One of the common models for participatory budgeting is \emph{budget proposal aggregation}, in which each voter declares his or her ideal budget allocation, and all these declarations are aggregated by some rule to yield the actual budget (see \Cref{sec:model} for formal model).
A simple aggregation rule is the arithmetic mean; it has good axiomatic properties \citep{intriligator1973probabilistic,elkind2023settling}, but provides strong incentives for voters to report false preferences in order to manipulate the outcome. This gave rise to more sophisticated aggregation algorithms that are \emph{truthful} --- cannot be profitably manipulated. Such algorithms typically use sophisticated variants of the median rule
\citep{moulin1980strategy,FREEMAN2021105234,caragiannis2022truthful,freeman2023project,deberg2024truthful}.
\\

Analyzing the properties of an aggregation rule requires some assumptions about the voters' preferences over non-ideal budgets. For example, suppose some voter thinks that the ideal allocation of a budget of $100$ among three issues is $(50,30,20)$. Which of the following non-ideal allocations would this voter prefer: $(41,30,29)$ or $(43,40,17)$ or $(43,26,31)$?

Different papers have different assumptions on this matter. Many papers assume that voters evaluate a non-ideal budget based on its distance from their ideal budget according to some metric, such as $\ell_1$ \citep{freeman2019truthful,caragiannis2022truthful} or $\ell_{\infty}$ \citep{freeman2023project}. In the above example, the $\ell_1$ distances are $18; 20; 22$, so a voter with $\ell_1$ preferences would prefer $(41,30,29)$. 
Different metrics may induce different preferences; the $\ell_2$ distances in the above example are $(\sqrt{162}, \sqrt{158}, \sqrt{186})$
so a voter with $\ell_2$ preferences would prefer $(43,40,17)$.%
\footnote{
With three issues, the $\ell_{\infty}$ distance is always $1/2$ of the $\ell_1$ distance, so the preferences induced by both metrics are the same. This is not true with four or more issues.
}\\

Recently, \citet{brandt2025optimal} have argued in favor of utility functions called \emph{Leontief}, that are not based on any metric: voters evaluate a non-ideal budget based on the smallest \emph{ratio} of the amount given to any issue to their ideal amount. The smallest ratios in the above example are $\frac{41}{50}= 0.82; \frac{17}{20}= 0.85; \frac{43}{50}=0.86$, so the voter would prefer $(43,26,31)$ as it provides the highest ratio.

Yet another way to compare distributions is the \textit{Kullback-Leibler (KL)} divergence, which measures how different one allocation is from another in terms of information loss. It is commonly used to compare probability distributions, but has been recently used in a social choice context \citep{brandl2024natural}.

In contrast, if voters have Leontief utilities, then the algorithm maximizing the Nash welfare (the product of utilities) is group-strategyproof (stronger than truthful), and satisfies core fair share (stronger than both Pareto-efficiency and proportionality).

These vastly different results invoke the question which is at the heart of the present research:
\begin{quote}
\emph{
 Which utility functions and properties are supported by human budget-allocation choices?
}
\end{quote}
As different people may have different utility functions, our aim is to construct a generic polling framework, that allows to check various properties of individual users' utility functions.

\subsection{Our contribution}
We present a framework for conducting opinion-polls based on \emph{pairwise comparisons}. 
This approach is inspired by \citet{keeney1993decisions}, who present case studies where pairwise comparisons were used to estimate utility functions of individuals and firms for multi-objective optimization.
In each poll, the user is shown a budget-allocation scenario (e.g. distributing state budget among government ministries, or distributing municipal budget among municipal departments), and asked ``what is your ideal budget allocation?''.
Then, each user is shown a list of pairs of non-ideal allocations, and asked to choose, in each pair, the allocation that he or she prefers. 
We focus on pairwise comparisons as they are simple, binary questions, reducing the cognitive burden on the participants to a minimum.
The main challenge in designing the polls is in constructing the list of pairs, such that the user's replies will provide us with meaningful information on his or her utility function.
Sample screenshots of the poll interface are shown in \Cref{app:poll_description}.
\\


We present various pair-generation algorithms, and report the outcomes of running the resulting polls on a representative sample of the voter population in Israel. The code for our polling framework is open-source and can easily be used by researchers elsewhere.

Our first algorithm accepts as input two utility models (e.g. $\ell_1$ and Leontief), and generates pairs that test whether the user consistently adheres to one of these models over the other one. 
Using this algorithm, we generated ten polls, corresponding to all pairwise comparisons among five common utility models: $\ell_1$, $\ell_2$, Leontief, Cosine Similarity and the KL divergence (see \Cref{sec:model} for the formal definitions).\\

Across the five utility models we tested, no single model consistently explained the choices of most participants: for each model, at least one pairwise comparison yielded fewer than $25\%$ of participants with at least $90\%$ consistency. KL and cosine similarity performed better than the other models, but even these models were highly consistent with only about one fifth to one quarter of participants in their weakest comparison.
(see \autoref{sub:comparing_two_utility_models} for complete results).

Following these negative results, we developed pair-generation algorithms for checking consistency with more general properties. Specifically, we checked whether subjects' utility functions are single-peaked, star-shaped, or peak-linear (see \Cref{sub:monotonicity-properties}). 
In these polls, the results were more positive: almost 90\% of the users replied consistently with star-shaped or single-peaked utilities, and  almost 80\% replied consistently with peak-linear utilities (a stronger condition than star-shaped). \\

Next, we aimed to check whether humans' utility functions are consistent with any $\ell_p$ metric. All $\ell_p$ metrics possess two types of symmetry: (1) \emph{Sign Symmetry} --- adding $x$ and subtracting $x$ from the ideal allocation contribute the same amount to the distance; (2) \emph{Issue Symmetry} --- adding $x$ to different issues contributes the same amount to the distance. For each symmetry type, we developed a pair-generation algorithm that tests whether subjects' utility functions exhibit this type of symmetry. Our results here were, again, negative: less than 10\% of the subjects showed at least 90\% symmetry in both respects (see \Cref{sub:symmetry} for more details).

Further analyses using generalized $\ell_p$ metrics with issue-specific weights or sign-specific weights revealed very limited consistency: fewer than 20\% of participants were fully consistent with issue-specific weights, and none with sign-specific weights. 
We also checked a satisfaction-based model, recently introduced by \citet{gourves2025satisfactory},  by which agents' utility is determined by the number of issues funded by at least their ideal amount. We found only limited support for this model, as about half the responses contradict it.

Note that for all reported proportions, we computed 95\% Wilson confidence intervals to quantify statistical uncertainty due to finite sample sizes.\\

Overall, these findings suggest that simple symmetric or weighted asymmetric $\ell_p$ metrics, as well as the satisfaction-based model, are insufficient to fully capture human preference patterns, highlighting the need for more flexible utility models.  
Detailed results for all of these properties and their analyses can be found in \Cref{sub:consistency-in-asymmetry}.

Finally, we wanted to check whether humans' utility functions are consistent with any norm-based metric, and particularly, whether they satisfy the triangle inequality. This turned out to be the most challenging check, as the triangle inequality involves a sum of two distances, $\|x+y\| \leq \|x\|+\|y\|$. To cope with this challenge, we asked the subjects to compare \emph{biennial budgets}. We conducted a preliminary poll, in which we found out that about 60\% of the subjects compare biennial budgets in a way that is consistent with adding utilities (\Cref{sub:biennial}). Among these subjects, we conducted another poll which checked whether their replies are consistent with the triangle inequality. A large majority of the subjects' replies were \emph{contrary} to the triangle inequality, i.e., they preferred the sum of distances $\|x\|+\|y\|$, to the sum $\|x+y\| + 0$. 

Taken together, our results indicate that utility functions based on metrics, particularly metrics that are symmetric with respect to sign and issue (such as $\ell_p$), are not very good for modeling human budget preferences. However, most humans' utility functions do belong to more general classes such as star-shaped or single-peaked or peak-linear. 
Future work could focus on these more general classes, and try to detect within them, the sub-classes that better fit actual utility functions. 

\section{Related Work}
Participatory Budgeting (PB) enables citizens to directly influence how public funds are allocated. What began as a social innovation has evolved into a computational problem of combining individual preferences. The central challenge is to aggregate individual preferences into a single collective decision that is efficient, truthful, and fair. 

\subsection{Utility models in Participatory Budgeting}
\subsubsection{Discrete participatory budgeting}
Most practical PB instances are based on 
\emph{project selection}: each project has a fixed cost, and voters simply indicate which projects they support, effectively casting binary yes/no votes. Participants do not control the exact level of funding; instead, aggregation rules determine which subset of projects is implemented.
See \citet{RSM25} for a recent survey of this setting.

In this setting, too, there are various assumptions regarding the voters' utility functions (also known as \emph{satisfaction functions}). The most common ones are: count-based utilities (a voter's utility is the number of supported projects that are funded), and cost-based utilities (a voter's utility is the total cost of supported projects that are funded). Intermediate utility models (such as the square-root of cost) are also studied \citep{faliszewski2018framework}. We are not aware of direct experiments testing which of these utility models, if any, reflects humans' real preferences.
The closest one we know of is by \citet{rosenfeld2021what}. They presented indirect evidence in favor of the count utilities: in several scenarios, they computed the utilitarian-optimal budget-allocation (the allocation that maximizes the sum of utilities) under five different utility models, and asked the subjects to choose which of the five resulting allocations they prefer. Most subjects prefer the budget that was utilitarian according to count-utilities.

\subsubsection{Continuous participatory budgeting}
In addition to the discrete PB model, a continuous PB model has also been studied, in which voters cast approval or cardinal ballots. 
This model was studied under the term \emph{fair mixing} \citep{aziz2019fair}. Later, the model was extended to the setting in which each voter is a donor, the budget is made of donations, and the goal is to coordinate the donations in an efficient and fair way \citep{brandl2021distribution,brandl2022funding,brandt2025coordinating}.
In this setting, too, different assumptions on the agents' utility functions lead to substantially different results. For example, when agents are assumed to have \emph{additive utilities} across issues, there is no rule that simultaneously satisfies Pareto-efficiency, truthfulness, and a very weak fairness requirement \citep{brandl2021distribution}. However, when agents are assumed th have \emph{Leontief utilities}, the Nash product rule attains strong versions of all these properties \citep{brandt2025coordinating}.
We are not aware of any empirical study in this setting.

\citet{garg2018iterative} propose the Iterative Local Voting (ILV) mechanism for voting in continuous spaces. In ILV, preferences are elicited dynamically through bounded local updates under different norms, with theoretical guarantees of convergence to socially optimal or median-based outcomes under structured utility assumptions. Empirical evidence shows that $\ell_\infty$-based updates yield particularly stable convergence and suggest decomposable utilities and presence of indifference regions.  

\citet{suksompong2026voting} presents a recent comprehensive survey of the different models and algorithms used for continuous PB, both in the project-selection model and in the budget-aggregation model.

\subsection{Empirical Research in Participatory Budgeting}
Empirical studies of PB distinguish between the \emph{frontend} which is the interface and elicitation format used to collect preferences, and the \emph{backend} --- the aggregation rule that combines these inputs. This distinction is useful for classifying experimental findings, as both issues shape voter experience and collective outcomes.

\subsubsection{Data Elicitation Formats}
The elicitation format strongly influences expressiveness, cognitive effort, and aggregation quality. Prior work has studied formats such as Knapsack Voting \citep{goel2019knapsack}, Cumulative Voting \citep{skowron2020participatory}, k-Approval Voting, Threshold Approval, and others \citep{benade2018efficiency,fairstein2023pbrealworld}. These studies highlight the trade-off between usability and expressiveness: simpler formats reduce cognitive effort but capture coarser preferences. 

For example, \citet{skedgel2013choosing} compares Discrete Choice Experiments (DCE) with Constant-Sum Paired Comparisons (CSPC). In their study, which focused on healthcare resource allocation, participants were asked to distribute a fixed budget among various health interventions or to choose between pairs of alternatives. The authors found that while DCE captures clear preferences with lower cognitive load, CSPC provides more detailed information about relative priorities and trade-offs between programs. Such findings underscore that elicitation design not only affects participant experience but also determines the quality of the data available for aggregation.
This insight is directly relevant to our poll format, described in \Cref{sec:experiment}, which similarly combines repeated paired comparisons with budget allocation tasks to capture participants’ nuanced priorities.


\subsubsection{Empirical Evaluation of Aggregation Rules}
On the backend, experiments evaluate how different aggregation rules perform in terms of fairness, efficiency, and robustness to strategic behavior. Studied rules include greedy algorithms, Equal Shares (MES) \citep{fairstein2023pbrealworld}, utilitarian aggregation, and the Nash-product rule \citep{rosenfeld2021what}. 
A further support is provided by recent experimental studies, which show how citizens perceive different aggregation rules in terms of fairness and legitimacy, highlighting important trade-offs for practical system design \citep{yang2024designing}.


\subsubsection{Other experiments on utility models}
Assumptions on utility functions are important not only in budget aggregation, but also in many other fields of economics and social choice. 

For example, in auction design, it is common to assume that the bidders have quasi-linear utilities (utility = item value minus monetary payments). 
\citet{castillo2023general} presents a laboratory experiment that provides some support for quasilinear utilities.
However, \citet{vasserman2021risk} present empirical evidence showing that quasi-linearity might not hold, due to risk-aversion effects.
\citet{bajari2005structural} also find that risk-aversion models are better at generating estimates of bidders' valuations.
These findings require to adapt the standard auction-design tools to accommodate for risk-averse bidders \citep{baisa2019efficient}.
Another related literature studies preferences over lotteries, where
decision-makers compare alternatives that are represented by probability
distributions over possible outcomes. In this literature, utility functions
are used to represent preferences between different distributions, for
example under expected-utility models \cite{cherchye2022revealed, feldman2022revealing}.
Our setting has a similar formal structure in that participants compare different distributions of a fixed resource. However, rather than comparing lotteries over uncertain outcomes, we ask participants to compare deterministic budget allocations across issues. The vector components represent the amounts assigned to different issues rather than probabilities of different
outcomes.

The most practical guide we have found so far for utility elicitation is the book by \citet{keeney1993decisions}, which presents a framework for eliciting people's preferences over multiple objectives, in the context of individual decision-making.

\section{Model and Notation}
\label{sec:model}
In a \emph{budget allocation} problem, 
there is a set $A$ of $m$ \emph{alternatives} (also called \emph{issues} or \emph{projects}).
The total budget is denoted by $B$. The set of all possible budget allocations is the simplex 
\begin{align*}
\Delta(B) := \{\mathbf{q}\in \mathbb{R}^m ~|~ \mathbf{q}\geq \mathbf{0} \text{ and } \sum_{j\in A} q_j = B \}.
\end{align*}
In our polls, we always assume $B=100$, meaning ``100\%'' (in other words, the numbers in our polls are interpreted as a percentage of the total budget). Hence, we represent the set of possible budget allocations simply by $\Delta$.
~
We assume that each person has a preference ranking over $\Delta$, which can be represented by a utility function $u: \Delta \to \mathbb{R}$.
We further assume that $u$ can be presented as $u(\other) = U(\peak,\other)$, where ---
\begin{itemize}
    \item $\peak$ is an \emph{ideal budget allocation} (also called the \emph{peak}) --- a unique vector in $\Delta$ which the person thinks is the best way to allocate the budget of $B$ among the $m$ issues.
    \item $U$ is a \emph{utility model function} --- a function from $\Delta\times\Delta$ to $\mathbb{R}$, that represents the utility of an agent with ideal budget allocation $\peak$ when the actual allocation is $\other$.
\end{itemize}

Whereas typically each person has a different utility function, we believe that different people may have similar utility \emph{model} functions; these are the functions we study in the present research.
Some common utility models are:
\begin{itemize}
    \item \emph{$\ell_1$ disutilities}: 
    $U(\peak,\other) = -\sum_{j\in A}|p_j-q_j|$;
    \item \emph{$\ell_2$ disutilities}: 
    $U(\peak,\other) = -\sqrt{\sum_{j\in A}(p_j-q_j)^2}$;
    \item \emph{$\ell_r$ disutilities}, for any $r\geq 1$: 
    $U(\peak,\other) = -(\sum_{j\in A}|p_j-q_j|^r)^{1/r}$ ($\ell_1$ and $\ell_2$ are special cases);
    \item \emph{Leontief utilities}:
    $U(\peak,\other) = \min_{j \in A} (\frac{q_j}{p_j})$;
    \item \emph{Kullback-Leibler divergence}:
    $U(\peak, \other) = - \sum_{j \in A} p_j \cdot \ln \left( \frac{p_{j}}{q_j} \right)$;
    \item \emph{Cosine Similarity}:
    $ U(\peak,\other) = \frac{\peak\cdot \other}{\|\peak\|_2\,\|\other\|_2}$.
\end{itemize}
KL divergence requires $q_j>0$ whenever $p_j>0$; our pair-generation algorithm for this model enforces strictly positive coordinates.

\section{Experimental Setup}
\label{sec:experiment}
\paragraph{Poll-generation framework}
We constructed a modular framework that lets one generate polls by combining several components:
\begin{itemize}
    \item \emph{Scenario} --- a textual description of the institution whose budget is allocated. We used national-government and municipal-budget scenarios.
    \item \emph{Issues} --- a list of $m$ issues among which the budget should be allocated (e.g., government ministries, municipal departments). In most polls we had $m=3$ issues, to reduce the cognitive burden to a minimum while keeping the problem multi-dimensional (the setting with $m=2$ is essentially one-dimensional). 
    In the pairwise-comparison polls we found out that with $m=3$ issues some models are too similar to be distinguished, so we used $m=4$.
    For comparison, we also generated polls with $m=5$.
    \item \emph{Pair-generation algorithm} --- a custom algorithm that takes as input the subject's ideal budget and returns a list of pairs. Each pair-generation algorithm is carefully designed to test specific properties of utility functions. \Cref{sec:poll-results} describes the various algorithms in detail.
    \item \emph{User filter} --- a custom filter that decides which users are suitable for a particular poll. In most polls, the filter only required that the ideal budget assigns positive amounts to at least two issues (as ideal budgets assigning everything to a single issue are degenerate and do not allow meaningful comparison of utility models). Some polls needed a stronger filter --- see \Cref{sec:poll-results}  for details.
    \item \emph{Language} --- all polls are available in English, but can be easily translated to the subjects' native language.
\end{itemize}
See \Cref{app:poll_description} for screenshots of the user interface, and 
\Cref{app:software_architecture} for a detailed system description and guidelines for reproducibility.

Each poll used several basic measures against behavioral biases:
\begin{itemize}
\item The order of vectors in each pair was randomized to avoid a primacy effect.
We also filtered out participants who consistently chose only the first or second option in all questions.
\item Each poll contained two \emph{attention checks}: two pairs in which one of the vectors was identical to the subject's ideal budget. Subjects who did not choose their ideal budget in one of these checks were filtered out of the results, as we suspected that they probably answered randomly, or just did not read the question correctly.
\item Subjects were forced to choose one option in each pair; there was no indifference option. This was intended to avoid the ``lazy'' choice of claiming that all non-ideal budgets are equally bad.
\item The pair-generation algorithms generally avoided allocations with
zero-valued coordinates, to reduce potential behavioral effects associated
with zero outcomes \citep{incekara2021zero,ludwig2024zero}.
\end{itemize}

\paragraph{Conducting the polls}
We conducted a total of 67 polls, comprising 5162 participant-poll observation. Subjects were recruited by Panel4All, a well-established Internet panel company with a large pool of participants that is frequently used for political opinion polling. At our request, the company 
aimed to recruit a demographically representative sample for each poll.
Participants were compensated by the polling company, based on their estimation of the time it should take to complete each poll.

Each poll included attention checks designed to identify participants who did not read the questions carefully or answered without sufficient attention. Of the 5162 participant-poll observations, 3239 passed the attention checks and were retained for analysis, whereas 1923 failed at least one attention check and were excluded (see \Cref{tab:attention_failures}. 
About 10\% of the subjects took part in more than one poll; these were evaluated separately in each poll. Therefore, the reported totals for ``participants'' refer to participant-poll pairs rather than to unique individuals.

The Internet panel company compensated participants with points redeemable for money, with the amount determined by the estimated time required to complete each poll. Participants who failed the attention checks received reduced compensation and were prevented from participating in subsequent polls. This policy provided an incentive to read the questions carefully and answer attentively. Since the polls elicited subjective preferences, however, participants could not be directly incentivized to answer "truthfully". As is standard in public-opinion research, we therefore relied on the assumption that participants wished to express their genuine preferences regarding public-budget allocations.

Importantly, all sample sizes and consistency measures reported throughout the paper are based exclusively on participant-poll observations that passed the attention checks. For each poll, participants who failed the screening procedure were excluded before any consistency measures or summary statistics were computed. Consequently, the sample size reported for each individual poll is the number of participants who passed the attention check. 

\section{Individual polls: algorithms and results}
\label{sec:poll-results}
In this section we describe in detail the pair-generation algorithms we used in each poll, as well as the poll results.
In the main paper we describe the algorithms informally and provide the main results; in the appendices we provide complete pseudo-code for each algorithm, as well as illustrative examples, and complete tables of results.

\subsection{Distribution of peak allocations}
\label{sub:distribution-of-peak-allocations}
Before examining the specific pairwise-comparison polls, we first analyze the distribution of participants' peak allocations across budgets with three, four, and five categories. Among the three-category budgets, the most frequent peak allocation is $[40,30,30]$ ($445$ responses), followed by $[50,25,25]$ ($284$), $[50,30,20]$ ($176$), and $[50,20,30]$ ($146$). The allocation $[60,20,20]$ is also relatively common, appearing in $131$ responses. Together, these results indicate a clear concentration around structured allocations in which one category receives a moderately larger share of the budget, while the remaining categories receive equal or nearly equal shares.

Additional common three-category allocations include $[40,40,20]$, $[35,35,30]$, $[35,30,35]$, $[30,40,30]$, $[30,35,35]$, and $[40,20,40]$. For four-category budgets, the most common peak is the equal allocation $[25,25,25,25]$ ($128$ responses), followed by $[40,20,20,20]$ ($96$) and $[50,20,20,10]$ ($58$). For five-category budgets, the equal allocation $[20,20,20,20,20]$ appears in $57$ responses (See \Cref{tab:peak_distribution} in \cref{app:distribution-of-peak-allocations} for a full breakdown).

A related phenomenon is the $1/N$ rule \citep{BenartziThaler2001}, according to which individuals tend to allocate resources evenly across available options, reflecting a preference for balanced and simple decisions.

\section{Individual polls: algorithms and results}
\label{sec:poll-results}
This section summarizes the main experiments and results. Pair-generation procedures are described at a high level. Complete pseudo-code, examples, and result tables are provided in \Cref{app:comparing_two_utility_models}.

We computed 95\% Wilson confidence intervals, which we denote in the text by CI[low,high].

\subsection{Comparing specific utility models}
\label{sub:comparing_two_utility_models}
In the first set of polls, we assumed, following theoretical works in social choice, that agents' utility models are one of $\ell_1$, $\ell_2$, Leontief, KL or Cosine (see \Cref{sec:model} for the formal definitions).
We aimed to check which of these five utility models is more prominent.
We conducted all $\binom{5}{2}=10$ pairwise comparisons between these models.

\subsubsection{Pair-generation algorithm}

Given two utility models $U_1$ and $U_2$, generate feasible budget allocations and identify pairs on which the two models induce opposite
rankings. To make these disagreements sufficiently salient, utilities are normalized to $[0,1]$, and each pair is scored by the smaller absolute difference between the two normalized utilities. The $k=10$ highest-scoring pairs are selected.
See \Cref{app:comparing_two_utility_models} for complete pseudo-code, implementation details and examples.


\subsubsection{Results}

Our original conjecture was that each participant would be consistent with one of the common utility models --- possibly a different model for each participant. If this were the case, we could use our polling framework in specific participatory budgeting instances as a preliminary step to determine the most common model in the relevant population, and use it to inform algorithm development.
~
Our results, however, did not support this conjecture. 
Across all $10$ four-issue model-comparison polls, only 41.8\% of participant-poll observations (CI $[0.38 , 0.45]$)) were consistent with at least one of the two models tested in that poll.
\footnote{
Because comparisons were deliberately selected to maximize disagreement between models, observed consistency should be interpreted as performance on model-discriminating comparisons rather than as general predictive accuracy.
}
Thus, 58.2\% of participant-poll observations were inconsistent with both models in the corresponding comparison.

Interestingly, among the $41.2\%$ consistent participants, the highest consistency rate was observed for Cosine similarity and for KL divergence: consistency with Cosine similarity was observed in 110/708 participants and consistency with KL divergence was observed in 64/708 participants. The rates were much lower for the other three models. This motivates further study of Cosine similarity and KL divergence as utility models for budget aggregation.

We next examined whether the observed inconsistencies could be explained by different types of choice error. We considered two classes of errors
(see 
\Cref{tab:summary_preferences_4} and \Cref{tab:noise_threshold_new_consistency_4} for analysis and confidence intervals).

\paragraph{Inattention errors.}
Such errors are captured, for example, by the trembling-hand model, in which the probability of choosing the inferior alternative is constant and does not depend on the strength of the underlying preference \citep{selten1975}. This type of fixed error remains a standard way of modeling behavioral variability in contemporary cognitive models \citep{zilker2022choice,seitz2025disentangling}. In particular, \citet{zilker2022choice} explicitly characterizes the trembling-hand choice rule as a constant-error model, and \citet{seitz2025disentangling} identify response-selection errors, including trembling-hand errors, as a source of behavioral variability.

We re-analyzed our polls allowing $k\in\{1,2,3,4\}$ inconsistent answers out of the ten questions in each poll. Thus, a participant is considered consistent with a model if at least $10-k$ of the answers conform to its predictions. 
Naturally, the number of consistent participants increases with $k$ (see
\Cref{tab:summary_preferences_4}, \Cref{app:comparing_two_utility_models}), but the increase is slow. For example. when allowing one mistake, only additional 10\% (71 participants) of the subjects were fully consistent with one of the models; $\approx 48\%$ of the participants were not even 90\% consistent with any of the five models.%
\footnote{
We use 90\% as a descriptive threshold for summarizing high consistency, rather than as a statistically validated cutoff for model adequacy. We therefore also report the underlying response-level consistency rates.
}

\paragraph{Utility noise.}
Such errors are captured, for example, 
by Weber's law \citep{shigemoto2003weber}, which relates
discriminability to relative rather than absolute differences.
To assess whether apparent inconsistencies can be explained by small
differences in predicted utility, we re-examine participant--poll observations with a dominant preference for one of the two compared models, excluding observations with an equal number of choices for both models. For each pair
of allocations $p$ and $q$, let $u_{\mathrm{best}}$ and $u_{\mathrm{chosen}}$ denote the utilities of the allocation preferred by
the tested model and the allocation chosen by the participant, respectively, where utility is computed according to the tested model.
For a model-inconsistent choice, we define the relative noise
\[
\epsilon :=
\frac{u_{\mathrm{best}}-u_{\mathrm{chosen}}}{u_{\mathrm{best}}}.
\]
We consider thresholds of $5\%$, $10\%$, and $15\%$.%
\footnote{
For thresholds of 20\% or higher, some of the polls contained only few questions with utility difference above the threshold, so we could not get meaningful conclusions for them.
}
A choice is considered consistent when the chosen alternative is within the corresponding threshold of the higher-utility alternative.

As can be seen in \Cref{tab:noise_threshold_new_consistency_4}, allowing up to 15\% noise does not substantially change the consistency analysis: over $50\%$ of participants remain inconsistent even when constrained to pairs with more than 15\% utility difference. 

These findings motivate the exploration of alternative utility models that may better capture the heterogeneity observed in participants' choices.

\subsection{Checking basic monotonicity properties}
\label{sub:monotonicity-properties}
Besides the five models we have tested in \Cref{sub:comparing_two_utility_models}, one could think of many other specific utility models to test. 
In order to narrow the search space, we have decided to design different kinds of polls, that check for more general properties of utility functions. 
We focused on three issues in the monotonicity tests for simplicity and ease of interpretation, as we did not observe any substantive differences in the results when using four or five issues.

We checked three basic properties:

\paragraph{1. Star-shaped}
 A utility-model function $U$ is called \emph{star-shaped} if the utility of an allocation  
strictly decreases as the allocation moves away from the agent's ideal allocation ("peak") in any direction. Formally \citep{brandt2025optimal}, 
for any distribution $\other \ne \peak$ and for all $\lambda \in (0,1)$,
\[
U(\peak,\peak) > U(\peak, \lambda \peak + (1 - \lambda) \other) > U(\peak, \other).
\]

Each of the following two properties is stronger than star-shaped.

\paragraph{2. Multi-dimensional single peaked (MDSP)}
Let $\other_1$ and $\other_2$ be two alternative distributions.  
We say that 
$\other_2$ is \emph{closer to $\peak$ than} $\other_1$ if for every issue $j$, either $q_{1j} \geq q_{2j} \geq p_{j}$ 
or $q_{1j} \leq q_{2j} \leq p_{j}$, and for at least one issue $j$, the inequality between 
$q_{1j}$ and $q_{2j}$ is strict.
A utility-model function $U$ is said to be \textit{multi-dimensional single-peaked} if whenever  
$\other_2$ is closer to $\peak$ than $\other_1$, it holds that $U(\peak,\other_2) > U(\peak,\other_1)$.
In \Cref{MDSP_is_star_shaped} we prove that MDSP is indeed stronger than star-shaped.

\paragraph{3. Peak-linear}
Peak-linearity captures the idea that moving halfway toward one’s ideal budget should yield exactly half the gain in utility compared to moving all the way.
\citet{brandt2025optimal} define a utility function as \emph{peak-linear} if for any distribution $\other$ and $\lambda \in [0,1]$,
$
U(\peak, \lambda \peak+(1-\lambda)\other) = \lambda U(\peak,\peak) + (1-\lambda) U(\peak,\other).
$
Their definition relies on the numeric value of the utility, which we have no way to measure. Hence we give a more general definition, which relies only on ordinal comparisons.
    We say that a utility function is \emph{peak-linear} if for any two distributions $\other_1,\other_2$ and $\lambda \in [0,1]$,
\begin{align*}
    U(\peak, \lambda \peak+(1-\lambda)\other_1) \geq 
    U(\peak, \lambda \peak+(1-\lambda)\other_2)
    \qquad
    \iff
    \qquad
    U(\peak, \other_1) \geq U(\peak,\other_2).
\end{align*}
It is easy to check that $\ell_p$ metrics are peak-linear according to both definitions. 
However, a utility function such as $\sum_j (p_j-q_j)^2$ is peak-linear according to our definition and not according to the definition in \citet{brandt2025optimal}, although such a utility function is clearly equivalent to $\ell_2$.

In \Cref{peak_linear_is_star_shaped} we prove that peak-linear is stronger than star-shaped.
However, peak-linearity and MDSP are independent --- none of them implies the other (See \Cref{app:monotonicity-properties-proofs} for proofs).

Leontief utilities are peak-linear too.
KL utilities are multi-dimensional single-peaked, but not peak-linear (see \Cref{app:monotonicity-properties-proofs} for full proofs).
Cosine similarity is star-shaped (\Cref{cosine_star_shaped}) but not peak linear (\Cref{cosine_peak_linear}).

\subsubsection{Pair-generation algorithms}
We present a pair-generation algorithm for each of the three monotonicity properties.

\paragraph{1. Star-shaped}
The algorithm generates $k=10$ pairs, one for each weight $\lambda \in \{0.1, 0.2, \dots, 0.9\}$. To standardize the experiment length and ensure exactly 10 questions per participant, the midpoint weight $\lambda = 0.5$ is tested twice using independently generated random vectors. For each $\lambda$, the algorithm generates a random budget-allocation vector $\other$, computes the convex combination $\other_{\lambda} := \lambda \peak+(1-\lambda)\other$, and adds $(\other,\other_{\lambda})$ to the list of pairs.
The pseudo-code for the algorithm is provided in \Cref{app:convex_algorithm}, and an example appears in \Cref{tab:examplecalc}, both in in \Cref{app:monotonicity-properties}.

For most values of $\lambda$, the convex combination has entries that are not multiples of $5\%$. We suspected that this might create some mental bias.
Therefore, we implemented a variant of this algorithm, which rounds all entries in the convex combination to the nearest multiple of $5$. Note that the rounded vectors are not exact convex-combinations anymore, but they are close to convex combinations. The algorithm and an example can be found in \Cref{round} and in \Cref{tab:round_example}.

\paragraph{2. Multi-dimensional single-peaked}
Initially we used a simple random-search algorithm: (1) Generate a random pair $(\other_1,\other_2)$; (2) If $\other_1$ is closer to $\peak$ than $\other_2$ or vice-versa, then add $(\other_1,\other_2)$ to the list of pairs; (3) Repeat until $k$ pairs have been added. This approach may result in a relatively high rejection rate, since most randomly generated pairs do not satisfy the required directional-closeness conditions. We therefore considered an alternative construction procedure that generates only valid MDSP pairs. The procedure first samples a random allocation $\other_{\text{far}}$, identifies one issue whose budget exceeds the participant's ideal allocation and one issue whose budget falls short of it, and then transfers a weighted fraction of the maximal feasible budget amount between these two issues. The resulting allocation $\other_{\text{near}}$ is therefore guaranteed to be weakly closer to the participant's ideal allocation in every issue and strictly closer in at least one issue. Finally, allocations are rounded to multiples of five and rebalanced to preserve the budget constraint. This construction eliminates the need for rejection sampling while maintaining the intended MDSP structure (see full pseudo-code in \Cref{alg:mdsp} in \Cref{app:monotonicity-properties}).

\paragraph{3. Peak-linear}
The pair-generation algorithm for peak-linearity is more involved, as peak-linearity requires utilities to be equal, and we cannot query numeric utility values using our pairwise-comparison framework. 
We coped with this issue in the following way.
(1) The algorithm generates three pairs comparing three extreme vectors. To avoid zero-bias, we chose ``extreme vectors'' without zero coordinates: $\extreme_1 = [10, 10, 80], \quad \extreme_2 = [10, 80, 10], \quad \extreme_3 = [80, 10, 10]$.
(2) For each weight $\lambda\in\{0.25, 0.5, 0.75\}$, the algorithm generates three convex combinations $\other_i := \lambda \peak + (1-\lambda) \extreme_i$ for $i\in\{1,2,3\}$, and adds the three pairwise comparisons among them.
This algorithm yields $12$ pairs overall. 
A subject with a peak-linear utility function should rank $\other_i$ and $q_j$ exactly the same as $\extreme_i$ and $\extreme_j$, for every $\lambda$.
The pseudo-code for this algorithm, as well as an illustrating example, appears in \cref{alg:peak-linear} in \Cref{app:monotonicity-properties}.

\subsubsection{Results}

\paragraph{Star-shapedness.}
$40$ of $84$ participants ($47.6\%$) showed $100\%$ consistency with star-shapedness.
Under the \textbf{Inattention errors} model, a further $23$ participants ($27.4\%$) were $90\%$ consistent. Together, $63$ participants ($75.0\%$)
were at least 90\% consistent with the star-shapedness comparisons included in this poll (CI $[0.65,0.83]$).

Under the \textbf{utility noise} model, 
since we do not assume here a specific utility model, we calculate the noise using the following relative-distance measure:
\[
\epsilon=\frac{d_{\ell_1}(\other_{\mathrm{far}},\other_{\mathrm{near}})}{d_{\ell_1}(\other_{\mathrm{far}},\peak)}.
\]
Allowing noise recovers relatively few additional participants: only $10$ participants become consistent at $20\%$ ($50$ participants, $59.5\%$).

We also examined the total fraction of answers consistent with star-shapedness
(\Cref{tab:consistency_by_weight}, \Cref{app:monotonicity-properties}). Overall, 88\% of the answers were consistent. For $\lambda=0.1$, overall consistency was somewhat lower than average ($77.46\%$ of answers); this is expected, since for such a small $\lambda$ the convex combination is very close to the random vector, and some participants may treat the two as equivalent. For larger values of $\lambda$, consistency remained high, though we did not observe a clear upward trend as $\lambda$ increased.
Both this finding and the per-subject analysis indicate that the noise-dependent model is not very good in explaining deviations from star-shapedness.


\paragraph{MDSP}
$66$ of $87$ participants ($75.9\%$) showed $100\%$ consistency with MDSP.
Under the \textbf{Inattention errors} model, a further $13$ participants ($14.9\%$) were $90\%$ consistent, so $79$ participants ($90.8\%$)
achieved at least 90\% consistency with the tested MDSP comparisons.

Under the \textbf{utility noise} model, computing the noise in a similar way to star-shaped, 
no additional participants became consistent at $5\%$ or $10\%$ noise; $1$ more became consistent at $15\%$ and $2$ more at $20\%$, yielding $69$ fully consistent participants ($79.3\%$) at the $20\%$ threshold - an increase of only $3$ participants over the noise-free baseline.


\paragraph{Star-shapedness vs. MDSP: different behavioral properties.}
Although MDSP may initially appear to be a stronger restriction, the two properties capture different aspects of preference behavior in our setting, and generate different types of movements in the budget space. In the star-shapedness test, the allocation moves toward the participant's ideal point, so that all coordinates typically change simultaneously as the allocation is transformed. In the MDSP test, by contrast, one coordinate is held fixed while the remaining budget is redistributed across the other coordinates. Higher consistency with MDSP should therefore not be read as evidence that MDSP is simply a weaker version of star-shapedness; rather, the two tests measure distinct behavioral properties.

\paragraph{Do utility noise errors explain the deviations?}
Comparing the two error models across both star-shapedness and MDSP suggests that the noise-dependent model does not, on its own, account for most of the observed deviations. Under a genuine noise-dependent process, errors should concentrate at small margins between alternatives, and allowing modest noise should recover most participants who fall just short of full consistency. Our results are not fully consistent with this pattern: for star-shapedness, the $23$ participants at exactly $90\%$ consistency are not necessarily explained by the noise-dependent analysis, since their single inconsistent response may involve a comparison with a relatively large margin. For MDSP, allowing noise up to $20\%$ adds only $3$ participants beyond those already perfectly consistent. Taken together, these results indicate that the observed errors are not primarily driven by the closeness of the alternatives, making the noise-independent error model the more natural description of the remaining deviations.

\paragraph{Peak-linearity.}
Our peak-linearity experiments use $\lambda\in\{0.25,0.5,0.75\}$, which yields vectors with large differences; hence, we did not apply the utility noise model here. Instead, we examine consistency, transitivity, and the stability of participants' induced preference ordering.

The design of this poll lets us reconstruct each participant's complete preference ordering separately for four groups of three alternatives: the core group of extreme allocations, and the three groups of transformed allocations corresponding to $\lambda=0.25$, $0.5$, and $0.75$. We use three complementary measures:

\begin{itemize}
    \item \emph{Overall consistency} is the fraction of all nine transformed pairwise comparisons (three values of $\lambda$, three pairs each) that agree with the ordering predicted by peak-linearity. For example, a participant who agrees with the predicted ordering in $1$ of $3$ comparisons at $\lambda=0.25$, $1$ of $3$ at $\lambda=0.5$, and $2$ of $3$ at $\lambda=0.75$ has an overall consistency of $\frac{1+1+2}{9}=\frac{4}{9}\approx44\%$.
    \item \emph{Transitivity} is evaluated separately within each of the four groups: a participant's choices are transitive if they can be represented by a complete ordering (e.g., preferring $A$ to $B$ and $B$ to $C$ implies preferring $A$ to $C$). A violation occurs when the three choices form a cycle, such as $A\succ B$, $B\succ C$, and $C\succ A$. \Cref{tab:intransitivity_example} shows an example of such a cycle drawn from poll 3.
    \item \emph{Order consistency} is stronger than transitivity: it requires that the same complete ranking be preserved across the core group \emph{and} all three transformed groups. A participant can be perfectly transitive within every group while still ranking the alternatives differently across groups. For instance, the rankings
    \[
    C\succ B\succ A,\qquad A\succ C\succ B,\qquad A\succ C\succ B,\qquad B\succ C\succ A
    \]
    are each individually transitive — yielding perfect transitivity — yet only two of the four groups induce the same ordering, so this participant exhibits only partial order consistency.
\end{itemize}

\begin{table}[H]
\centering
\small
\caption{Example of an intransitive preference cycle from poll 3.}
\label{tab:intransitivity_example}
\begin{tabular}{cccc}
\toprule
\textbf{Pair} & \textbf{Option 1} & \textbf{Option 2} & \textbf{Chosen Allocation} \\
\midrule
1 & $[42,32,26]$ & $[25,50,25]$ & $[25,50,25]$ \\
2 & $[42,32,26]$ & $[25,32,43]$ & $[42,32,26]$ \\
3 & $[25,50,25]$ & $[25,32,43]$ & $[25,32,43]$ \\
\bottomrule
\end{tabular}
\end{table}

Among the $76$ participants, $69$ ($90.8\%$, CI $[0.81,0.95]$) exhibited perfect transitivity in every group. For overall consistency, $24$ participants were fully consistent with peak-linearity across all three values of $\lambda$, and a further $22$ were consistent at two of the three values; together these $46$ participants make up $60.5\%$ of the sample (CI $[0.49,0.71]$). Order consistency - the strongest of the three measures - held for only $26$ participants ($34.2\%$, CI $[0.24,0.45]$).

These results reveal a distinction between local and global consistency: most participants made transitive choices \emph{within} each group, but far fewer preserved the \emph{same} ordering across transformation levels. Intransitive behavior was relatively rare, while changes in the relative ranking of alternatives across values of $\lambda$ were considerably more common. Complete results are reported in \Cref{tab:peak_linear_ci}, \Cref{app:monotonicity-properties}.

Average agreement with peak-linearity also increased with $\lambda$: consistency was $74\%$, $83\%$, and $85\%$ for $\lambda=0.25$, $0.5$, and $0.75$ respectively, averaging $81\%$ overall across all values of $\lambda$ and all pairwise comparisons. Under our construction, larger $\lambda$ places more weight on the extreme vectors, so the transformed alternatives are more clearly separated and better preserve the ordering of the original extremes - making consistent ranking easier for participants. \Cref{tab:inconsistency} gives an illustrative example of an inconsistent response pattern, and complete results are reported in \Cref{tab:peaklinear_consistency}.

\paragraph{Results for 4 and 5 issues for MDSP}
Increasing the number of issues from 3 to 4 or 5 does not lead to a meaningful change in overall consistency: in all three settings, a clear majority of participants exhibit very high consistency, with a large fraction reaching $100\%$. A mild trend does emerge as more dimensions are added, however: higher-dimensional settings introduce a small number of participants with intermediate consistency levels (e.g., $50\%$, $60\%$, or $70\%$) that do not appear in the 3-issue case - a pattern consistent with a modest increase in cognitive difficulty when evaluating higher-dimensional budget vectors, even among attentive participants. See \Cref{tab:mdsp_results_topics} in \Cref{app:monotonicity-properties}, for complete results.

\subsection{Checking symmetry}
\label{sub:symmetry}
In the next set of polls, our aim was to check whether people's utility functions are compatible with \emph{any} $\ell_p$ metric.
The $\ell_p$ metrics are \emph{symmetric} in the sense that
only the size of the deviation matters --- not where in the budget it occurs, nor whether it represents an increase or decrease in funding. This symmetry consists of two independent properties:
\begin{enumerate}
	\item \emph{Issue Symmetry} --- Adding $x$ to issue $i$ is equivalent to adding $x$ to issue $j$, for all issues $i,j$;
	\item \emph{Sign Symmetry} --- Adding $x$ in issue $i$ is equivalent to subtracting $x$ from $i$, for all issues $i$.
\end{enumerate}
A numeric example is given in \Cref{tab:symmetry-example} in the appendix.

\subsubsection{Pair-generation algorithms}
We designed an algorithm for each of the two symmetry properties. Both algorithms follow the same idea: construct sets $S_1,S_2$ of vectors, such that the vectors in each $S_i$ should have the same utility according to the tested symmetry property; then compare vectors in $S_1$ with corresponding vectors in $S_2$. A subject whose utility function satisfies the tested symmetry property should rank all pairs $\other_1\in S_1$ and $\other_2\in S_2$ in the same way.

\begin{enumerate}
	\item For Issue Symmetry, the sets $S_i$ were constructed in the following way. (1) Choose a random budget-allocation vector $\other$. (2) Compute the difference vector $\diff := \other-\peak$. (3) Compute all $m-1$ rotations of $\diff$ (e.g. for the distance-vector $[20,-15,-5]$, the rotations would be $[-5,20,-15]$ and $[-15,-5,20]$).
	(4) For each rotated distance vector $\diff'$, add 	$\diff'+\peak$ to the set $S_i$. 
	We constructed $4$ pairs of sets; each set-pair yielded $m=3$ pairwise vector comparisons, for a total of $12$ pairs per poll.
	See \Cref{project_symmetry_algorithm} for pseudo-code.
	
	\item For \textit{Sign Symmetry}, 
	the sets $S_i$ were constructed as follows. (1) Choose a random budget-allocation vector $\other$. (2) Compute the difference vector $\diff := \other-\peak$. (3) Add both $\other =  \diff+\peak$ and $\other' = -\diff+\peak$ to the set $S_i$. 
	We constructed $6$ pairs of sets; each set-pair yielded $2$ pairwise vector comparisons, for a total of $12$ pairs per poll.
	See \Cref{sign_symmetry} for pseudo-code.
\end{enumerate}

\begin{example}
	\label[example]{exm:issue-symmetry}
Suppose the ideal budget is $\peak = [30, 30, 40]$.
The pair-generation algorithm for Issue Symmetry could generate the following pairs.
\begin{table}[H]
	\centering
	\small
	\begin{tabular}{c|c|c}
		\toprule
		& $\other_1$ & $\other_2$ \\
		\midrule
		Original & [50, 34, 16] & [20, 25, 55] \\
		Deviation & [20, 4, -24] & [-10, -5, 15] \\
		Rotated Deviation 1 & [-24, 20, 4] & [15, -10, -5] \\
		Rotation 1 & [6, 50, 44] & [45, 20, 35] \\
		Rotated Deviation 2 & [4, -24, 20] & [-5, 15, -10] \\
		Rotation 2 & [34, 6, 60] & [25, 45, 30] \\
		\bottomrule
	\end{tabular}
\end{table}
\end{example}

\paragraph{Example of issue inconsistency}
An example of project inconsistency is the following, given the ideal budget $[30, 30, 40]$:\\
\begin{table}[H]
\centering
\small
\begin{tabular}{c|c|c|c|c|c}
\toprule
Pair & Option A & Deviation A & Option B & Deviation B & Participant Choice \\
\midrule
Pair 1 & [66, 5, 29] & [+36, -25, -11] & [6, 60, 34] & [-24, +30, -6] & A \\
Pair 2 & [5, 19, 76] & [-25, -11, +36] & [60, 24, 16] & [+30, -6, -24] & B \\
\bottomrule
\end{tabular}
\label{tab:project_incon}
\end{table}

This example demonstrates issue-level inconsistency. In both pairs, Option A exhibits the same deviation pattern, differing only by a permutation of deviations across issues. Likewise, Option B also follows an identical deviation pattern across pairs, again differing only in the assignment of deviations to specific issues.

Despite this structural equivalence, the participant chooses Option A in the first pair and Option B in the second pair. That is, for the same deviation patterns, applied to different issues, the participant’s preference reverses.

\subsubsection{Results}
Our results for both symmetry properties were negative (See \Cref{tab:sign_consistency} and \Cref{tab:issue_consistency} in \Cref{app:symmetry} for complete results):
\begin{enumerate}
	\item In the Issue Symmetry poll, only $11\%$ of the subjects (8 out of 73)  answered consistently in all four groups; only an additional $22\%$ answered consistently in three groups (CI $[0.23 , 0.44]$).
	\item In the Sign Symmetry poll, 3 subjects answered consistently in all six groups, and 14 more out of 61 subjects answered consistently in five groups (CI $[0.18, 0.4]$).
\end{enumerate}

~
This inconsistency suggest that the $\ell_p$ model, as well as any other utility model that treats issues symmetrically or treats increases and decreases symmetrically, may not adequately represent people's preferences.

\subsubsection{Is the asymmetry in issues caused by asymmetry in amounts?}
The asymmetry among issues could be explained in two ways: (1) People assign different values for cuts in different issues (e.g. a cut of 10 in Defense is different than a cut of 10 in Education). 
(2) People assign different values for cuts in different initial amounts (e.g. a cut of 10 in an ideal budget of 20 is different than a cut of 10 in an ideal budget of 30).
The second explanation would imply the following slight generalization of the $\ell_p$ utility model:
$U(\peak,\other) = \sum_{j=1}^m D(p_j, |p_j - q_j|)^r$, where $D$ is a ``deformation function'' that modifies the difference $|p_j - q_j|$ based on the initial amount.

To test this possibility, we conducted an additional targeted poll. In this poll, we filtered and retained only participants whose allocations assigned \emph{identical amounts} to two different issues. The budget of the third issue was fixed at its original value, while the two identical budgets were systematically varied across alternatives.
The formal procedure used to construct the comparison pairs is described in \Cref{alg:issue-symmetry-with-identical-allocations} in \Cref{app:symmetry}.

As an example, a subject with an ideal budget of $[40,30,30]$ could be asked to compare $[40,15,45]$ with $[40,45,15]$. A subject with a ``deformation-based'' utility model would be indifferent between these two vectors. In general, such a subject would be indifferent between adding $x$ to issue 2 and subtracting $x$ to issue 3, and vice-versa. Hence, over 10 pairs, such a subject's answers would be close to random (near $50\%$ percent supporting an increase in issue 2).

The actual results were quite different (can be seen in \Cref{asymmetric_results} in \Cref{app:symmetry}): Out of the $31$ participants, $16$ ($51\%$) exhibited more than $90\%$ consistency (CI $[0.35,0.68]$), including $13$ participants who demonstrated perfect ($100\%$) consistency across all questions. In contrast, only $6$  ($19\%$) participants showed low consistency levels, with less than $60\%$ consistent choices. 
These results suggest that subjects' decisions are influenced not only by the magnitude of budget changes but also by the specific issue being modified. In particular, the observed asymmetric preferences indicate that symmetric utility models, such as $\ell_1$, may be insufficient to fully capture participants' behavior in this setting, even when accounting for a "deformation effect".

\subsection{Checking Consistency in Asymmetry}
\label{sub:consistency-in-asymmetry}
Following the negative results of \Cref{sub:symmetry}, we checked whether subjects' utility functions are consistent with a generalization of an $\ell_p$ metric, which allows asymmetry in issues or signs.
\begin{enumerate}
	\item A utility-model function such as $U(\peak,\other) = \sum_{j=1}^m a_j \cdot |p_j - q_j|^r$ exhibits a consistent asymmetry between issues, represented by the issue-specific weights $a_j$;
	\item A utility-model function such as $U(\peak,\other) = \sum_{j=1}^m a \cdot \max(p_j - q_j,0)^r + b\cdot \max(q_j - p_j,0)^r$ exhibits a consistent asymmetry between signs, represented by the sign-specific weights $a,b$;
\end{enumerate}

\subsubsection{Pair-generation algorithms}
We designed an algorithm for each of the weighted utility models. 

\paragraph{1. Issue-specific weights}
The algorithm picks a positive value $x$ and generated $m=3$ difference vectors varying by rotation, namely $\diff_1 = [2x, -x, -x]$ and $\diff_2 = [-x, 2x, -x]$ and $\diff_3 = [-x, -x, 2x]$. The weighted utility corresponding to $\diff_1$ is $a_1 (2 x)^r + a_2 (x)^r + a_3 (x)^r = x^r\cdot  (a_1 2^r + a_2 + a_3)$. Similarly, the weighted utility corresponding to $\diff_2$ is $x^r\cdot  (a_1 + 2^r a_2 + a_3)$, and 
	the weighted utility corresponding to $\diff_3$ is $x^r\cdot  (a_1 + a_2 + 2^r a_3)$. Hence, the ranking between these three difference vectors should be the same regardless of $x$.
In other words, if a subject prefers a concentrated increase in issue 1 ($[2x,-x,-x]$) to a concentrated increase in issue 2 ($[-x,2x,-x]$), then the same should hold for any $x$.

	We generated 4 triplets of vectors, corresponding to two positive and two negative values of $x$. These values were selected such that all resulting vectors would have all-positive coordinates.
	
\begin{figure}[H]
    \centering
    \fbox{\includegraphics[width=0.8\textwidth]{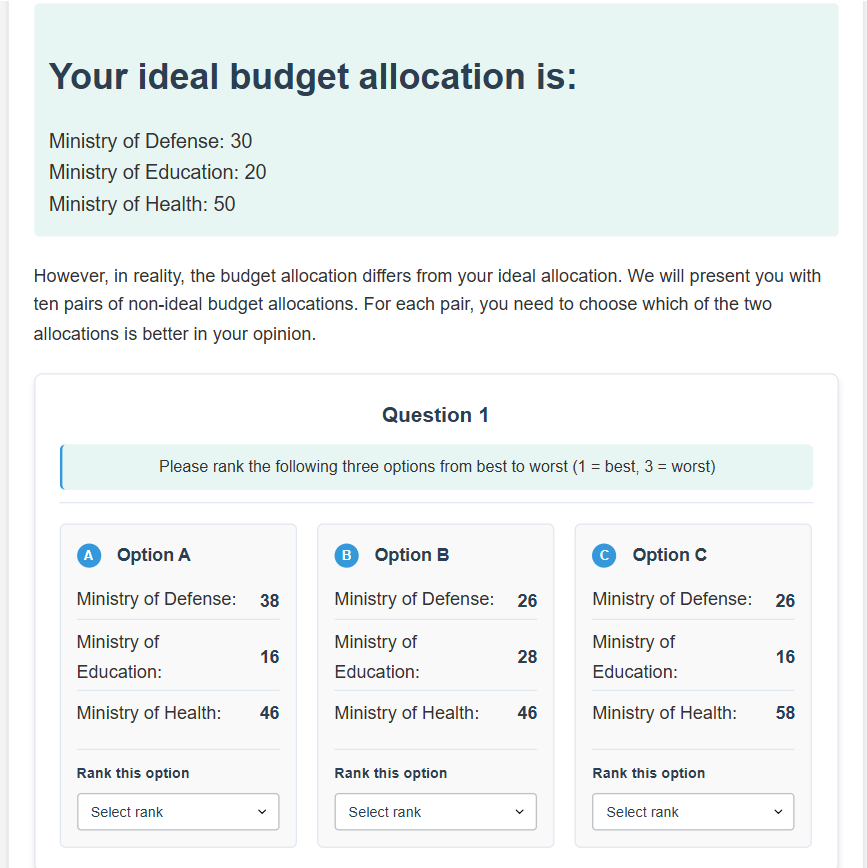}}
     \Description{The poll's ranking task. After viewing their ideal budget, the participant is presented with three different budget allocations (Options A, B, and C) and must rank them from 1 (best) to 3 (worst).}
    \caption{A question where participants are required to rank the three options.
    \label{fig:rank}}
\end{figure}

	Instead of asking the subjects three questions per triplet (one for each pairwise comparison), we decided it was simpler to ask them to rank the three vectors; see \Cref{fig:rank} for GUI example.
	See \Cref{alg:consistency-in-issue-asymmetry} for the pseudo-code and  \Cref{tab:generated_options} for a numeric example of the resulting allocations.
	
\paragraph{2. Sign-specific weights}
The algorithm picks a positive value $x$, and generated two difference vectors varying by sign, namely $\diff_1 = [x, x, -2x]$ and $\diff_2 = [-x, -x, 2x]$.
	The weighted utility corresponding to $\diff_1$ is $a (2 x)^r + b (x)^r + b (x)^r = x^r\cdot  (2^r \cdot a + 2 b)$. Similarly, the weighted utility corresponding to $\diff_2$ is $x^r\cdot (2^r \cdot b + 2 a)$. Hence, subjects with this utility model should rank this pair in the same way for all $x$. 
	We generated $6$ pairs of vectors, corresponding to two different values of $x$ and $m=3$ rotations of the difference vectors, for a total of $12$ pairwise comparisons.
See \Cref{alg:consistency-in-sign-asymmetry} for the pseudo-code.
Note that the algorithm can fail to find valid budget-allocation vectors, particularly when an ideal budget is close to an extreme. For these cases we have a fallback procedure, which is detailed in \Cref{alg:fallback}.
		
In this poll, participants who allocated a budget of zero to any issue were excluded and prevented from proceeding to the comparison questions, as the algorithm requires non-zero values for all the issues to generate valid pairs.
	

\subsubsection{Results}
\paragraph{1. Issue-specific weights}
In this poll, a total of 65 participants took part. 
Consistency with a utility-model with issue-specific weights would imply
that, for each pair of $i,j$ of issues (1 vs 2, 2 vs 3, 3 vs 1),
the subject would consistently prefer a concentrated increase in $i$ to a concentrated increase in $j$, or consistently prefer the other way around.
In fact, Only $12$ (less than $20\%$ of the subjects) showed a consistent ranking among all three pairs.
See \Cref{consistent_number} for complete results.

\begin{figure}[H]
    \centering
        \caption{A participant's preference matrix showing full consistency choices for each row.
    \label{fig:full_consis}}
    \includegraphics[width=0.8\textwidth]{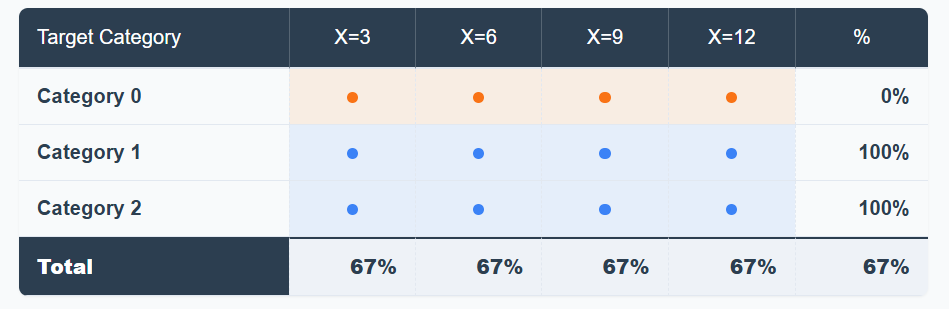}
\end{figure}

\begin{figure}[H]
    \centering
        \caption{A participant's preference matrix showing consistent choices for category 1 and 2 but inconsistent yet monotonic choices for category 0 across different budget change magnitudes.
    \label{fig:mono}}
    \includegraphics[width=0.8\textwidth]{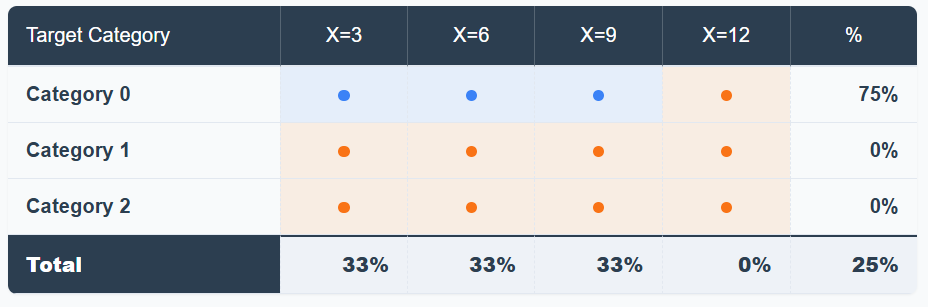}
\end{figure}

\begin{figure}[H]
    \centering
        \caption{A participant's preference matrix showing consistent choices for Defense, but inconsistent choices for Education and Health across different budget change magnitudes.
    \label{fig:not_mono}}
    \includegraphics[width=0.8\textwidth]{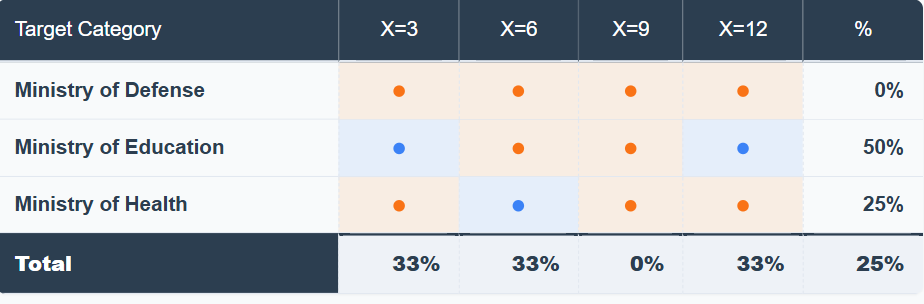}
\end{figure}

\paragraph{2. Sign-specific weights}
For each participant, we represented the responses using a preference matrix. The rows of the matrix correspond to the three issues, and the columns correspond to four different magnitudes of budget change. Each cell indicates whether the participant preferred the concentrated change or the distributed change for the corresponding issue and magnitude. Examples of such preference matrices are shown in \Cref{fig:full_consis,fig:mono,fig:not_mono}.

A utility model with fixed sign-specific weights would require a participant to make the same choice between the concentrated and distributed alternatives across all three issues and all four magnitudes. However, none of the 65 participants exhibited such full consistency across all 12 comparisons. See \Cref{tab:concentrated_vs_distributed} for the complete results.

\subsubsection{Row-wise consistency and monotonicity}

Although no participant was fully consistent across all 12 comparisons, participants' responses often exhibited a structured pattern within each issue. We therefore examined whether, for each issue separately, the participant's preference between the concentrated and distributed alternatives remained stable or changed monotonically as the magnitude of the budget change increased.

First, 15 of the 65 participants exhibited \emph{row-wise consistency}: for each of the three issues, they made the same choice across all four magnitudes. Their preferred type of change could differ between issues, but within each issue it remained constant. Thus, these participants were not necessarily fully consistent across the entire matrix, but each of their rows was internally consistent.

An additional 20 participants exhibited \emph{row-wise monotonicity}. For each issue, their preference changed at most once as the magnitude increased, and for at least one issue it changed exactly once, from a concentrated change to a distributed change or vice versa. These participants therefore displayed a threshold-like response pattern rather than an arbitrary sequence of choices.

Combining these two groups, 35 out of 65 participants ($53.8\%$) exhibited a monotonic pattern in every issue: their preference either remained constant across all four magnitudes or switched direction at most once. The remaining 30 participants ($46.2\%$) exhibited a non-monotonic pattern in at least one issue, changing their preference more than once as the magnitude increased.

These findings suggest that, although fixed sign-specific weights do not adequately describe the participants' responses, a more flexible model incorporating issue-dependent thresholds may capture a substantial portion of the observed behavior.

\subsubsection{Satisfaction-based model}
\citet{gourves2025satisfactory} present a \emph{satisfaction-based}  utility model. According to their model, the utility of an agent with ideal budget $\peak$ from actual budget $\other$ is determined by the number of issues $j$ for which $q_j\geq p_j$. 
Such a user would always prefer a large decrease and two small increases, over a large increase and two small decreases.
Our results provide only weak support for this model: In our poll, the overall summary of participants’ choices showed that 50.5\% of responses corresponded to concentrated decreases, while 49.5\% corresponded to distributed increases. 

\subsubsection{An even more general utility model}
In this section, we tested utility model functions that are asymmetric in sign or in issue, but not both. A more general utility-model function, that allows both types of asymmetry, is   $U(\peak,\other) = 
\sum_{j=1}^m
[
a_j \cdot \max(0, p_j-q_j)^r
+
b_j \cdot \max(0, q_j-p_j)^r
]$.
Currently, we do not know how to test if subjects' utility functions are consistent with this form. We leave this question to future work.

\subsection{Biennial Budgets and the Triangle Inequality}
\label{sub:biennial}
Our motivation for the next poll was to test whether participants’ preferences correspond to \emph{any} metric. One of the defining properties of a metric is the \emph{triangle inequality}, which says that, for any three points $A,B,C$,  the distance from $A$ to $B$ plus the distance from $B$ to $C$ is at least as large as the distance from $A$ to $C$. In terms of utility (the negative of distance), this would imply that
\begin{align}
	\label{eq:triangle-inequality-1}
U(\other_A,\other_B) + U(\other_B,\other_C) \leq U(\other_A,\other_C).
\end{align}
Unfortunately, we cannot check \eqref{eq:triangle-inequality-1} directly in our framework, as we can only compare utilities computed with respect to a single fixed vector for each subject (the subject's peak $\peak$). 
Therefore, we focus on \emph{norm-based metrics}.

Recall that a \emph{norm} is a function from a vector space to $\mathbb{R}_+$, usually denoted by $\|\cdot \|$, that satisfies three conditions:
(a) Homogeneity: $\|s \cdot x\| = |s|\cdot \|x\|$ for every vector $x$ and scalar $s$;
(b) Positiveness: $\|x\| = 0$ if and only if $x=0$;
(c) Triangle inequality: for every two vectors $x$ and $y$,
$
\|x + y\| \leq \|x\| + \|y\|.
$
A metric is called \emph{norm-based} if there exists some norm $\|\cdot\|$ such that the distance between every two point $A$ and $B$ is equal to the norm of the difference vector, $\|A-B\|$.
Every $\ell_p$ metric is norm-based (based on the so-called \emph{$\ell_p$ norm}).

Suppose the utility-model function $U$ is based on a norm-based metric with norm $\|\cdot\|$. Then $U(\peak,\other) = -\|\peak-\other\|$ for each vector $\other$.
Let $\diff_A, \diff_B$ be two difference-vectors (vectors whose components sum up to $0$), and let $\diff_C := \diff_A+\diff_B$. 
Let $q_j = \peak + \diff_j$ for all $j\in\{A,B,C\}$. Then $U(\peak, q_j) = -\|\diff_j\|$. The triangle inequality implies that
$\|\diff_A\| + \|\diff_B\| \geq \|\diff_C\|$. This implies the following for utilities:
\begin{align}
\label{eq:triangle-inequality-2}
	U(\peak, \other_A) + U(\peak, \other_B) \leq U(\peak, \other_C).
\end{align}
Inequality \eqref{eq:triangle-inequality-2} is more convenient to test than \eqref{eq:triangle-inequality-1}, as it involves only utilities with respect to the same peak $\peak$. However, it still requires to compute a sum of utilities.

We consider the problem of comparing sums of utilities a major challenge for future research, and do not claim to have an adequate solution for it.
We present a preliminary idea for addressing this challenge.

The idea is to consider the budget over two consecutive years. Suppose the budget in year A is $\other_A$ and the budget in year B is $\other_B$. If a subject evaluates each year independently of the other year, then the subject's total utility from the two years will be the sum $U(\peak, \other_A) + U(\peak, \other_B)$.

The assumption of independence between years is a strong one. An alternative reasonable assumption is that a subject consider consecutive years to be \emph{complementary}: if in year A the budget deviated from $\peak$ to one direction, then the subject would prefer the budget in year B to deviate from $\peak$ in the opposite direction, so that the two-year average equals the ideal budget $\peak$.

Therefore, before actually testing the triangle inequality, we conducted a preliminary experiment in which we compared the above two assumptions: independence (implying additivity) versus complementarity.

\subsubsection{Biennial budgets: independent or complementary?}
The preliminary poll consisted of a simple repeated choice task. In each task, participants were asked to choose which budget they preferred for the current year, while the not-chosen budget would be automatically allocated in the subsequent year.

This poll contains $12$ questions generated by three distinct algorithms. 
The order of the questions follows a cyclic pattern with respect to the type of algorithm.
\paragraph{Sub-poll 1:}
In this setting, the subject chooses between receiving their ideal budget in the first year (and a random budget in the second year) or receiving the ideal budget in the second year (and the same random budget in the first year). This sub-poll is not directly related to the question of independence vs. complementarity, but rather comes to check whether there are systematic present-preferences or future-preferences.
\paragraph{Sub-poll 2:}  
Here, the first-year budget is fixed in advance. The participant chooses the budget for the second year: 
either their exact ideal budget, or an alternative budget such that the two-year average (between the first- 
and second-year budgets) equals the participant's ideal budget.
\paragraph{Sub-poll 3:}  
In this case, the second-year budget is fixed in advance. The participant chooses the budget for the first year: 
Either their exact ideal budget or an alternative budget that ensures the two-year average is equal to the participant's ideal budget.
The poll generation algorithm and an example of questions for a participant are provided in \Cref{alg:biennial}.

Our results here are mixed (see \Cref{tab:subsurveys_results} for detailed results).

In sub-polls 2 and 3, over 60\% of the subjects (27 and 26 out of 39) consistently preferred to get their ideal budget, rather than a balancing budget. This indicates that, at least for these subjects, a biennial budget can be used to compare sums of utilities.

On the other hand, in sub-poll 1, over 60\% of the subjects (24 out of 39) consistently (in all 4 questions) preferred their ideal budget to be implemented in year 1. This hints that utilities in different years are not additive using a simple sum, but may be additive using a weighted sum.


\subsubsection{Triangle Inequality}
\label{triangle-inequality}
Based on these results, we want to examine the primary objective of the triangle inequality.
We construct a new poll consisting of 14 questions. The purpose of the first two questions is to filter out participants who balance the budgets across the two years --- that is, the poll will only include individuals who do \textbf{not} balance between years (meaning they choose their ideal budget in one year and a random budget in the other, rather than selecting a random budget in one year and a budget that balances to the average in the other).


After restricting the poll to these participants, we present the remaining 12 questions of the poll. 
Each question asks to compare two biennial budgets:
\begin{itemize}
    \item \textbf{Option 1:} A \emph{concentrated change}, where the entire deviation from the ideal budget occurs within a single year.
    \item \textbf{Option 2:} A \emph{distributed change}, where the same overall deviation is divided evenly between the two years, so that each year deviates only partially from the ideal.
\end{itemize}

The algorithm tests the triangle inequality by presenting choices that compare a single concentrated change to a split change.  
Let \(\diff_C\) be a difference-vector and let us decompose it as \(\diff_C=\diff_A+\diff_B\). The subjects are asked to compare the following two biennial budgets
$
(\peak, \peak+\diff_C)  \text{ vs. } (\peak+\diff_A, \peak+\diff_B).
$ The reply reveals whether
\[
\|0\| + \|\diff_C\| \le \|\diff_A\|+\|\diff_B\|,
\]
which is exactly the triangle inequality.
~
To demonstrate, assume the ideal budget is $\peak = [30,30,40]$ and the base difference vector is
$
\diff_C = [-20, 10, 10]
$, which is decomposed as $\diff_C=\diff_A+\diff_B$, where 
$
\diff_A = [-10,\ 10,\ 0], 
\diff_B = [-10,\ 0,\ 10].
$
Then:
\begin{itemize}
	\item Option 1 (concentrated change) is:
$
\text{Year 1: } \peak = [30,30,40], 
\text{Year 2: } \peak + \other_C = [10,40,50].
$
	\item Option 2 (distributed change) is:
$
\text{Year 1: } \peak + \other_A = [20,40,40],
\text{Year 2: } \peak + \other_B = [20,30,50].
$
\end{itemize}

The generation procedure, including the precise algorithm, is provided in Appendix~\ref{appendix:triangle_algorithm}.  
Among the 88 participants, 44 ($50\%$) exhibited consistency levels of at least $75\%$. Across these participants, $72.9\%$ of the choices favored the \emph{distributed change}, compared with $27.1\%$ favoring the concentrated change. The preference for distributed changes was observed at every reported consistency level: $65.6\%$ among participants with $75\%$ consistency, $83.3\%$ at $83.3\%$ consistency, and $91.7\%$ at $91.7\%$ consistency. Among fully consistent participants, the distributed option was still preferred in $60\%$ of the choices.
Only $17$ out of $88$ participants satisfied the triangle inequality criterion, corresponding to a 95\% Wilson confidence interval of $[0.12 , 0.29]$.

Full results are in 
\Cref{tab:concentrated_vs_distributed_by_consistency} in \Cref{appendix:triangle_algorithm}.

These results indicate that, in general, most subjects' utility models do \emph{not} satisfy the triangle inequality. Hence, any norm-based metric might not be a good representation of agents' utilities.
Interestingly, the majority shows convexity in preferences over the distances from the ideal budget: two small changes are preferable to one large change.

Finally, as a point of comparison, we note that in the municipal budgeting setting, preferences were approximately evenly split between concentrated and distributed options. A possible explanation is that, when decisions involve less critical domains, participants tend to exhibit more indifferent behavior. They are more willing to accept concentrated budgets, even when this entails substantial losses in specific issues.

\section{Story effects}
\label{sec:story}

Throughout the paper, we focused on national budget allocation, but we also conducted experiments with municipal budget allocation.
The results were qualitatively similar (particularly, a large majority of the experiments subjects showed consistency with star-shaped, MDSP and peak-linear preferences).
However, there were interesting quantitative differences (e.g., in the number of subjects showing Sign Symmetry or Issue Symmetry). 
Some of these differences may be explained by the perception that municipal budgets feel less critical or consequential to participants compared to national budgets.

A more detailed comparison further reveals that the narrative framing can affect not only consistency levels but also the relative preference between utility models. In particular, in the comparison between KL and $\ell_2$ in a 3-issue setting, we observe a reversal in preferences: while in the national-budget context a majority of consistent participants preferred KL, in the municipal context the majority preferred $\ell_2$. At the same time, consistency levels increased substantially in the municipal setting. A similar trend appears in higher-dimensional comparisons between $\ell_2$ and Leontief, where increasing the number of issues leads to higher consistency but a gradual weakening in the preference for $\ell_2$, possibly due to increased cognitive load and a shift toward more balance-oriented evaluations.

These findings suggest that in municipal contexts, participants may be more willing to accept extreme trade-offs (e.g., strongly underfunding certain issues), whereas in national contexts they tend to avoid allocations that severely harm essential domains, favoring more balanced outcomes.
(See \Cref{municipal_vs_gov} for a detailed comparison).

Overall, while the narrative framing leads to some quantitative differences and may induce different underlying utility functions across subjects, the qualitative conclusions of our study remain unchanged: participants’ preferences exhibit similar structural properties and satisfy the same core axioms across both settings.
The question of how exactly the narrative framework of the poll influences participant responses is worth further study and is left for future work.

\section{Discussion and Future Work}
We introduce a systematic polling methodology that uses structured pairwise comparisons to elicit preferences over budget allocations directly. This approach enables us to empirically evaluate a wide range of utility function properties, including distance metrics, symmetry assumptions, and intertemporal preferences, while maintaining methodological rigor and minimizing cognitive burden on participants.


\paragraph{Geographic consideration}
So far, we have only executed the polls with subjects from a single country. Hence, we cannot claim that our results hold globally. 
It is theoretically possible that in some individual city that uses participatory budgeting, most citizens are completely consistent with $\ell_1$ utilities,%
\footnote{
As an anecdote, one of the authors of a paper that assumes $\ell_1$ utilities took our $\ell_1$ vs Leontief poll, and was found out to be 100\% consistent with $\ell_1$ utilities.
}
so that a mechanism based on the $\ell_1$ assumption can safely be used; our polling framework can be used to verify this assumption in each individual city.
The main message of our work is that assumptions on utility functions can and should be tested empirically.

\paragraph{Demographic Considerations}
It is possible that demographic characteristics of participants, such as age, education level, or political orientation, influence how they interpret budget scenarios or perceive similarity between allocations. 

\paragraph{Other properties of utility models}
Various other properties of utility-model functions, besides the ones studied here, could possibly be tested. 
For example, it is possible that subjects care most only about the ranking of allocations among different issues (which issue gets the largest amount, the second-largest, etc.). Testing such conjectures using our framework requires developing new pair-generation algorithms.

Our work invokes the following question about the limitation of our pairwise-comparison framework, which we find interesting also from a theoretical-mathematical perspective:
\begin{quote}
	\emph{What properties of utility-model functions can be checked by pairwise comparisons?}
\end{quote}
In parallel, it is interesting to explore other possible types of questions. For example, it is possible to directly ask the subjects ``why did you prefer option A to option B?''. Such a poll would be harder to analyze automatically, but might yield interesting insights if analyzed manually.

\bibliographystyle{ACM-Reference-Format}
\bibliography{main,file-from-another-project,quasilinearity}

@misc{RSM25,
      title={The (Computational) Social Choice Take on Indivisible Participatory Budgeting}, 
      author={Simon Rey and Felicia Schmidt and Jan Maly},
      journal={arXiv preprint arXiv:2303.00621},  
      year={2025}
}

@article{FREEMAN2021105234,
title = {Truthful aggregation of budget proposals},
journal = {Journal of Economic Theory},
volume = {193},
pages = {105234},
year = {2021},
issn = {0022-0531},
doi = {https://doi.org/10.1016/j.jet.2021.105234},
url = {https://www.sciencedirect.com/science/article/pii/S002205312100051X},
author = {Rupert Freeman and David M. Pennock and Dominik Peters and Jennifer {Wortman Vaughan}},
}

@inproceedings{aziz2019fair,
  title={Fair mixing: the case of dichotomous preferences},
  author={Aziz, Haris and Bogomolnaia, Anna and Moulin, Herv{\'e}},
  booktitle={Proceedings of the 2019 ACM Conference on Economics and Computation},
  pages={753--781},
  year={2019}
}

@inproceedings{brandl2021distribution,
  title={Distribution rules under dichotomous preferences: two out of three ain't bad},
  author={Brandl, Florian and Brandt, Felix and Peters, Dominik and Stricker, Christian},
  booktitle={Proceedings of the 22nd ACM Conference on Economics and Computation},
  pages={158--179},
  year={2021}
}

@article{brandl2022funding,
  title={Funding public projects: A case for the Nash product rule},
  author={Brandl, Florian and Brandt, Felix and Greger, Matthias and Peters, Dominik and Stricker, Christian and Suksompong, Warut},
  journal={Journal of Mathematical Economics},
  volume={99},
  pages={102585},
  year={2022},
  publisher={Elsevier}
}

@article{brandt2025coordinating,
  title={Coordinating charitable donations with Leontief preferences},
  author={Brandt, Felix and Greger, Matthias and Segal-Halevi, Erel and Suksompong, Warut},
  journal={Journal of Economic Theory},
  pages={106096},
  year={2025},
  publisher={Elsevier}
}

@inproceedings{suksompong2026voting,
  title={Voting in Divisible Settings: {A} Survey},
  author={Suksompong, Warut and Teh, Nicholas},
  booktitle={Proceedings of AAAI Conference on Artificial Intelligence (AAAI-2026)},
  year={2026}
}

@inproceedings{freeman2019truthful,
  title={Truthful Aggregation of Budget Proposals},
  author={Freeman, Rupert and Pennock, David M. and Peters, Dominik and Wortman Vaughan, Jennifer},
  booktitle={Proceedings of the 2019 ACM Conference on Economics and Computation},
  pages={751--752},
  year={2019},
  publisher={Association for Computing Machinery},
  address={New York},
  doi={10.1145/3328526.3329557},
  isbn={978-1-4503-6792-9},
  url={https://doi.org/10.1145/3328526.3329557},
  note={arXiv:1905.00457}
}

@article{caragiannis2022truthful,
  title={Truthful Aggregation of Budget Proposals with Proportionality Guarantees},
  author={Caragiannis, Ioannis and Christodoulou, George and Protopapas, Nicos},
  journal={Proceedings of the AAAI Conference on Artificial Intelligence},
  volume={36},
  number={5},
  pages={4917--4924},
  year={2022},
  doi={10.1609/aaai.v36i5.20421},
  issn={2374-3468},
  url={https://ojs.aaai.org/index.php/AAAI/article/view/20421},
  note={arXiv:2203.09971}
}

@misc{freeman2023project,
  title={Project-Fair and Truthful Mechanisms for Budget Aggregation},
  author={Freeman, Rupert and Schmidt-Kraepelin, Ulrike},
  year={2023},
  eprint={2309.02613},
  archivePrefix={arXiv},
  primaryClass={cs.GT}
}

@inproceedings{elkind2023settling,
  title={Settling the Score: Portioning with Cardinal Preferences},
  author={Elkind, Edith and Suksompong, Warut and Teh, Nicholas},
  booktitle={ECAI 2023},
  series={Frontiers in Artificial Intelligence and Applications},
  pages={621--628},
  year={2023},
  publisher={IOS Press},
  doi={10.3233/FAIA230324},
  isbn={9781643684369},
  note={arXiv:2307.15586}
}

@article{intriligator1973probabilistic,
  title={A Probabilistic Model of Social Choice},
  author={Intriligator, M. D.},
  journal={The Review of Economic Studies},
  volume={40},
  number={4},
  pages={553--560},
  year={1973},
  doi={10.2307/2296588},
  jstor={2296588},
  issn={0034-6527}
}

@article{moulin1980strategy,
  title={On strategy-proofness and single peakedness},
  author={Moulin, H.},
  journal={Public Choice},
  volume={35},
  number={4},
  pages={437--455},
  year={1980},
  doi={10.1007/BF00128122},
  issn={1573-7101}
}

@article{goel2019knapsack,
  title={Knapsack Voting for Participatory Budgeting},
  author={Goel, Ashish and Krishnaswamy, Anilesh K. and Sakshuwong, Sukolsak and Aitamurto, Tanja},
  journal={ACM Transactions on Economics and Computation},
  volume={7},
  number={2},
  pages={8:1--8:27},
  year={2019},
  doi={10.1145/3340230},
  issn={2167-8375},
  note={arXiv:2009.06856}
}

@inproceedings{benade2018efficiency,
  title = {Efficiency and Usability of Participatory Budgeting Methods},
  author = {Gerdus Benadè and Nevo Itzhak and Nisarg Shah and Ariel D. Procaccia and Ya’akov Gal},
  booktitle = {Empirical Studies in Participatory Budgeting},
  series = {Proceedings of the PB Conference / Workshops},
  year = {2018},
  publisher = {Unpublished manuscript / Academic preprint},
  pages = {1--8},
  note = {Empirical study with over 1,200 voters comparing input formats for PB},
  url = {https://procaccia.info/wp-content/uploads/2018/03/pb19.pdf}
}

@article{skedgel2013choosing,
  author = {Skedgel, Chris D. and Wailoo, Allan J. and Akehurst, Ron L.},
  title = {Choosing vs. Allocating: Discrete Choice Experiments and Constant-Sum Paired Comparisons for the Elicitation of Societal Preferences},
  journal = {Health Expectations},
  volume = {18},
  number = {5},
  pages = {1227--1240},
  year = {2015},
  doi = {10.1111/hex.12098},
  url = {https://doi.org/10.1111/hex.12098}
}

@article{skowron2020participatory,
  author = {Piotr Skowron and Arkadii Slinko and Stanis{\l}aw Szufa and Nimrod Talmon},
  title = {Participatory Budgeting with Cumulative Votes},
  journal = {arXiv preprint arXiv:2009.02690},
  year = {2020},
  url = {https://arxiv.org/abs/2009.02690}
}

@article{gourves2025satisfactory,
  title={Satisfactory Budget Division},
  author={Gourvès, Laurent and Lampis, Michael and Melissinos, Nikolaos and Pagourtzis, Aris},
  journal={arXiv preprint arXiv:2502.00484},
  year={2025},
  url={https://arxiv.org/abs/2502.00484}
}

@article{fairstein2023pbrealworld,
  title={Participatory Budgeting Design for the Real World},
  author={Fairstein, Roy and Benadè, Gerdus and Gal, Kobi},
  year={2023},
  journal={arXiv preprint arXiv:2302.13316},
  url={https://arxiv.org/abs/2302.13316}
}

@article{rosenfeld2021what,
  title = {What Should We Optimize in Participatory Budgeting? An Experimental Study},
  author = {Rosenfeld, Ariel and Talmon, Nimrod},
  journal = {CoRR},
  volume = {abs/2111.07308},
  year = {2021},
  eprint = {2111.07308},
  archivePrefix= {arXiv},
  primaryClass = {cs.MA},
  url = {https://arxiv.org/abs/2111.07308}
}

@article{deberg2024truthful,
  author = {Mark de Berg and Rupert Freeman and Ulrike Schmidt-Kraepelin and Markus Utke},
  title = {Truthful Budget Aggregation: Beyond Moving-Phantom Mechanisms},
  journal = {arXiv preprint arXiv:2405.20303},
  year = {2024},
  url = {https://arxiv.org/abs/2405.20303}
}

@article{yang2024designing,
  title={Designing Digital Voting Systems for Citizens: Achieving Fairness and Legitimacy in Participatory Budgeting},
  author={Yang, Joshua C. and Hausladen, Carina I. and Peters, Dominik and Pournaras, Evangelos and H{\"a}nggli Fricker, Regula and Helbing, Dirk},
  journal={Digital Government: Research and Practice},
  volume={5},
  number={1},
  pages={1--16},
  year={2024},
  publisher={ACM},
  doi={10.1145/3665332},
  url={https://dl.acm.org/doi/10.1145/3665332}
}

@article{brandl2024natural,
  title={A natural adaptive process for collective decision-making},
  author={Brandl, Florian and Brandt, Felix},
  journal={Theoretical Economics},
  volume={19},
  number={2},
  pages={667--703},
  year={2024},
  publisher={Wiley Online Library}
}

@article{garg2018iterative,
  title={Iterative Local Voting for Collective Decision-making in Continuous Spaces},
  author={Garg, Nikhil and Kamble, Vijay and Goel, Ashish and Marn, David and Munagala, Kamesh},
  journal={Journal of Artificial Intelligence Research},
  volume={64},
  pages={315--355},
  year={2019}
}

@article{brandt2025optimal,
  title={Optimal budget aggregation with star-shaped preference domains},
  author={Brandt, Felix and Greger, Matthias and Segal-Halevi, Erel and Suksompong, Warut},
  journal={Mathematics of Operations Research},
  year={2025},
  publisher={INFORMS}
}

@article{faliszewski2018framework,
	title={A framework for approval-based budgeting methods},
	author={Faliszewski, Piotr and Talmon, Nimrod},
	journal={arXiv preprint arXiv:1809.04382},
	year={2018}
}

@book{keeney1993decisions,
  title={Decisions with multiple objectives: preferences and value trade-offs},
  author={Keeney, Ralph L and Raiffa, Howard},
  year={1993},
  publisher={Cambridge university press}
}

@article{BenartziThaler2001,
  author = {Benartzi, Shlomo and Thaler, Richard H.},
  title = {Naive Diversification Strategies in Defined Contribution Savings Plans},
  journal = {American Economic Review},
  year = {2001},
  volume = {91},
  number = {1},
  pages = {79--98}
}

@article{shigemoto2003weber,
  title={Weber-Fechner's Law and Demand Function},
  author={Shigemoto, K.},
  journal={arXiv preprint physics/0303118},
  year={2003}
}

@article{Selten1975,
  author  = {Selten, Reinhard},
  title   = {Reexamination of the Perfectness Concept for Equilibrium Points in Extensive Games},
  journal = {International Journal of Game Theory},
  volume  = {4},
  number  = {1},
  pages   = {25--55},
  year    = {1975},
  doi     = {10.1007/BF01766400}
}

@article{incekara2021zero,
  author    = {Incekara-Hafalir, Elif and Kim, Eungsik and Stecher, Jack D.},
  title     = {Is the Allais paradox due to appeal of certainty or aversion to zero?},
  journal   = {Experimental Economics},
  volume    = {24},
  number    = {3},
  pages     = {751--771},
  year      = {2021},
  doi       = {10.1007/s10683-020-09678-4}
}

@article{ludwig2024zero,
  author    = {Ludwig, Jonas and Jaudas, Alexander and Achtziger, Anja},
  title     = {The Zero Effect: An Eye-Tracking Study of Affect and Motivation in Risky Choices},
  journal   = {Journal of Behavioral Decision Making},
  volume    = {37},
  number    = {3},
  pages     = {e2400},
  year      = {2024},
  doi       = {10.1002/bdm.2400}
}

@article{zilker2022choice,
  author  = {Zilker, Veronika},
  title   = {Choice Rules Can Affect the Informativeness of Model Comparisons},
  journal = {Computational Brain \& Behavior},
  volume  = {5},
  pages   = {397--421},
  year    = {2022},
  doi     = {10.1007/s42113-022-00142-5}
}

@article{seitz2025disentangling,
  author  = {Seitz, Florian I. and Jarecki, Jana B. and Rieskamp, J{\"o}rg and von Helversen, Bettina},
  title   = {Disentangling Perceptual and Process-Related Sources of Behavioral Variability in Categorization},
  journal = {Perspectives on Psychological Science},
  volume  = {20},
  number  = {3},
  pages   = {590--617},
  year    = {2025},
  doi     = {10.1177/17456916241264259}
}

@article{cherchye2022revealed,
  author  = {Cherchye, Laurens and Demuynck, Thomas and De Rock, Bram and Freer, Mikhail},
  title   = {Revealed preference analysis of expected utility maximization under prize-probability trade-offs},
  journal = {Journal of Mathematical Economics},
  volume  = {99},
  pages   = {102607},
  year    = {2022},
  doi     = {10.1016/j.jmateco.2021.102607}
}

@article{feldman2022revealing,
  author  = {Feldman, Paul and Rehbeck, John},
  title   = {Revealing a preference for mixtures: An experimental study of risk},
  journal = {Quantitative Economics},
  volume  = {13},
  number  = {2},
  pages   = {761--786},
  year    = {2022},
  doi     = {10.3982/QE1694}
}

@article{bajari2005structural,
	title={Are Structural Estimates of Auction Models Reasonable? Evidence from Experimental Data},
	author={Bajari, Patrick and Hortaçsu, Ali},
	journal={Journal of Political Economy},
	volume={113},
	number={4},
	pages={703--741},
	year={2005},
	publisher={University of Chicago Press}
}

@article{vasserman2021risk,
	title={Risk Aversion and Auction Design: Theoretical and Empirical Evidence},
	author={Vasserman, Shoshana and Watt, Mitchell},
	journal={International Journal of Industrial Organization},
	volume={79},
	pages={102780},
	year={2021},
	publisher={Elsevier}
}

@article{castillo2023general,
	title={A general revealed preference test for quasilinear preferences: theory and experiments},
	author={Castillo, Marco and Freer, Mikhail},
	journal={Experimental Economics},
	volume={26},
	number={3},
	pages={673--696},
	year={2023},
	publisher={Cambridge University Press \& Assessment}
}

@article{baisa2019efficient,
	title={Efficient ex post implementable auctions and English auctions for bidders with non-quasilinear preferences},
	author={Baisa, Brian and Burkett, Justin},
	journal={Journal of Mathematical Economics},
	volume={82},
	pages={227--246},
	year={2019},
	publisher={Elsevier}
}

\newpage
\appendix
\onecolumn

\section*{APPENDIX}
\section{Poll Interface and Question Design}
\label{app:poll_description}


\begin{figure}[H]
    \centering
    \fbox{\includegraphics[width=1\textwidth]{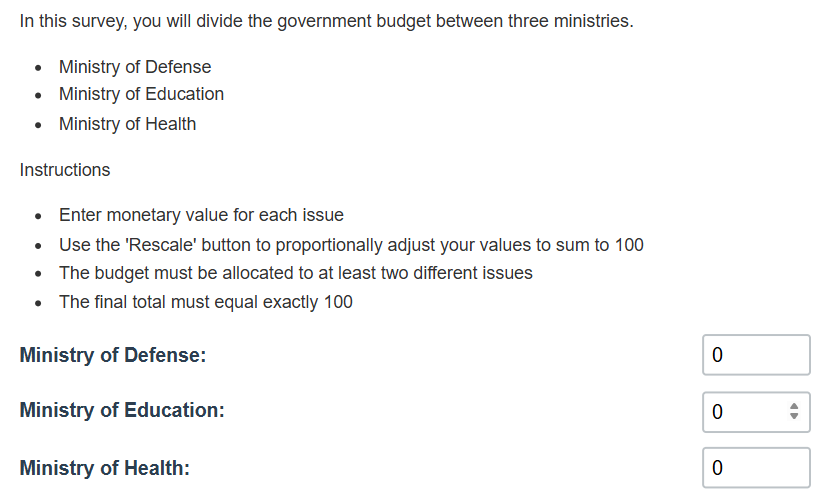}}
    \Description{The poll's initial screen, where participants allocate a budget. It lists three ministries: Defense, Education, and Health, each with a numerical input field, initially set to zero.}
    \caption{Initial screen where participants enter their ideal budget allocation.
    \label{fig:intro}}
\end{figure}

\begin{figure}[H]
	\centering
	\fbox{\includegraphics[width=1 \textwidth]{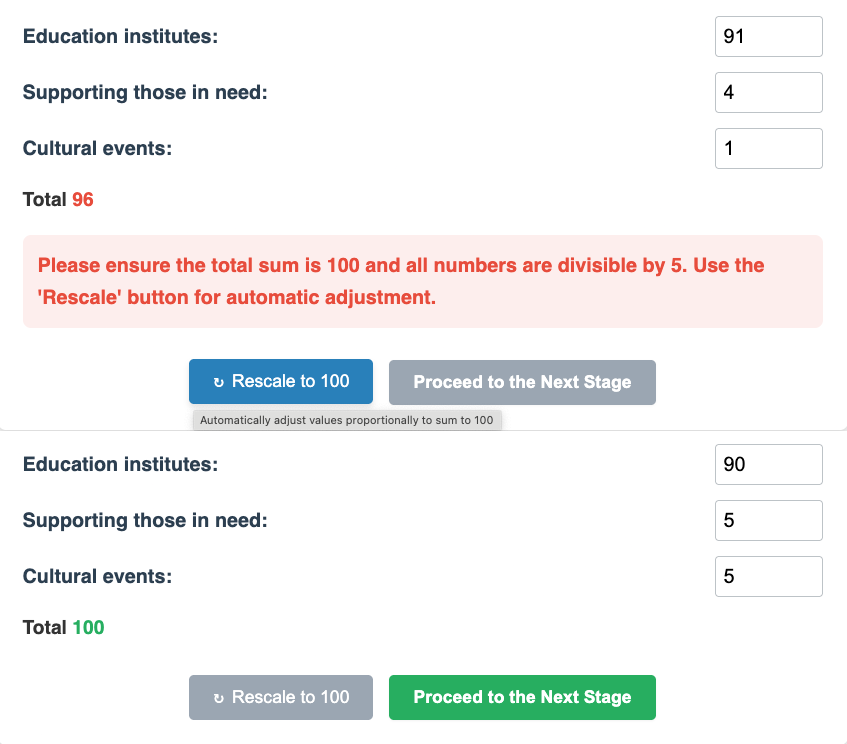}}
	\Description{A before and after image of the 'Rescale' feature. The top panel shows an initial budget entry (91, 4, 1) that incorrectly sums to 96. The bottom panel shows the corrected allocation after rescaling, where the values have been automatically adjusted to [90, 5, 5] to sum to 100.}
	\caption{An example of the automatic budget rescaling feature. \textbf{(Top)} A participant’s initial allocation that does not sum to 100. \textbf{(Bottom)} The allocation after using the ``Rescale'' button, which automatically adjusts the values to meet the poll's constraints while preserving the user's proportional intent.}
	
	\label{fig:rescale}
\end{figure}

\begin{figure}[H]
    \centering
    \fbox{\includegraphics[width=0.8\textwidth]{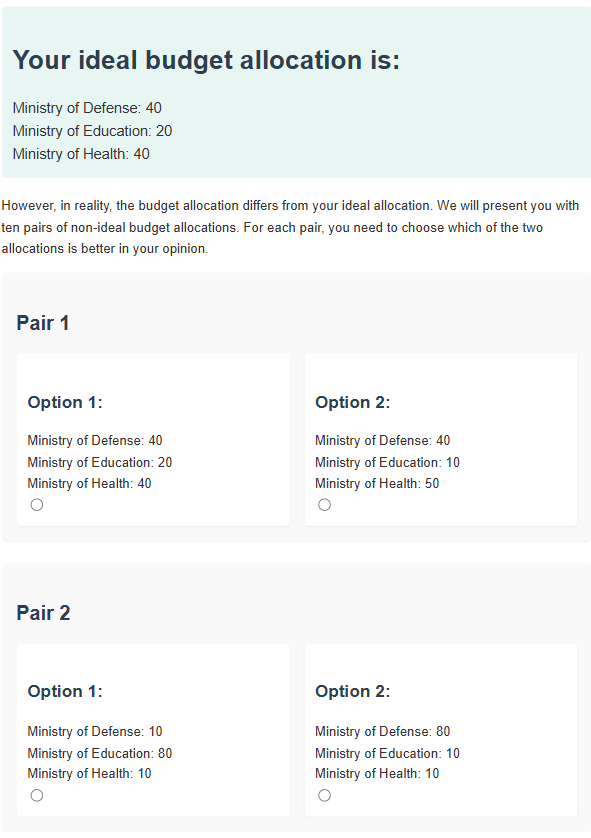}}
    \Description{A poll interface showing a participant's ideal budget for Defense, Education, and Health: [40, 20, 40]. Below this, Pair 1 is an attention check where one option matches the ideal budget. Pair 2 presents a choice between two distinct, non-ideal budget allocations.}
    \caption{Example of a pairwise comparison question, where participants are asked to choose between two alternative allocations.
    Pair 1 is an attention check: Option 1 is identical to the ideal budget, so a user choosing Option 2 will be filtered out. Although in the first attention check question the budget was displayed directly above the question, approximately 45\% of the participants who failed the attention checks failed this particular question.
    \label{fig:pairwise}}
\end{figure}

\section{Appendix to \Cref{sub:distribution-of-peak-allocations}: Distribution of Peak Allocations}
\label[appendix]{app:distribution-of-peak-allocations}

\begin{table}[H]
\centering
\small
\caption{Most frequent peak allocations}
\label{tab:peak_distribution}
\begin{tabular}{lcc}
\toprule
\textbf{Optimal allocation} & \textbf{Dimension} & \textbf{Frequency} \\
\midrule
$[40,30,30]$          & 3 & 445 \\
$[50,25,25]$          & 3 & 284 \\
$[50,30,20]$          & 3 & 176 \\
$[50,20,30]$          & 3 & 146 \\
$[60,20,20]$          & 3 & 131 \\
$[25,25,25,25]$       & 4 & 128 \\
$[40,40,20]$          & 3 & 119 \\
$[40,20,20,20]$       & 4 & 96 \\
$[35,35,30]$          & 3 & 92 \\
$[35,30,35]$          & 3 & 90 \\
$[30,40,30]$          & 3 & 81 \\
$[30,35,35]$          & 3 & 71 \\
$[40,20,40]$          & 3 & 69 \\
$[50,20,20,10]$       & 4 & 58 \\
$[20,20,20,20,20]$    & 5 & 57 \\
\bottomrule
\end{tabular}
\end{table}

\section{Appendix to \Cref{sub:comparing_two_utility_models}: Comparing Specific Utility Models}
\label[appendix]{app:comparing_two_utility_models}

\begin{algorithm}[H]
    \caption{Pair generation for comparing two specific utility models
    \label{alg:algo_comparing_two_utility_models}}
    \KwIn{Utility models $U_1, U_2$; number of projects $m$; ideal budget $\peak$; number of pairs $k$; minimum allocation per project $\ell$}
    
    \KwOut{A set $S$ of $k$ informative comparison pairs}

    $V \gets$ GenerateFeasibleBudgets($m, \ell$)\; 

    \For{each $v \in V$}{
        $u_1(v) \gets U_1(\peak,v)$\; 
        $u_2(v) \gets U_2(\peak,v)$\;
    }

    $r_1 \gets$ RankNormalize($\{u_1(v)\}_{v \in V}$)\; 
    $r_2 \gets$ RankNormalize($\{u_2(v)\}_{v \in V}$)\; \tcp*{Values in $[0,1]$} 

    Initialize empty list $P$\;     

    \For{each pair $(v_i, v_j)$ with $i < j$}{ 
        \If{$r_1(v_i) > r_1(v_j)$ \textbf{and} $r_2(v_i) < r_2(v_j)$}{
            $\text{score} \gets \min\!\big(r_1(v_i)-r_1(v_j),\; r_2(v_j)-r_2(v_i)\big)$\; 
            Add $(v_i, v_j, \text{score})$ to $P$\; 
            }
        \ElseIf{$r_1(v_i) < r_1(v_j)$ \textbf{and} $r_2(v_i) > r_2(v_j)$}{
            $\text{score} \gets \min\!\big(r_1(v_j)-r_1(v_i),\; r_2(v_i)-r_2(v_j)\big)$\; 
            Add $(v_j, v_i, \text{score})$ to $P$\; 
        }
    }

    Sort $P$ by decreasing score\; 
    $S \gets$ first $k$ pairs in $P$\; 

    \Return $S$\;
\end{algorithm}

\begin{itemize}
    \item \textsc{GenerateFeasibleBudgets}$(m,\ell)$ generates the set $V$ of all feasible budget vectors over $m$ issues. Each budget allocates a percentage to every issue such that (i) the total allocation sums to $100$, (ii) each allocation is a multiple of $5$, and (iii) each issue receives at least $\ell$.

    \item \textsc{UtilityByU1}$(v,\peak)$ and \textsc{UtilityByU2}$(v,\peak)$ compute the utility of a budget vector $v$ relative to the agent’s ideal budget $\peak$ according to utility models $U_1$ and $U_2$, respectively.

    \item \textsc{RankNormalize}$(\{u(v)\}_{v \in V})$ takes the utilities of all budget vectors in $V$, ranks them from lowest to highest, and maps these ranks linearly to the interval $[0,1]$. The least-preferred vector receives value $0$, the most-preferred vector receives value $1$, and intermediate vectors are assigned proportionally spaced values. This normalization makes utilities from different models comparable while preserving ordinal preferences.
\end{itemize}

\subsection{Illustrative Example}
\label{utility_models_example}
To illustrate the algorithm, consider a participant whose ideal allocation is \textbf{$\peak = [40, 30, 30]$}. 
They might be presented with the following pair of alternative allocations with Pair Score of 0.34.
Note that Pair Score = min(L1 advantage, Leontief advantage).
Advantage = how much better one vector is than the other in each metric.
Higher score is a clearer choice.

\begin{table}[h!]
\centering
\small
\begin{tabular}{@{}lcc@{}}
\toprule
\textbf{Poll} & \textbf{Allocation A: [55, 35, 10]} & \textbf{Allocation B: [20, 15, 65]} \\ \midrule
$\ell_1$ vs. Leontief & $\ell_1 = 40$, Leontief = 0.5 & $\ell_1 = 70$, Leontief = 0.33 \\
\bottomrule
\end{tabular}
\end{table}

\subsection{Aggregate Preference Results}
\label{tab:summary_preferences}

\Cref{tab:summary_preferences_4} summarizes the results of the primary four-issue model-comparison experiments. This setting includes all ${5\choose2}$ pairwise comparisons among the five utility models and therefore serves as the basis for the main analysis and model ranking presented in the paper. For each comparison, the table reports the distribution of participants according to their exact consistency level with each model.
Note that in the Cosine Similarity vs. $\ell_1$ poll, the sample includes 134 participants because the polling company recruited more participants than requested.

\begin{table}[H]
\centering
\small
\caption{Summary of participant preferences by model comparison and consistency threshold, for \textbf{four} issues.}
\label{tab:summary_preferences_4}
\resizebox{\textwidth}{!}{%
\begin{tabular}{lccccccc}
\toprule
\textbf{Comparison} & \textbf{100\%} & \textbf{90\%} & \textbf{80\%} & \textbf{70\%} & \textbf{60\%} & \textbf{50\%} & \textbf{Total} \\
\midrule
KL over Leontief & 44.4\% (28) & 14.2\% (9) & 12.6\% (7) & 18.8\% (8) & 3.1\% (2) & \multirow{2}{*}{3.2\% (2)} & \multirow{2}{*}{63} \\
Leontief over KL & 3.1\% (2) & 1.5\% (1) & 0.0\% (0) & 3.1\% (2) & 3.1\% (2) & & \\
\midrule
KL over $\ell_1$ & 19.1\% (13) & 7.3\% (5) & 13.2\% (9) & 17.4\% (11) & 13.2\% (9) & \multirow{2}{*}{13.2\% (9)} & \multirow{2}{*}{68} \\
$\ell_1$ over KL & 1.4\% (1) & 1.4\% (1) & 4.4\% (3) & 2.9\% (2) & 7.3\% (5) & & \\
\midrule
KL over $\ell_2$ & 21.6\% (13) & 15.0\% (9) & 4.4\% (3) & 7.3\% (5) & 13.3\% (8) & \multirow{2}{*}{8.3\% (5)} & \multirow{2}{*}{60} \\
$\ell_2$ over KL & 1.6\% (1) & 1.6\% (1) & 10.0\% (6) & 6.6\% (4) & 8.3\% (5) & & \\
\midrule
$\ell_1$ over Leontief & 32.3\% (22) & 10.2\% (7) & 13.2\% (9) & 10.2\% (7) & 10.2\% (7) & \multirow{2}{*}{10.3\% (7)} & \multirow{2}{*}{68} \\
Leontief over $\ell_1$ & 5.8\% (4) & 1.4\% (1) & 1.4\% (1) & 1.4\% (1) & 2.9\% (2) & & \\
\midrule
$\ell_1$ over $\ell_2$ & 3.1\% (2) & 0.0\% (0) & 3.1\% (2) & 0.0\% (0) & 4.6\% (3) & \multirow{2}{*}{3.1\% (2)} & \multirow{2}{*}{64} \\
$\ell_2$ over $\ell_1$ & 34.3\% (22) & 12.5\% (8) & 12.5\% (8) & 10.9\% (7) & 14.7\% (10) & & \\
\midrule
$\ell_2$ over Leontief & 45.1\% (28) & 8.0\% (5) & 4.8\% (3) & 11.2\% (7) & 3.2\% (2) & \multirow{2}{*}{6.5\% (4)} & \multirow{2}{*}{62} \\
Leontief over $\ell_2$ & 16.1\% (6) & 3.2\% (2) & 1.6\% (1) & 1.6\% (1) & 4.8\% (3) & & \\
\midrule
Cosine Similarity over KL & 17.4\% (11) & 7.9\% (5) & 6.3\% (4) & 7.9\% (5) & 6.3\% (4) & \multirow{2}{*}{11.1\% (7)} & \multirow{2}{*}{63} \\
KL over Cosine Similarity & 15.8\% (10) & 3.1\% (2) & 9.5\% (6) & 7.9\% (5) & 6.3\% (4) & & \\
\midrule
Cosine Similarity over $\ell_2$ & 39.0\% (25) & 0.0\% (0) & 4.7\% (3) & 4.7\% (3) & 7.8\% (5) & \multirow{2}{*}{9.4\% (6)} & \multirow{2}{*}{64} \\
$\ell_2$ over Cosine Similarity & 21.8\% (14) & 0.0\% (0) & 4.7\% (3) & 4.7\% (3) & 3.1\% (2) & & \\
\midrule
Cosine Similarity over $\ell_1$ & 46.9\% (51) & 6.2\% (13) & 3.1\% (9) & 4.7\% (8) & 4.7\% (9) & \multirow{2}{*}{10.4\% (14)} & \multirow{2}{*}{134} \\
$\ell_1$ over Cosine Similarity & 7.8\% (11) & 6.2\% (5) & 4.7\% (4) & 0.0\% (4) & 4.7\% (6) & & \\
\midrule
Cosine Similarity over Leontief & 37.1\% (23) & 11.3\% (7) & 11.3\% (7) & 8.1\% (5) & 6.4\% (4) & \multirow{2}{*}{4.8\% (3)} & \multirow{2}{*}{62} \\
Leontief over Cosine Similarity & 14.5\% (9) & 0.0\% (0) & 1.6\% (1) & 4.8\% (3) & 0.0\% (0) & & \\
\midrule
\textbf{Overall consistent}
& \textbf{41.8\% (296)}
& \textbf{11.4\% (81)}
& \textbf{12.6\% (89)}
& \textbf{12.9\% (91)}
& \textbf{13.0\% (92)}
& \textbf{8.3\% (59)}
& \textbf{708} \\
\midrule
\textbf{95\% Wilson confidence interval}
& \textbf{[38.2\%, 45.5\%]}
& \textbf{[9.3\%, 14.0\%]}
& \textbf{[10.3\%, 15.2\%]}
& \textbf{[10.6\%, 15.5\%]}
& \textbf{[10.7\%, 15.7\%]}
& \textbf{[6.5\%, 10.6\%]}
& -- \\
\bottomrule
\end{tabular}
}
\end{table}

\begin{table}[H]
\centering
\small
\caption{Participants becoming fully consistent with each utility model at different relative-distance noise thresholds, for \textbf{four} issues. Each participant is counted only at the first threshold at which they become fully consistent with the corresponding model.}
\label{tab:noise_threshold_new_consistency_4}
\resizebox{\textwidth}{!}{%
\begin{tabular}{lcccccc}
\toprule
\textbf{Comparison} & \textbf{0\%} & \textbf{5\%} & \textbf{10\%} & \textbf{15\%} & \textbf{Not fully consistent at 15\%} & \textbf{Total} \\
\midrule
Leontief over KL & 3.2\% (2) & 0.0\% (0) & 0.0\% (0) & 0.0\% (0) & \multirow{2}{*}{52.4\% (33)} & \multirow{2}{*}{63} \\

KL over Leontief & 44.4\% (28) & 0.0\% (0) & 0.0\% (0) & 0.0\% (0) & & \\
\midrule
KL over $\ell_1$ & 19.1\% (13) & 0.0\% (0) & 0.0\% (0) & 0.0\% (0) & \multirow{2}{*}{79.4\% (54)} & \multirow{2}{*}{68} \\

$\ell_1$ over KL & 1.5\% (1) & 0.0\% (0) & 0.0\% (0) & 0.0\% (0) & & \\
\midrule
KL over $\ell_2$ & 21.7\% (13) & 0.0\% (0) & 0.0\% (0) & 0.0\% (0) & \multirow{2}{*}{66.7\% (40)} & \multirow{2}{*}{60} \\

$\ell_2$ over KL & 1.7\% (1) & 0.0\% (0) & 0.0\% (0) & 10.0\% (6) & & \\
\midrule
$\ell_1$ over Leontief & 32.4\% (22) & 0.0\% (0) & 0.0\% (0) & 1.5\% (1) & \multirow{2}{*}{60.3\% (41)} & \multirow{2}{*}{68} \\

Leontief over $\ell_1$ & 5.9\% (4) & 0.0\% (0) & 0.0\% (0) & 0.0\% (0) & & \\
\midrule
$\ell_1$ over $\ell_2$ & 3.1\% (2) & 0.0\% (0) & 0.0\% (0) & 4.7\% (3) & \multirow{2}{*}{51.6\% (33)} & \multirow{2}{*}{64} \\

$\ell_2$ over $\ell_1$ & 34.4\% (22) & 0.0\% (0) & 0.0\% (0) & 6.2\% (4) & & \\
\midrule
$\ell_2$ over Leontief & 45.2\% (28) & 0.0\% (0) & 0.0\% (0) & 0.0\% (0) & \multirow{2}{*}{45.2\% (28)} & \multirow{2}{*}{62} \\

Leontief over $\ell_2$ & 9.7\% (6) & 0.0\% (0) & 0.0\% (0) & 0.0\% (0) & & \\
\midrule
Cosine Similarity over KL & 17.5\% (11) & 0.0\% (0) & 0.0\% (0) & 1.6\% (1) & \multirow{2}{*}{65.1\% (41)} & \multirow{2}{*}{63} \\

KL over Cosine Similarity & 15.9\% (10) & 0.0\% (0) & 0.0\% (0) & 0.0\% (0) & & \\
\midrule
Cosine Similarity over $\ell_2$ & 39.1\% (25) & 0.0\% (0) & 0.0\% (0) & 1.6\% (1) & \multirow{2}{*}{31.3\% (20)} & \multirow{2}{*}{64} \\

$\ell_2$ over Cosine Similarity & 21.9\% (14) & 1.6\% (1) & 3.1\% (2) & 1.6\% (1) & & \\
\midrule
Cosine Similarity over $\ell_1$ & 38.1\% (51) & 1.5\% (2) & 0.0\% (0) & 0.7\% (1) & \multirow{2}{*}{50.0\% (67)} & \multirow{2}{*}{134} \\

$\ell_1$ over Cosine Similarity & 8.2\% (11) & 0.0\% (0) & 0.0\% (0) & 1.5\% (2) & & \\
\midrule
Cosine Similarity over Leontief & 37.1\% (23) & 0.0\% (0) & 1.6\% (1) & 0.0\% (0) & \multirow{2}{*}{46.8\% (29)} & \multirow{2}{*}{62} \\

Leontief over Cosine Similarity & 14.5\% (9) & 0.0\% (0) & 0.0\% (0) & 0.0\% (0) & & \\
\midrule
\textbf{Overall consistent} & \textbf{41.8\% (296)} & \textbf{0.4\% (3)} & \textbf{0.4\% (3)} & \textbf{2.8\% (20)} & \textbf{54.5\% (386)} & \textbf{708} \\
\midrule
\textbf{95\% Wilson confidence interval}
& [38.23\%,45.48\%]
& [0.14\%,1.24\%]
& [0.14\%,1.24\%]
& [1.84\%,4.32\%]
& [50.84\%,58.15\%]
& -- \\
\bottomrule
\end{tabular}%
}
\end{table}

\subsection{Confidence Intervals for Four-Issue Model Comparisons}
\label{CI_appendix_utility_function}

To quantify the uncertainty arising from the relatively modest sample sizes in the four-issue model-comparison polls, we computed Wilson $95\%$ confidence intervals for the proportion of participants exhibiting at least $90\%$ consistency with each model 
. 
The results are reported in \Cref{tab:confidence_intervals_utility}.

\begin{table}[H]
\centering
\small
\caption{Estimated proportions and Wilson $95\%$ confidence intervals for participants exhibiting at least $90\%$ consistency with the indicated model in the four-issue comparisons (that is, the sum of the $100\%$ and $90\%$ columns in \Cref{tab:summary_preferences_4}), listed in descending order.}
\label{tab:confidence_intervals_utility}
\begin{tabular}{lcc}
\toprule
\textbf{Property} & \textbf{Fraction} & \textbf{95\% CI} \\
\midrule
KL preferred over Leontief & $37/63$ & $[0.46,0.70]$ \\

$\ell_2$ preferred over Leontief & $33/62$ & $[0.41,0.65]$ \\

Cosine Similarity preferred over $\ell_1$ & $34/64$ &  $[0.41 , 0.65]$ \\

Cosine Similarity preferred over Leontief  & $30/62$ &  $[0.36 , 0.61]$ \\

$\ell_2$ preferred over $\ell_1$ & $30/64$ & $[0.35,0.59]$ \\

$\ell_1$ preferred over Leontief & $29/68$ & $[0.32,0.54]$ \\

Cosine Similarity preferred over $\ell_2$  & $25/64$ &  $[0.28 , 0.51]$ \\

KL preferred over $\ell_2$ & $21/60$ & $[0.24,0.48]$ \\

KL preferred over $\ell_1$ & $18/68$ & $[0.17,0.38]$ \\

Cosine Similarity preferred over KL & $16/63$ & $[0.16 , 0.37]$ \\

$\ell_2$ preferred over Cosine Similarity &  $14/64$& $[0.14 , 0.33]$ \\

KL preferred over Cosine Similarity & $12/63$ & $[0.11 , 0.3]$ \\ 

Leontief preferred over Cosine Similarity & $9/62$ & $[0.078 , 0.25]$ \\

$\ell_1$ preferred over Cosine Similarity & $9/64$ &  $[0.076 , 0.25]$ \\

Leontief preferred over $\ell_2$ & $8/62$ & $[0.067 , 0.23]$ \\

Leontief preferred over KL & $3/63$ & $[0.016 , 0.13]$ \\

Leontief preferred over $\ell_1$ & $5/68$ & $[0.032 , 0.16]$ \\

$\ell_2$ preferred over KL & $2/60$ & $[0.009,0.11]$ \\

$\ell_1$ preferred over $\ell_2$ & $2/64$ & $[0.0086 , 0.11]$ \\

$\ell_1$ preferred over KL & $2/68$ & $[0.005,0.11]$ \\

\bottomrule
\end{tabular}
\end{table}

\subsection{Aggregate Preference Results for Three- and Five-Issue Budgets}
\label{tab:summary_preferences_3_5}

\Cref{tab:summary_preferences_3} and \Cref{tab:summary_preferences_5} report the results of the model-comparison polls conducted with three- and five-issue budgets, respectively.

\begin{table}[H]
\centering
\small
\caption{Summary of participant preferences by model comparison and consistency threshold, for \textbf{three }issues (percentages out of all participants in each comparison).
We did these experiments on small samples, and later decided to focus on four-issue budgets.
}
\label{tab:summary_preferences_3}
\begin{tabular}{lcccccc}
\toprule
\textbf{Comparison} & \textbf{60\%} & \textbf{70\%} & \textbf{80\%} & \textbf{90\%} & \textbf{100\%} & \textbf{Total Participants} \\
\midrule
$\ell_1$ over $\ell_2$       & 6.5\% (2)  & 12.9\% (4) & 6.5\% (2)  & 3.2\% (1)  & 0.0\% (0)        & 31 \\
$\ell_2$ over $\ell_1$       & 22.6\% (7) & 12.9\% (4) & 19.4\% (6) & 3.2\% (1)  & 0.0\% (0)        & 31 \\
\midrule
$\ell_1$ over Leontief       & 15.6\% (5) & 21.9\% (7) & 9.4\% (3)  & 12.5\% (4) & 21.9\% (7) & 32 \\
Leontief over $\ell_1$       & 12.5\% (4) & 6.3\% (2)  & 0.0\% (0)  & 0.0\% (0)  & 0.0\% (0)  & 32 \\
\midrule
KL over $\ell_1$             & 19.4\% (6) & 16.1\% (5) & 16.1\% (5) & 3.2\% (1)  & 0.0\% (0)        & 31 \\
$\ell_1$ over KL             & 12.9\% (4) & 3.2\% (1)  & 3.2\% (1)  & 3.2\% (1)  & 0.0\% (0)        & 31 \\
\midrule
KL over $\ell_2$             & 12.9\% (4) & 9.7\% (3)  & 16.1\% (5) & 6.5\% (2)  & 3.2\% (1) & 31 \\
$\ell_2$ over KL             & 19.4\% (6) & 19.4\% (6) & 6.5\% (2)  & 0.0\% (0)  & 0.0\% (0) & 31 \\
\midrule
$\ell_2$ over Leontief       & 10.0\% (3) & 13.3\% (4) & 20.0\% (6) & 20.0\% (6) & 26.7\% (8) & 30 \\
Leontief over $\ell_2$       & 3.3\% (1)  & 0.0\% (0)  & 3.3\% (1)  & 0.0\% (0)  & 0.0\% (0)  & 30 \\
\midrule
KL over Leontief             & 6.3\% (2)  & 12.5\% (4) & 18.8\% (6) & 18.8\% (6) & 28.1\% (9) & 32 \\
Leontief over KL             & 6.3\% (2)  & 3.1\% (1)  & 0.0\% (0)  & 0.0\% (0)  & 0.0\% (0)  & 32 \\
\midrule
Cosine Similarity over $\ell_1$    & 12.9\% (4)  & 0\% (0) & 6.4\% (2) & 6.4\% (2) & 38.7\% (12) & 31 \\
$\ell_1$ over Cosine Similarity    & 0\% (0)  & 6.4\% (2)  & 0.0\% (0)  & 6.4\% (2)  & 22.5\% (7)  & 31 \\
\midrule
Cosine Similarity over $\ell_2$    & 3.1\% (1)  & 0\% (0) & 6.2\% (2) & 0.0\% (0) & 40.6\% (13) & 32 \\
$\ell_2$ over Cosine Similarity    & 3.1\% (1)  & 3.1\% (1)  & 0.0\% (0)  & 9.3\% (3)  & 31.2\% (10)  & 32 \\
\midrule
Cosine Similarity over Leontief    & 6.6\% (2)  & 10.0\% (3) & 10.0\% (3) & 13.3\% (4) & 50.0\% (15) & 30 \\
Leontief over Cosine Similarity    & 0\% (0)  & 3.3\% (1)  & 0.0\% (0)  & 0.0\% (0)  & 0.0\% (0)  & 30 \\
\midrule
\bottomrule
\end{tabular}
\end{table}

\begin{table}[H]
\centering
\small
\caption{Summary of participant preferences by model comparison and consistency threshold, for \textbf{five }issues.}
\label{tab:summary_preferences_5}
\begin{tabular}{lcccccc}
\toprule
\textbf{Comparison} & \textbf{60\%} & \textbf{70\%} & \textbf{80\%} & \textbf{90\%} & \textbf{100\%} & \textbf{Total} \\
\midrule
KL over Leontief & 9.4\% (3) & 6.3\% (2) & 12.5\% (4) & 9.4\% (3) & 31.3\% (10) & 32 \\
Leontief over KL & 0.0\% (0) & 9.4\% (3) & 3.1\% (1) & 6.3\% (2) & 0.0\% (0) & 32 \\
\midrule
KL over $\ell_1$ & 6.3\% (2) & 12.5\% (4) & 18.8\% (6) & 9.4\% (3) & 25.0\% (8) & 32 \\
$\ell_1$ over KL & 3.1\% (1) & 3.1\% (1) & 9.4\% (3) & 3.1\% (1) & 3.1\% (1) & 32 \\
\midrule
KL over $\ell_2$ & 28.1\% (9) & 9.4\% (3) & 9.4\% (3) & 0.0\% (0) & 12.5\% (4) & 32 \\
$\ell_2$ over KL & 9.4\% (3) & 12.5\% (4) & 3.1\% (1) & 3.1\% (1) & 3.1\% (1) & 32 \\
\midrule
$\ell_1$ over Leontief & 10.0\% (3) & 13.3\% (4) & 10.0\% (3) & 23.3\% (7) & 16.7\% (5) & 30 \\
Leontief over $\ell_1$ & 3.3\% (1) & 0.0\% (0) & 3.3\% (1) & 3.3\% (1) & 6.7\% (2) & 30 \\
\midrule
$\ell_1$ over $\ell_2$ & 3.2\% (1) & 3.2\% (1) & 0.0\% (0) & 0.0\% (0) & 0.0\% (0) & 31 \\
$\ell_2$ over $\ell_1$ & 12.9\% (4) & 19.4\% (6) & 16.1\% (5) & 6.5\% (2) & 35.5\% (11) & 31 \\
\midrule
$\ell_2$ over Leontief & 10.0\% (3) & 10.0\% (3) & 10.0\% (3) & 10.0\% (3) & 33.3\% (10) & 30 \\
Leontief over $\ell_2$ & 0.0\% (0) & 6.7\% (2) & 0.0\% (0) & 6.7\% (2) & 10.0\% (3) & 30 \\
\bottomrule
\end{tabular}
\end{table}

\begin{table}[H]
\centering
\small
\caption{Summary of attention check failures across polls with different numbers of issues.
Polls on more issues put a higher cognitive burden on participants, leading to higher failure rates.}
\label{tab:attention_failures}
\begin{tabular}{lccc}
\toprule
\textbf{Poll} & \textbf{Total Participants} & \textbf{Failed} & \textbf{Exclusion Rate} \\
\midrule
Polls on 3 issues & 2845 & 886 & 31.1\% \\
Polls on 4 issues & 1592   & 691 & 43.4\% \\
Polls on 5 issues & 725   & 346 & 47.7\% \\
\bottomrule
\end{tabular}
\end{table}

\subsection{Extending Budget Vectors Beyond Three Issues}
\label{beyond_3_categories}
\label{sec:large-m-computational-scalability}
As the number of issues increases, the computational complexity of generating and evaluating comparison pairs grows substantially. In particular, naively computing distances or identifying informative pairs across all alternatives induces a quadratic dependence on the number of issues, resulting in an $O(n^2)$ complexity (where $n$ is the number of vectors in the simplex), that quickly becomes computationally heavy in higher dimensions.
To address this challenge, we developed an algorithm that significantly reduces the effective computational burden by avoiding exhaustive pairwise comparisons. This approach enables scalable distance evaluation and pair generation even as the dimensionality of the budget vector increases, thereby preserving the practical feasibility of the framework for settings with many issues.

\paragraph{Finding the $k$ Most Different Pairs}%
\label[appendix]{Finding_the_k_most_diff_pairs}
\footnote{
We are grateful to 
Siddhanth Ramakrishnan from computer science stackexchange ( https://cs.stackexchange.com/a/175997/1342) for this algorithm.
}

We are given $n$ items, where each item $i$ is associated with two real-valued attributes $(a_i,b_i)$.  
For each pair $(i,j)$, the difference score is defined as
\[
d(i,j) := \min\left(|a_i-a_j|,\;|b_i-b_j|\right).
\]
The goal is to identify the $k \ll n^2$ pairs with the highest difference scores, without explicitly enumerating all $\binom{n}{2}$ pairs.

\paragraph{Key idea.}
Instead of ranking all pairs, the algorithm searches for the largest threshold $D$ such that there exist at least $k$ pairs $(i,j)$ satisfying
\[
d(i,j) \ge D.
\]
Equivalently, such pairs must satisfy both $|a_i-a_j|\ge D$ and $|b_i-b_j|\ge D$.

\paragraph{Binary search over $D$.}
The algorithm performs a binary search over possible values of $D$. For each candidate $D$, it checks whether the number of pairs with $d(i,j)\ge D$ is at least $k$. Since there are at most $O(n^2)$ distinct values of $d(i,j)$, this requires $O(\log n)$ iterations.

\paragraph{Counting pairs for a fixed $D$.}
Items are first sorted by their $a$-values. Using a two-pointers technique, for each item $i$ we maintain a set of items $j$ such that $a_j-a_i \ge D$. These items are stored in a balanced search tree ordered by their $b$-values.
For each $i$, we count how many such $j$ satisfy either
\[
b_j \le b_i - D \quad \text{or} \quad b_j \ge b_i + D.
\]
To count efficiently, we maintain, in each node in the tree, the number of elements smaller and larger than its element.
Then, we search for $b_i-D$ and $b_i+D$ in the tree.
Each query and update takes $O(\log n)$ time, yielding a total running time of $O(n\log n)$ for counting pairs for a fixed $D$.

\paragraph{Selecting the pairs.}
After finding the maximal threshold $D^*$ such that at least $k$ pairs satisfy $d(i,j)\ge D^*$, the same procedure is run again to explicitly enumerate qualifying pairs. Any $k$ of these pairs may be returned.

\paragraph{Complexity.}
The overall running time of the algorithm is
\[
O(n\log^2 n + k),
\]
which is significantly faster than the naive $O(n^2)$ approach when $k \ll n^2$.

\newpage

\section{Appendix to \Cref{sub:monotonicity-properties}: Monotonicity properties}
\label[appendix]{app:monotonicity-properties}
\label[appendix]{app:convex_algorithm}

\begin{algorithm}[H]
    \caption{Pair generation for checking star-shapedness.}

    \KwIn{Participant's ideal budget $\peak$; weights $\Lambda = \{0.1, 0.2, \dots, 0.9\}$}
    \KwOut{A set $S$ of budget allocation questions }

    Initialize empty set of questions $S \gets \emptyset$\;

    \For{each $\lambda \in \Lambda$}{
        Let $\other$ be a random vector representing a budget allocation \;
    
        Construct convex combination: $\other_\lambda \gets \lambda \peak + (1-\lambda) \other$ \;
    
        Add $(\other, \other_\lambda)$ to the poll set $S$ \;
    }

    \Return $S$ \;
\end{algorithm}

\subsection*{Example Calculation}
\begin{table}[H]
\centering
\caption{An example calculation for a participant with an ideal allocation of $[30, 40, 30]$.
\label{tab:examplecalc}}
\begin{tabular}{c c c c}
\toprule
$\lambda$ & $q$ & Calculation & Weighted vector \\
\midrule
0.1 & $[20,60,20]$ & $0.1\cdot[30,40,30] + 0.9\cdot[20,60,20]$ & $[21,58,21]$ \\
0.2 & $[25,35,40]$ & $0.2\cdot[30,40,30] + 0.8\cdot[25,35,40]$ & $[26,36,38]$ \\
0.3 & $[40,20,40]$ & $0.3\cdot[30,40,30] + 0.7\cdot[40,20,40]$ & $[37,26,37]$ \\
0.4 & $[10,70,20]$ & $0.4\cdot[30,40,30] + 0.6\cdot[10,70,20]$ & $[18,58,24]$ \\
0.5 & $[50,30,20]$ & $0.5\cdot[30,40,30] + 0.5\cdot[50,30,20]$ & $[40,35,25]$ \\
0.5 & $[60,15,25]$ & $0.5\cdot[30,40,30] + 0.5\cdot[60,15,25]$ & $[45,27.5,27.5]$ \\
0.6 & $[35,45,20]$ & $0.6\cdot[30,40,30] + 0.4\cdot[35,45,20]$ & $[32,42,26]$ \\
0.7 & $[40,50,10]$ & $0.7\cdot[30,40,30] + 0.3\cdot[40,50,10]$ & $[33,43,24]$ \\
0.8 & $[20,40,40]$ & $0.8\cdot[30,40,30] + 0.2\cdot[20,40,40]$ & $[28,40,32]$ \\
0.9 & $[45,25,30]$ & $0.9\cdot[30,40,30] + 0.1\cdot[45,25,30]$ & $[31.5,38.5,30]$ \\
\bottomrule
\end{tabular}
\end{table}
\label{app:star-shaped_example}

\begin{algorithm}[H]
\caption{Pair generation for checking star-shapedness; rounded values.
\label{alg:star-shaped-rounded}
\label{round}
}

    \KwIn{Participant's ideal budget $\peak$; weights $\Lambda = \{0.1, 0.2, \dots, 0.9\}$}
    \KwOut{A set $S$ of questions consisting of random vectors and their adjusted convex combinations}

    Initialize empty set of questions $S \gets \emptyset$\;

    \For{each $\lambda \in \Lambda$}{
        Let $\other$ be a random vector representing the budget allocation\;
    
        Construct convex combination: $\other_\lambda \gets \lambda p + (1-\lambda) \other$\; 
    
        \For{$i = 1$ \KwTo $m$}{
        Round $\other_\lambda[i]$ to the nearest multiple of $5$\; (In the case of a tie, round to the nearest even multiple of 5)
        } 
    
        \If{$\sum_{i=1}^m \other_\lambda[i] \neq 100$}{
            Let $j \gets \arg\max_{i} \other_\lambda[i]$\;
            Set $\other_\lambda[j] \gets 100 - \sum_{i \neq j} \other_\lambda[i]$\;
        }
    
        Add $(\other, \other_\lambda)$ to the poll set $S$\; 
    }

    \Return $S$\;
\end{algorithm}

An example demonstrating how the convex combination vector $\other_\lambda$ was constructed for a participant with a given ideal and random vector. The process illustrates how intermediate adjustments ensure the total sum equals 100 and values are rounded to meaningful units.

\begin{table}[H]
\centering
\small
\caption{Example illustrating the construction of $\other_\lambda$ for a peaked participant.
\label{tab:round_example}}
\begin{tabular}{l|c}
\toprule
 & Vector \\
\midrule
Ideal vector $p$ & $[50,30,20]$ \\
Random vector $\other$ & $[23,35,42]$ \\
$\lambda=0.5$ convex combination & $[36.5,32.5,31]$ \\
Rounded to nearest multiple of 5 & $[35,30,30]$ \\
Adjusted largest project to sum 100 & $[40,30,30]$ \\
Final $\other_\lambda$ & $[40,30,30]$ \\
\bottomrule
\end{tabular}
\end{table}

\subsection*{Distribution of participants by level of consistency star-shaped preferences (rounded)}

\begin{table}[h]
\centering
\caption{Distribution of participants by consistency level for Star-shaped.
\label{tab:consistency_distribution}}
\begin{tabular}{c c c}
\hline
Consistency Level & \# of Participants & Percentage of Participants \\
\hline
60\%  & 4  & 4.8\%  \\
70\%  & 6  & 7.1\%  \\
80\%  & 7  & 8.3\%  \\
90\%  & 23 & 27.4\% \\
100\% & 40 & 47.6\% \\
\hline
\end{tabular}
\end{table}

\subsection*{Consistency results per $\lambda$ (rounded)}
\begin{table}[H]
\centering
\caption{Consistency results per $\lambda$
\label{tab:consistency_by_weight}}
\begin{tabular}{ccc}
\toprule
\textbf{$\lambda$} & \textbf{Average Consistency (\%)} & \textbf{Total Pairs} \\
\midrule
0.1  & 77.46 & 71  \\
0.2  & 90.14 & 71  \\
0.3  & 92.96 & 71  \\
0.4  & 84.51 & 71  \\
0.5  & 91.55 & 142 \\
0.6  & 91.55 & 71  \\
0.7  & 90.14 & 71  \\
0.8  & 88.73 & 71  \\
0.9  & 91.55 & 71  \\
\bottomrule
\end{tabular}
\end{table}

We note that this analysis includes 71 responses out of the 84 participants who completed the full poll. The reason for this is that 13 participants completed an earlier version of the poll, prior to our ability to extract data for the $\lambda$ distribution.\\

\subsection{Confidence Intervals for Star Shapedness}
\label{CI_appendix_star_shaped}

\begin{table}[H]
\centering
\small
\caption{Estimated proportions and Wilson 95\% confidence intervals}
\label{tab:confidence_intervals_star_shaped}
\begin{tabular}{lcc}
\hline
Property & Fraction & 95\% CI \\
\hline
Star-shaped & $63/84$ & $[0.65,0.83]$ \\
\hline
\end{tabular}
\end{table}

\begin{algorithm}[H]
\caption{Weighted pair generation for testing Multi-Dimensional Single-Peakedness (MDSP)}
\label{alg:mdsp}
    
\KwIn{Participant's ideal allocation $\peak$; number of questions $k$; weights $\Lambda = \{0.1, 0.2, \dots, 0.9\}$}
\KwOut{A set $S$ of MDSP-consistent question pairs}

Initialize empty set of questions $S \gets \emptyset$\;

\While{$|S| < k$}{

    Sample a random allocation $\other_{\text{far}}$\;

    Identify:
    \[
    O=\{i:\other_{\text{far}}^i>\peak^i\},
    \qquad
    U=\{j:\other_{\text{far}}^j<\peak^j\}
    \]
    
    Compute
    \[
    \Delta_{\max}
    =
    \min\!\left(
    \max_{i\in O}(\other_{\text{far}}^i-\peak^i),\;
    \max_{j\in U}(\peak^j-\other_{\text{far}}^j)
    \right)
    \]
    
    Select the next $\lambda$ from the sequence $\Lambda$;
    Set
    \[
    \Delta = \lambda \cdot \Delta_{\max}
    \]

    Initialize
    $\other_{\text{near}} \gets \other_{\text{far}}$\;

    Set
    $i \gets \argmax_{i\in O}(\other_{\text{far}}^i-\peak^i)$\;

    Set
    $j \gets \argmax_{j\in U}(\peak^j-\other_{\text{far}}^j)$\;

    Transfer budget:
    \[
    \other_{\text{near}}^i
    \gets
    \other_{\text{near}}^i-\Delta
    \]

    \[
    \other_{\text{near}}^j
    \gets
    \other_{\text{near}}^j+\Delta
    \]

    Round $\other_{\text{near}}$ to multiples of 5 and rebalance so that
    $\sum_\ell \other_{\text{near}}^\ell =100$\;

    \If{$\other_{\text{near}}=\other_{\text{far}}$}{
        continue\;
    }

    Add ordered pair
    $(\other_{\text{far}},\other_{\text{near}})$
    to $S$\;
}

\Return $S$\;

\end{algorithm}

Importantly, closeness is defined directionally: the preferred allocation must lie on the same side of the peak in every dimension and move weakly toward it, rather than merely being closer in absolute distance.

\paragraph{Numerical example.}
Consider a participant whose ideal allocation is
\[
\peak = [40,30,30].
\]
Suppose the algorithm samples the allocation
\[
\other_{\mathrm{far}}=[5,55,40].
\]
We first identify the coordinates that are above and below the ideal allocation:
\[
O=\{2,3\},
\qquad
U=\{1\}.
\]
The corresponding deviations from the ideal allocation are
\[
\other_{\mathrm{far}}^2-\peak^2=55-30=25,
\qquad
\other_{\mathrm{far}}^3-\peak^3=40-30=10,
\]
and
\[
\peak^1-\other_{\mathrm{far}}^1=40-5=35.
\]
Therefore,
\[
\Delta_{\max}
=
\min(25,35,10)
=
10.
\]

For the weight $\lambda=0.2$, the transferred amount is
\[
\Delta=\lambda\Delta_{\max}=0.2\cdot10=2.
\]
The coordinates selected for the transfer are
\[
i=\argmax_{r\in O}(\other_{\mathrm{far}}^r-\peak^r)=2,
\qquad
j=\argmax_{r\in U}(\peak^r-\other_{\mathrm{far}}^r)=1.
\]
Thus, before rounding,
\[
\other_{\mathrm{near}}
=
[7,53,40].
\]
Since allocations are restricted to multiples of $5$, the resulting vector is rounded and rebalanced, yielding
\[
\other_{\mathrm{near}}=[10,50,40].
\]
Hence, the ordered pair generated by the algorithm is
\[
\boxed{
\left(
[5,55,40],
[10,50,40]
\right)
}
\]
where the first allocation is the \emph{far} option and the second is the \emph{near} option.

\ayelet{new MDSP results}
\subsection*{Consistency Results for Multi-Dimensional Single-Peaked Preferences}
\begin{table}[h]
\centering
\caption{Consistency results supporting the multi-dimensional single-peaked for three dimensions
\label{tab:mdsp_results}}
\begin{tabular}{l c}
\hline
Result Category & Percentage \\
\hline
Perfect consistency ($100\%$)  & 76.5\% \\
Consistency $90\%$ & 11.7\% \\
Closer vector chosen & 95.9\% \\
Farther vector chosen & 4.1\% \\
\hline
\end{tabular}
\end{table}

\subsection*{MDSP results per number of topics:}
\begin{table}[H]
\centering
\small
\caption{MDSP Results for 3, 4, and 5 Topics: Overall Survey Statistics
\label{tab:mdsp_results_topics}
}
\begin{tabular}{llcccc}
\toprule
\textbf{\# Topics} & \textbf{Consistency Level} & \textbf{\# of Users} & \textbf{Far Vector} & \textbf{Near Vector} \\
\midrule

\multirow{3}{*}{3 Topics}
 & 90.0\%  & 13  & 10.0\% & 90.0\% \\
 & 100.0\% & 66 & 0.0\%  & 100.0\% \\
 & Total   & 87 & 5.6\%  & 94.4\% \\
\midrule
\multirow{5}{*}{4 Topics}
 & 60.0\%  & 1  & 40.0\% & 60.0\% \\
 & 70.0\%  & 2  & 30.0\% & 70.0\% \\
 & 80.0\%  & 5  & 20.0\% & 80.0\% \\
 & 90.0\%  & 9  & 10.0\% & 90.0\% \\
 & 100.0\% & 13 & 0.0\%  & 100.0\% \\
 & Total   & 30 & 9.7\%  & 90.3\% \\
\midrule
\multirow{5}{*}{5 Topics}
 & 70.0\%  & 3  & 30.0\% & 70.0\% \\
 & 80.0\%  & 2  & 20.0\% & 80.0\% \\
 & 90.0\%  & 10  & 10.0\% & 90.0\% \\
 & 100.0\% & 16 & 0.0\%  & 100.0\% \\
 & Total   & 33 & 10.3\%  & 89.7\% \\
\bottomrule
\end{tabular}
\end{table}


\label{peak_linear_algorithm}
\begin{algorithm}[H]
\caption{Pair generation for testing peak-linearity.
\label{alg:peak-linear}
}

    \KwIn{Participant's ideal budget $p$; weights $\Lambda = \{0.25, 0.5, 0.75\}$}
    \KwOut{A set $S$ of comparison questions based on convex combinations}

    Initialize empty set of questions $S \gets \emptyset$\;

    Define three extreme vectors: 
    $\extreme_A = [10, 10, 80], \quad \extreme_B = [10, 80, 10], \quad \extreme_C = [80, 10, 10]$\;

    Add $(\extreme_A,\extreme_B)$, $(\extreme_A,\extreme_C)$, $(\extreme_B,\extreme_C)$ to $S$\;

    \For{each $\lambda \in \Lambda$}{
        Compute convex combinations: \;
        $\other_A \gets \lambda p + (1 - \lambda) \extreme_A$\;
        $\other_B \gets \lambda p + (1 - \lambda) \extreme_B$\;
        $\other_C \gets \lambda p + (1 - \lambda) \extreme_C$\;
    
        Add $(\other_A,\other_B)$, $(\other_A,\other_C)$, $(\other_B,\other_C)$ to $S$\;
    }

    \Return $S$\;
\end{algorithm}

\begin{table}[H]
\centering
\caption{Consistency of pairwise comparisons for different weight percentiles
\label{tab:peaklinear_consistency}}
\setlength{\fboxsep}{4pt}
\setlength{\fboxrule}{0.8pt}
\fbox{%
\resizebox{\linewidth}{!}{%
\begin{tabular}{lcccc}
\hline
\textbf{Weight ($\lambda$)} & \textbf{A vs. B} & \textbf{A vs. C} & \textbf{B vs. C} & \textbf{Average Consistency} \\
\hline
25\% ($\lambda=0.25$) & 71\% (54/76) & 76\% (58/76) & 74\% (56/76) & 74\% (168/228) \\
50\% ($\lambda=0.5$)  & 83\% (63/76) & 86\% (65/76) & 82\% (62/76) & 83\% (190/228) \\
75\% ($\lambda=0.75$) & 87\% (66/76) & 86\% (65/76) & 83\% (63/76) & 85\% (194/228) \\
\textbf{All percentiles} & 80\% (183/228) & 82\% (188/228) & 79\% (181/228) & \textbf{81\% (552/684)} \\
\hline
\end{tabular}%
}%
}
\end{table}

To illustrate the algorithm, we present below an example of answers that were inconsistent with peak-linearity.
\begin{table}[H]
\caption{An example of inconsistency is a participant whose ideal budget is $[30, 20, 50]$.
This example illustrates an inconsistent choice pattern, as the participant’s preferences over the $\lambda$-weighted averages do not consistently mirror the ranking of the original extreme vectors, contrary to what would be expected under a peak-linear utility function.
\label{tab:inconsistency}
}
\centering
\begin{tabular}{c|c|c|c}
\toprule
Pair / $\lambda$ & Option A & Option B & Participant Choice \\
\midrule
Extreme Vectors & [10, 10, 80] & [10, 80, 10] & A \\
$\lambda=0.25$ & [24, 18, 58] & [25, 35, 40] & B \\
$\lambda=0.5$ & [20, 15, 65] & [20, 50, 30] & A \\
$\lambda=0.75$ & [16, 12, 72] & [15, 65, 20] & B \\
\bottomrule
\end{tabular}
\end{table}

\subsection*{Peak-linear Consistency Rates and 95\% Confidence Intervals}

\begin{table}[H]
\centering
\small
\caption{Observed proportions and 95\% confidence intervals for the main consistency measures.}
\label{tab:peak_linear_ci}
\begin{tabular}{l c c}
\toprule
Metric & Rate & 95\% CI \\
\midrule
Overall consistency ($\geq90\%$) & $46/77$ & $[0.49 , 0.7]$ \\
Transitivity & $69/77$ & $[0.81,0.95]$ \\
Perfect order consistency & $26/77$ & $[0.24 , 0.45]$ \\
\bottomrule
\end{tabular}
\end{table}

\subsection*{Consistency Across Participant Groups}

\begin{table}[H]
\centering
\small
\caption{Consistency metrics across all participants and among those preferring weighted vectors
\label{tab:consistency_weighted}}
\begin{tabular}{l c c}
\toprule
Metric & All users & Star-shaped users \\
\midrule
Users & $44$ & $22$ \\
Overall consistency & $78.3\%$ & \textbf{84.1\%} \\
Transitivity rate & \textbf{96\%} & $94\%$ \\
Order consistency & $70.1\%$ & \textbf{78.6\%} \\
\bottomrule
\end{tabular}
\end{table}

\begin{table}[H]
\centering
\small
\caption{Distribution of transitivity consistency levels in poll 3.}
\begin{tabular}{lccc}
\toprule
\textbf{Transitivity Level} & 100\% & 75\% & 50\% \\
\midrule
\ All users & 39 & 5 & 1 \\
\ Star-Shaped users & 18 & 3 & 1 \\
\bottomrule
\end{tabular}
\end{table}


\section{Relations between Monotonicity Properties: Proofs}
\label[appendix]{app:monotonicity-properties-proofs}
\subsection{Peak-Linearity as a Stronger Condition than Star-Shapedness}
\label[appendix]{peak_linear_is_star_shaped}
\begin{proposition}
If a utility function \(U\) is \emph{peak-linear} around the peak \(\peak\), then it is \emph{star-shaped} around \(\peak\).
The converse does not hold: there exist star-shaped utility functions that are not peak-linear.
\end{proposition}

\begin{proof}
Assume by contradiction that $U$ is peak-linear and continuous, but not strictly star-shaped. 
Since $U$ is not strictly star-shaped, there exists a distribution $\other \neq \peak$ and a scalar $\lambda \in (0,1)$ such that moving towards $\peak$ does not strictly increase the utility:
$$U(\peak, \lambda \peak + (1-\lambda)\other) \leq U(\peak, \other).$$

Let us define a sequence of distributions $\{\other_n\}_{n=0}^{\infty}$ recursively:
\begin{align*}
    \other_0 &= \other \\
    \other_{n+1} &= \lambda \peak + (1-\lambda)\other_n
\end{align*}
By our initial assumption, $U(\peak, \other_1) \leq U(\peak, \other_0)$.

According to the definition of peak-linearity, for any two distributions $x, y$ and a scalar $\alpha \in (0,1)$, we have 
$$U(\peak, \alpha \peak + (1-\alpha)x) \leq U(\peak, \alpha \peak + (1-\alpha)y) \iff U(\peak, x) \leq U(\peak, y).$$

Applying this property recursively for all $n$, we obtain a monotonically non-increasing sequence of utilities:
$$U(\peak, \other_0) \geq U(\peak, \other_1) \geq U(\peak, \other_2) \geq \dots \geq U(\peak, \other_n) \geq \dots$$

Notice that the distance between $\other_n$ and $\peak$ shrinks by a factor of $(1-\lambda)$ at each step. Since $\lambda \in (0,1)$, as $n \to \infty$, the sequence of distributions $\other_n$ converges to the peak $\peak$.

Now, substitute these specific terms into the definition of peak-linearity. By setting $\alpha = \lambda$, $x = \other_1$, and $y = \other_0$, the left-hand side of the equivalence yields:
$$U(\peak, \lambda \peak + (1-\lambda)\other_1) \leq U(\peak, \lambda \peak + (1-\lambda)\other_0)$$

Here, examining the arguments within the utility function and recalling the recursive construction of our sequence, we can see that the argument on the right-hand side is exactly the definition of $\other_1$. Similarly, the argument on the left-hand side represents a further contraction from $\other_1$ towards the peak, which is precisely the definition of $\other_2$. Therefore, the inequality translates directly to:
$$U(\peak, \other_2) \leq U(\peak, \other_1)$$

We have thus established that $\other_2$ yields a utility less than or equal to that of $\other_1$. We can now repeat this exact process: by substituting $\other_2$ and $\other_1$ as $x$ and $y$ respectively in the peak-linearity definition, we obtain $U(\peak, \other_3) \leq U(\peak, \other_2)$. Applying this property recursively chains these inequalities into a single monotonically non-increasing sequence:
$$U(\peak, \other_0) \geq U(\peak, \other_1) \geq U(\peak, \other_2) \geq \dots \geq U(\peak, \other_n) \geq \dots$$

Because the utility function $U(\peak, \cdot)$ is continuous, the limit of the utilities must equal the utility of the limit point:
$$\lim_{n \to \infty} U(\peak, \other_n) = U(\peak, \peak).$$

Since the sequence $U(\peak, \other_n)$ is monotonically non-increasing to this limit, every element in the sequence must be greater than or equal to the limit. In particular for the first element:
$$U(\peak, \other) \geq U(\peak, \peak).$$

However, $\peak$ is the unique global maximum, meaning $U(\peak,\peak) > U(\peak,\other)$ for all $\other \neq \peak$. This is a direct contradiction. 
Therefore, our assumption must be false, and $U$ must be strictly star-shaped.
\end{proof}

\subsection{The Role of Continuity.} 
\begin{example}
Consider an indicator utility function where the agent only derives utility from their exact ideal distribution:
$$
U(\peak, \other) = \begin{cases} 
1 & \text{if } \other = \peak \\
0 & \text{if } \other \neq \peak 
\end{cases}
$$
This function is peak-linear: For any $\other_1, \other_2$ and $\lambda \in (0,1)$, the condition holds trivially. For instance, if $\other_1 = \peak$ and $\other_2 \neq \peak$, both sides of the equivalence evaluate to $1 \geq 0$. If both $\other_1, \other_2 \neq \peak$, both sides evaluate to $0 \geq 0$. 

However, this function violates the strict star-shaped property. For any $\other \neq \peak$ and $\lambda \in (0,1)$, the intermediate distribution $\lambda \peak + (1-\lambda)\other$ is strictly not equal to $\peak$. Thus:
$$ U(\peak, \lambda \peak + (1-\lambda)\other) = 0 \ngtr 0 = U(\peak, \other). $$
Moving strictly closer to the peak does not strictly increase utility.
\end{example}

\medskip
\begin{proof}
\textbf{Star-shaped does not imply peak-linear.}
Consider the utility function
\[
U(\other) = -\big( ( \other_1-\peak_1 )^2 + ( \other_2-\peak_2 )^4 \big).
\]

This function is strictly maximized at $\peak$.

\textbf{Star-shapedness.}
For any $\lambda\in(0,1)$,
\[
U(\lambda \peak + (1-\lambda)\other)
=
-\big( (\lambda \peak_1+(1-\lambda)\other_1-\peak_1 )^2 + (\lambda \peak_2+(1-\lambda)\other_2-\peak_2 )^4 \big)
\]

\[
=
-\big( ((\lambda-1) \peak_1+(1-\lambda)\other_1 )^2 + ((\lambda-1) \peak_2+(1-\lambda)\other_2)^4 \big)
\]

\[
=
-\big( ((1-\lambda)(\other_1-\peak_1) )^2 + ((1-\lambda)(\other_2-\peak_2) )^4 \big)
=
-\big((1-\lambda)^2(\other_1-\peak_1)^2
+
(1-\lambda)^4(\other_2-\peak_2)^4\big).
\]

Since $(1-\lambda)^2<1$ and $(1-\lambda)^4<1$, we obtain
\[
U(\lambda \peak + (1-\lambda)\other) > U(\other),
\]
for every $\other \neq \peak$.
Hence $U$ is star-shaped.

\textbf{Failure of peak-linearity.}
Fix $\peak=(0,0)$ and consider
\[
\other^{(1)}=(0,2),
\qquad
\other^{(2)}=(1,1).
\]

Then
\[
U(\other^{(1)})=-16,
\qquad
U(\other^{(2)})=-2,
\]
so
\[
U(\other^{(2)}) > U(\other^{(1)}).
\]

However, taking $\lambda=0.9$,
\[
U(\lambda\peak+(1-\lambda)\other^{(1)})=-0.0016,
\qquad
U(\lambda\peak+(1-\lambda)\other^{(2)})=-0.0101,
\]
and therefore
\[
U(\lambda\peak+(1-\lambda)\other^{(1)})
>
U(\lambda\peak+(1-\lambda)\other^{(2)}),
\]

Thus peak-linearity fails.
\end{proof}

\subsection{Multi-Dimensional Single-Peaked as a Stronger Condition than Star-Shapedness}
\label[appendix]{MDSP_is_star_shaped}
\begin{proposition}
If the utility function \(U\) is \emph{multi-dimensional single-peaked} around the peak \(\peak\), then \(U\) is also \emph{star-shaped} around \(\peak\).
\end{proposition}
\begin{proof}
For any allocation \(\other\) and any coefficient \(\alpha \in [0,1]\), define
\[
\other^\alpha := \peak + \alpha(\other - \peak).
\]
To prove star-shapedness around \(\peak\), we must show that for every \(\other \neq \peak\) and every \(\alpha \in (0,1)\),
\[
U(\other^\alpha) > U(\other),
\]
That is, moving toward the peak strictly increases utility.

Recall the definition of \emph{multi-dimensional single-peakedness}:  
\(\other_2\) is said to be \emph{closer to \(\peak\) than} \(\other_1\) if for every dimension \(j\),
either
\[
q_{1j} \ge q_{2j} \ge p_{j}
\quad\text{or}\quad
q_{1j} \le q_{2j} \le p_{j},
\]
and in at least one dimension, the inequality between \(q_{1j}\) and \(q_{2j}\) is strict.  
Whenever \(\other_2\) is closer to \(\peak\) than \(\other_1\), multi-dimensional single-peakedness requires that
\[
U(\other_2) > U(\other_1).
\]

Now take \(\other_1 = \other\) and \(\other_2 = \other^\alpha\) for some \(\alpha \in (0,1)\).  
Fix any coordinate \(j\). There are three cases:
\begin{enumerate}
    \item If \(q_j = p_{j}\), then \(\other^\alpha_j = p_{j}\) as well.
    \item If \(q_j > p_{j}\), then
    \[
    q_j - p_{j} > 0
    \quad\text{and}\quad
    \other^\alpha_j - p_{j} = \alpha(q_j - p_{j}),
    \]
    so
    \[
    q_j > \other^\alpha_j > p_{j}.
    \]
    \item If \(q_j < p_{j}\), then
    \[
    q_j - p_{j} < 0
    \quad\text{and}\quad
    \other^\alpha_j - p_{j} = \alpha(q_j - p_{j}),
    \]
    so
    \[
    q_j < \other^\alpha_j < p_{j}.
    \]
\end{enumerate}

Thus, for every coordinate \(j\), \(\other^\alpha_j\) lies weakly between \(q_j\) and \(p_{j}\) in the same direction from the peak, and whenever \(q_j \neq p_{j}\) the inequality between \(q_j\) and \(\other^\alpha_j\) is strict.  
If \(\other \neq \peak\), there is at least one such coordinate, and therefore  
\(\other^\alpha\) is \emph{closer to \(\peak\)} than \(\other\) in the sense of the definition.

By multi-dimensional single-peakedness, it follows that
\[
U(\other^\alpha) > U(\other).
\]

Since this holds for every \(\other \neq \peak\) and every \(\alpha \in (0,1)\), we conclude that utility strictly increases as one moves along the line segment from \(\other\) toward \(\peak\).  
This is exactly the definition of a \emph{star-shaped} utility function around \(\peak\).

\end{proof}

\subsection{Logical Relation Between Peak-Linear and Multi-Dimensional Single-Peakedness}
\label[appendix]{MDSP_vs_peak_linear}
\begin{proposition}
Peak-linearity and Multi-Dimensional Single-Peakedness are independent: neither property implies the other in general.
\end{proposition}
Below, we give a short justification and counterexample:

\paragraph{MDSP does \emph{not} imply Peak-linear:} 

Let $\peak\in\mathbb{R}^2$ be fixed and define
\[
U(q) = -\Big((\other_1-\peak_1)^2 + (\other_2-\peak_2)^4\Big).
\]

\emph{(1) $U$ satisfies MDSP.}

Suppose $\other^{(1)}$ and $\other^{(2)}$ lie on the same orthant relative to $\peak$
and that $\other^{(2)}$ is coordinatewise closer to $\peak$ than $\other^{(1)}$,
with strict inequality in at least one coordinate.
Then
\[
|\other^{(2)}_1-\peak_1| < |\other^{(1)}_1-\peak_1|,
\qquad
|\other^{(2)}_2-\peak_2| \le |\other^{(1)}_2-\peak_2|.
\]
Since both $x\mapsto x^2$ and $x\mapsto x^4$ are strictly increasing on $\mathbb{R}_{\ge 0}$,
we obtain
\[
(\other^{(2)}_1-\peak_1)^2 + (\other^{(2)}_2-\peak_2)^4
<
(\other^{(1)}_1-\peak_1)^2 + (\other^{(1)}_2-\peak_2)^4,
\]
and therefore
\[
U(\other^{(2)}) > U(\other^{(1)}).
\]
Thus $U$ satisfies MDSP.

\emph{(2) $U$ is not peak-linear.}

Let $\peak=(0,0)$ and consider
\[
\other^{(1)}=(0,2),
\qquad
\other^{(2)}=(1,1).
\]
Then
\[
U(\other^{(1)})=-16,
\qquad
U(\other^{(2)})=-2,
\]
so
\[
U(\other^{(2)}) > U(\other^{(1)}).
\]

Now take $\lambda=0.9$.
Then
\[
U(\lambda \peak + (1-\lambda)\other^{(1)})=-0.0016,
\qquad
U(\lambda \peak + (1-\lambda)\other^{(2)})=-0.0101,
\]
and hence
\[
U(\lambda \peak + (1-\lambda)\other^{(1)})
>
U(\lambda \peak + (1-\lambda)\other^{(2)}),
\]
so the ordering reverses.
Therefore, the ordinal equivalence required by peak-linearity fails.

\paragraph{Peak-linear does \emph{not} imply MDSP:}
We present a counterexample in which the utility function is peak-linear yet violates multi-dimensional single-peakedness:

Let $\peak\in\mathbb{R}^m$ be fixed and define
\[
U(\other) = -\max_{j=1,\dots,m} |q_j-p_{j}|.
\]

\emph{(1) $U$ is peak-linear.}

For any $\other$ and any $\lambda\in[0,1]$,
\[
\lambda \peak + (1-\lambda)\other - \peak
= (1-\lambda)(\other-\peak),
\]
and therefore
\[
U(\lambda \peak + (1-\lambda)\other)
=
-\max_j |(1-\lambda)(q_j-p_{j})|
=
(1-\lambda)U(\other).
\]
Hence, for any $\other{(1)},\other^{(2)}$,
\[
U(\other^{(1)}) \ge U(\other^{(2)})
\iff
U(\lambda \peak + (1-\lambda)\other^{(1)})
\ge
U(\lambda \peak + (1-\lambda)\other^{(2)}),
\]
so $U$ satisfies peak-linearity.

\emph{(2) $U$ violates MDSP.}

Let $m=2$ and take
\[
\peak=(0,0),
\qquad
\other^{(1)}=(2,2),
\qquad
\other^{(2)}=(1,2).
\]
Then $\other^{(2)}$ is coordinatewise closer to $\peak$ than $\other^{(1)}$
with strict inequality in the first coordinate.
However,
\[
U(\other^{(1)})=-2,
\qquad
U(\other^{(2)})=-2.
\]
Thus utility does not strictly increase when moving closer in every coordinate,
and MDSP fails.

Therefore,
\[
\text{Peak-linear} \;\not\Rightarrow\; \text{MDSP}.
\]

Overall, these two properties are independent: neither one implies the other.

\subsection{Leontief utilities are peak-linear}
\label{leontief_peak_linear}
\begin{proof}
Let $U(p,q) = \min_{j \in A} \left( \frac{q_j}{p_j} \right)$.
For any alternative distribution $q$ and $\lambda \in [0,1)$, we evaluate the utility of the mixed distribution:
\[
U(p, \lambda p + (1-\lambda)q) = \min_{j \in A} \left( \frac{\lambda p_j + (1-\lambda)q_j}{p_j} \right) = \min_{j \in A} \left( \lambda + (1-\lambda)\frac{q_j}{p_j} \right).
\]

Since $\lambda$ and $(1-\lambda)$ are non-negative constants, this affine transformation preserves the order of the elements inside the minimum operator. Thus, we can extract the constants:
\[
U(p, \lambda p + (1-\lambda)q) = \lambda + (1-\lambda) \min_{j \in A} \left( \frac{q_j}{p_j} \right) = \lambda + (1-\lambda) U(p,q).
\]

Now, for any two distributions $q_1, q_2$ and $\lambda \in (0,1)$:
\[
U(p,q_1) \ge U(p,q_2) \iff (1-\lambda) U(p,q_1) \ge (1-\lambda) U(p,q_2) \iff \lambda + (1-\lambda)U(p,q_1) \ge \lambda + (1-\lambda)U(p,q_2).
\]

\[
U(p, \lambda p + (1-\lambda)q_1) \ge U(p, \lambda p + (1-\lambda)q_2) \iff \lambda + (1-\lambda)U(p,q_1) \ge \lambda + (1-\lambda)U(p,q_2).
\]

Because $1-\lambda > 0$, we can subtract $\lambda$ and divide both sides by $1-\lambda$ without changing the inequality's direction. This simplifies exactly to $U(p,q_1) \ge U(p,q_2)$, perfectly satisfying the ordinal definition of peak-linearity.
\end{proof}

\subsection{KL-Divergence Utility implies Multi-Dimensional Single-Peak}
\label{KL_MDSP_proof}
\begin{proof}
Recall that the KL-based utility is given by:
\[
U(\peak,\other) = - \sum_{j \in A} p_j \cdot \ln \left( \frac{p_j}{q_j} \right) = \sum_{j \in A} p_j \ln(q_j) - \sum_{j \in A} p_j \ln(p_j).
\]

Note that for the KL utility function to be well-defined, we assume that all allocations and peaks are strictly positive (i.e., $q_j > 0$ and $p_j > 0$ for all $j \in A$). This means we are avoiding division by zero and undefined logarithmic values.

Since the natural logarithm is a strictly concave function, $U$ is strictly concave with respect to $\other$. 
Recall the gradient inequality for a strictly concave function $f$: for any two distinct points $x$ and $y$, we have:
\[
f(x) < f(y) + \nabla f(y) \cdot (x - y)
\]

Rearranging this inequality to isolate the difference gives:
\[
f(y) - f(x) > \nabla f(y) \cdot (y - x)
\]

Let $\other_1, \other_2 \in \Delta$ be two distinct allocations, where $\other_2$ is closer to $\peak$ than $\other_1$ according to the MDSP definition. By substituting $x = \other_1$ and $y = \other_2$ into our inequality, we obtain:
\[
U(\peak, \other_2) - U(\peak, \other_1) > \sum_{j \in A} \frac{\partial U}{\partial q_j}(\other_2) \cdot (q_{2j} - q_{1j}) = \sum_{j \in A} \frac{p_{j}}{q_{2j}} (q_{2j} - q_{1j}).
\]

Because both $\other_1$ and $\other_2$ are valid allocations, the sum of their coordinates must equal the total budget $B$. Therefore, the sum of their component-wise differences is zero: 
\[
\sum_{j \in A} (q_{2j} - q_{1j}) = 0.
\]

We can subtract this sum (which is exactly zero) from our right-hand side without changing its value:
\[
\sum_{j \in A} \frac{p_j}{q_{2j}} (q_{2j} - q_{1j}) - \sum_{j \in A} 1 \cdot (q_{2j} - q_{1j}) = \sum_{j \in A} \left( \frac{p_j}{q_{2j}} - 1 \right) (q_{2j} - q_{1j}).
\]

Recall the definition of MDSP:
\textit{Let $\other_1$ and $\other_2$ be two alternative distributions.  
We say that 
$\other_2$ is \emph{closer to $\peak$ than} $\other_1$ if for every issue $j$, either $q_{1j} \geq q_{2j} \geq p_j$ 
or $q_{1j} \leq q_{2j} \leq p_j$, and for at least one issue $j$, the inequality between 
$q_{1j}$ and $q_{2j}$ is strict. A utility-model function $U$ is said to be \textit{multi-dimensional single-peaked} if whenever  
$\other_2$ is closer to $\peak$ than $\other_1$, it holds that $U(\peak,\other_2) > U(\peak,\other_1)$.}

This means that for every coordinate $j$, one of the following cases holds:
\begin{itemize}
    \item $q_{1j} \le q_{2j} \le p_{j}$: In this case, $(q_{2j} - q_{1j}) \ge 0$. Furthermore, since $q_{2j} \le p_{j}$, we have $\frac{p_{j}}{q_{2j}} \ge 1$, which implies $\left( \frac{p_{j}}{q_{2j}} - 1 \right) \ge 0$. The product of two non-negative terms is non-negative.
    \item $q_{1j} \ge q_{2j} \ge p_{j}$: In this case, $(q_{2j} - q_{1j}) \le 0$. Furthermore, since $q_{2j} \ge p_{j}$, we have $\frac{p_{j}}{q_{2j}} \le 1$, which implies $\left( \frac{p_{j}}{q_{2j}} - 1 \right) \le 0$. The product of two non-positive terms is non-negative.
\end{itemize}

In all cases, every term in the summation is non-negative. Therefore, the entire sum is greater than or equal to zero:
\[
\sum_{j \in A} \left( \frac{p_{j}}{q_{2j}} - 1 \right) (q_{2j} - q_{1j}) \ge 0.
\]

Combining this non-negative sum with our strict gradient inequality from earlier, we finally get:
\[
U(\peak, \other_2) - U(\peak, \other_1) > \sum_{j \in A} \left( \frac{p_{j}}{q_{2j}} - 1 \right) (q_{2j} - q_{1j}) \ge 0 \implies U(\peak, \other_2) > U(\peak, \other_1).
\]
This proves that the KL utility model is multi-dimensional single-peaked.
\end{proof}

\subsection{KL-Divergence Utility is not Peak-Linear}
\label{KL_peak_linear}
\begin{proof}
Let $m=2$ and assume a peak:
\[
\peak = (0.8, 0.2).
\]

Take two alternative allocations:
\[
\other_1 = (0.5, 0.5), \qquad \other_2 = (0.97, 0.03), \qquad \lambda = 0.9.
\]

Using the KL-based utility
\[
U(\peak,\other) = -\sum_{j \in A} p_{j} \ln\!\left(\frac{p_{j}}{q_j}\right)
= \sum_{j} p_{j} \ln(q_j) - \sum_{j} p_{j} \ln(p_{j}),
\]
the constant term for our peak is $\sum p_{j} \ln(p_{j}) = 0.8\ln(0.8) + 0.2\ln(0.2) \approx -0.5004$.
We compute the utilities for $\other_1$ and $\other_2$:

\[
U(\peak,\other_1) = (0.8\ln(0.5) + 0.2\ln(0.5)) - (-0.5004) \approx -0.1927
\]

\[
U(\peak,\other_2) = (0.8\ln(0.97) + 0.2\ln(0.03)) - (-0.5004) \approx  -0.2253
\]

Hence, initially,
\[
U(\peak,\other_1) > U(\peak,\other_2).
\]

We compute the interpolated allocations with $\lambda = 0.9$:
\[
\tilde{\other}_1 = \lambda\peak + (1-\lambda)\other_1 = 0.9(0.8, 0.2) + 0.1(0.5, 0.5) = (0.77, 0.23)
\]

\[
\tilde{\other}_2 = \lambda\peak + (1-\lambda)\other_2 = 0.9(0.8, 0.2) + 0.1(0.97, 0.03) = (0.817, 0.183)
\]

Now, we recalculate the utilities for the interpolated points:
\[
U(\peak,\tilde{\other}_1) = (0.8\ln(0.77) + 0.2\ln(0.23)) - (-0.5004) \approx -0.002624
\]

\[
U(\peak,\tilde{\other}_2) = (0.8\ln(0.817) + 0.2\ln(0.183)) - (-0.5004) \approx -0.000944
\]

Thus,
\[
U(\peak,\tilde{\other}_2) > U(\peak,\tilde{\other}_1),
\]
which is a strict reversal of the original preference.

We have shown that
\[
U(\peak,\other_1) > U(\peak,\other_2) \quad \text{but} \quad U(\peak,\lambda\peak+(1-\lambda)\other_2) > U(\peak,\lambda\peak+(1-\lambda)\other_1).
\]
This contradicts the definition of peak-linearity. Hence, KL utilities are not peak-linear.
\end{proof}

\subsection{Cosine-Similarity Utility is Star-Shaped}
\label{cosine_star_shaped}

\begin{proof}
Let the utility function be defined by cosine similarity:
$U(\peak,\other)=\frac{\peak\cdot\other}{|\peak|_2|\other|_2}$

Take any allocation $\other\neq\peak$ and any $\lambda\in(0,1)$. Define
\[
\tilde{\other}=\lambda\peak+(1-\lambda)\other.
\]

We prove that
\[
U(\peak,\peak)>U(\peak,\tilde{\other})>U(\peak,\other).
\]

First, since all budget allocations have nonnegative coordinates, both
$U(\peak,\tilde{\other})$ and $U(\peak,\other)$ are nonnegative. Therefore, it is sufficient to compare their squares.

We have
\[
\peak\cdot\tilde{\other}=\lambda|\peak|_2^2+(1-\lambda)(\peak\cdot\other),
\]
and
\[
|\tilde{\other}|_2^2=\lambda^2|\peak|_2^2+2\lambda(1-\lambda)(\peak\cdot\other)+(1-\lambda)^2|\other|_2^2.
\]

Using these expressions, the difference between the squared cosine similarities is
\begin{align*}
 &
U(\peak,\tilde{\other})^2-U(\peak,\other)^2
\\
&=
\frac{\left(\peak\cdot\tilde{\other}\right)^2}
{|\peak|_2^2|\tilde{\other}|_2^2}
-
\frac{\left(\peak\cdot\other\right)^2}
{|\peak|_2^2|\other|_2^2}
\\
&=
\frac{\left(\lambda|\peak|_2^2+(1-\lambda)(\peak\cdot\other)\right)^2}
{|\peak|_2^2|\tilde{\other}|_2^2}
-
\frac{(\peak\cdot\other)^2}
{|\peak|_2^2|\other|_2^2}
\\
&=
\frac{\left|\other|_2^2(\lambda|\peak|_2^2+(1-\lambda)(\peak\cdot\other)\right)^2 - |\tilde{\other}|_2^2(\peak\cdot\other)^2}
{|\peak|_2^2|\tilde{\other}|_2^2|\other|_2^2}
\\
&=
\frac{\lambda\left(\lambda|\peak|_2^2+2(1-\lambda)(\peak\cdot\other)\right)\left(|\peak|_2^2|\other|_2^2-
(\peak\cdot\other)^2
\right)
}{
|\peak|_2^2
|\tilde{\other}|_2^2
|\other|_2^2
}
\end{align*}

The denominator is positive. Moreover,
\[\lambda\left(\lambda|\peak|_2^2+2(1-\lambda)(\peak\cdot\other)\right)> 0,
\]
since $\lambda>0$, $|\peak|_2^2>0$, and $\peak\cdot\other\geq0$.

By the Cauchy--Schwarz inequality,
\[
|\peak|_2^2|\other|_2^2-(\peak\cdot\other)^2\geq0.
\]
Equality holds only if $\other$ is a scalar multiple of $\peak$. However, $\peak$ and $\other$ belong to the same budget simplex and therefore have the same coordinate sum. Hence, if $\other$ is a scalar multiple of $\peak$, the scalar must be $1$, implying $\other=\peak$, contrary to our assumption. Therefore,
\[
|\peak|_2^2|\other|_2^2 - (\peak\cdot\other)^2 > 0.
\]

It follows that
\[
U(\peak,\tilde{\other})^2 >U(\peak,\other)^2,
\]
and since both utilities are nonnegative,
\[
U(\peak,\tilde{\other})>U(\peak,\other).
\]

Finally,
\[
U(\peak,\peak)=1.
\]
Also, $\tilde{\other}\neq\peak$, since $\lambda\in(0,1)$ and
$\other\neq\peak$. By the strict form of the Cauchy--Schwarz inequality,
\[
U(\peak,\tilde{\other})<1.
\]
Therefore,
\[
U(\peak,\peak)>U(\peak,\tilde{\other})>U(\peak,\other).
\]

Thus, cosine-similarity utility is strictly star-shaped on the budget simplex.
\end{proof}

\subsection{Cosine-Similarity Utility is not Peak-Linear}
\label{cosine_peak_linear}

\begin{proof}
Let $m=4$ and assume $\peak=(40,30,20,10)$.

Take two alternative allocations:
\[
\other_1=(5,20,30,45),
\qquad
\other_2=(5,85,5,5),
\qquad
\lambda=\frac{1}{2}.
\]

Using the cosine-similarity utility, we compute the utilities of $\other_1$ and $\other_2$:

\[
\operatorname{CosSim}(\peak,\other_1)
=\frac{40\cdot5+30\cdot20+20\cdot30+10\cdot45}
{\sqrt{40^2+30^2+20^2+10^2}\,\sqrt{5^2+20^2+30^2+45^2}} 
\approx
0.5836
\]

Similarly,
\[
\operatorname{CosSim}(\peak,\other_2)
=\frac{40\cdot5+30\cdot85+20\cdot5+10\cdot5}
{\sqrt{40^2+30^2+20^2+10^2}\,\sqrt{5^2+85^2+5^2+5^2}} 
\approx
0.6197
\]

Hence, initially,
\[
U(\peak,\other_2)>U(\peak,\other_1).
\]

We compute the interpolated allocations with $\lambda=\frac{1}{2}$:
\[
\tilde{\other}_1 =\lambda\peak+(1-\lambda)\other_1
\]

\[
\frac{1}{2}(40,30,20,10) +
\frac{1}{2}(5,20,30,45) = (22.5,25,25,27.5),
\]

and
\[
\tilde{\other}_2 = \lambda\peak+(1-\lambda)\other_2
\]

\[
\frac{1}{2}(40,30,20,10) + \frac{1}{2}(5,85,5,5) = (22.5,57.5,12.5,7.5).
\]

Now, we recalculate the utilities of the interpolated allocations:

\[
U(\peak,\tilde{\other}_1)= \frac{40\cdot22.5+30\cdot25+20\cdot25+10\cdot27.5}{\sqrt{3000}\sqrt{22.5^2+25^2+25^2+27.5^2}} \approx 0.8833
\]

Similarly,
\[
U(\peak,\tilde{\other}_2)=\frac{40\cdot22.5+30\cdot57.5+20\cdot12.5+10\cdot7.5}{
\sqrt{3000}\sqrt{22.5^2+57.5^2+12.5^2+7.5^2}} \approx 0.8489
\]

Thus,
\[
U(\peak,\tilde{\other}_1) > U(\peak,\tilde{\other}_2),
\]
which is a strict reversal of the original preference.

We have shown that
\[
U(\peak,\other_2)>U(\peak,\other_1)\]
but
\[U\left(\peak,\lambda\peak+(1-\lambda)\other_1\right) > U\left(\peak,\lambda\peak (1-\lambda)\other_2 \right).
\]

This contradicts the definition of peak-linearity. Hence, cosine-similarity utilities are not peak-linear.
\end{proof}

\newpage

\section{Appendix to \Cref{sub:symmetry}: Checking Symmetry}
\label[appendix]{app:symmetry}
\subsection*{Symmetry Example}
Let the ideal budget be $[27, 33, 40]$. The table below shows two illustrative cases: one for Issue Symmetry and one for Sign Symmetry. In each case, the alternatives have identical $\ell_p$ distances from the ideal allocation, despite differences in the positions or signs of the deviations.
\begin{table}[h]
\centering
\begin{tabular}{lcccc}
\toprule
\textbf{Symmetry Type} & \textbf{Alternative} & \textbf{Allocation} & \textbf{Deviations from Ideal} \\
\midrule
Issue Symmetry & A & $[36, 33, 31]$ & $[+9, 0, -9]$ \\
                   & B & $[27, 42, 31]$ & $[0, +9, -9]$ \\
Sign Symmetry      & A & $[26, 30, 44]$ & $[-1, -3, +4]$ \\
                   & B & $[28, 36, 36]$ & $[+1, +3, -4]$ \\
\bottomrule
\end{tabular}
\label{tab:symmetry-example}
\end{table}

\begin{algorithm}[H]
    \caption{Pair generation for testing Issue Symmetry.
\label{project_symmetry_algorithm}
}
    
    \KwIn{An ideal budget $\peak$; number of poll sets $k$; number of issues $m$} 
    \KwOut{A set $S$ of budget allocation poll sets}
    
    Set $S \gets \emptyset$\;
    
    \While{$|S| < mk$}{
        Let $\other_1, \other_2$ be random vectors representing budget allocations\; 
        Compute differences from peak: $\diff_1 \gets \other_1 - \peak$, $\diff_2 \gets \other_2 - \peak$\;

        \For{$j = 1$ \KwTo $m-1$}{
            Generate $j$-th rotation of differences: $\diff_1^{(j)}, \diff_2^{(j)}$\;
            Compute shifted allocations: $\other_1^{(j)} \gets \peak + \diff_1^{(j)}$, $\other_2^{(j)} \gets \peak + \diff_2^{(j)}$\; 
        
            \If{all components of $\other_1^{(j)}$ and $\other_2^{(j)} \ge 0$}{ 
                Append $(\other_1^{(j)},\other_2^{(j)})$ to the poll set $S$.
            }
        }
    
        Add the original pair $(\other_1, \other_2)$ to $S$\;
        }
    \Return $S$\;
\end{algorithm}

\begin{algorithm}[H]
    \caption{Pair generation for testing Sign Symmetry.
    \label{sign_symmetry}
}

    \KwIn{An ideal budget $\peak$; number of poll sets $k$}
    \KwOut{A set $S$ of poll sets containing original and negated deviations}
    
    Set $S \gets \emptyset$\;
    
    \While{$|S| < k$}{
        Let $\other_1, \other_2$ be random vectors representing budget allocations\; 
        Compute deviations from peak: $\diff_1 \gets \other_1 - \peak$, $\diff_2 \gets \other_2 - \peak$\; 
        
        Generate negated deviations: $\diff_1' \gets -\diff_1, \quad \diff_2' \gets -\diff_2$\;
        
        \If{$\peak + \diff_i' \ge 0 \;\;\; \forall i \in \{1,2\}$}{ 
            Set negated allocations: $\other_1' \gets \peak + \diff_1'$, $\other_2' \gets \peak + \diff_2'$\;
            Add the allocations $(\other_1, \other_2)$ and $(\other_1', \other_2')$ as a poll set to $S$\;
        }
    }
    \Return $S$\;
\end{algorithm}

\subsection*{Analysis of Symmetry in Preferences}
The following tables presents the proportion of participants whose responses demonstrated project or Sign Symmetry at different levels of consistency.
\begin{table}[H]
\centering
\small
\caption{Number of participants by consistency level (Sign Symmetry)}
\begin{tabular}{lccccccc}
\toprule
\textbf{Consistency} & 0 & 1/6 & 2/6 & 3/6 & 4/6 & 5/6 & 6/6 \\
\midrule
\textbf{\# Participants} & 1 (1.6\%) & 4 (6.5\%) & 4 (6.5\%) & 20 (32.7\%) & 16 (26.2\%) & 13 (21.4\%) & 3 (4.9\%) \\
\bottomrule
\end{tabular}
\label{tab:sign_consistency}
\end{table}

\begin{table}[H]
\centering
\small
\caption{Number of participants by consistency level (Issue Symmetry)}
\begin{tabular}{lccccc}
\toprule
\textbf{Consistency} & 0 & 1/4 & 2/4 & 3/4 & 4/4 \\
\midrule
\textbf{\# Participants} & 9 (12.3\%) & 18 (24.6\%) & 22 (30.1\%) & 16 (21.9\%) & 8 (10.9\%) \\
\bottomrule
\end{tabular}
\label{tab:issue_consistency}
\end{table}

\section{Confidence Intervals for Symmetry}
\label{CI_appendix_symmetry}

\begin{table}[H]
\centering
\small
\caption{Estimated proportions and Wilson 95\% confidence intervals for symmetry}
\label{tab:confidence_intervals_symmetry}
\begin{tabular}{lcc}
\hline
Property & Fraction & 95\% CI \\
\hline
Sign-symmetric & $16/61$ & $[0.18, 0.4]$ \\
Issue-symmetric & $24/73$ & $[0.23, 0.44]$ \\
\hline
\end{tabular}
\end{table}

\begin{algorithm}[H]
\caption{
Pair generation for testing issue symmetry among issues with identical allocations
\label{alg:issue-symmetry-with-identical-allocations}
}

\KwIn{Participant's budget vector $\peak \in \mathbb{R}^M$; number of poll questions $K$}
\KwOut{A set $S$ of comparison pairs}

Initialize empty set of questions $S \gets \emptyset$\;

Identify two issues $i,j$ such that $\peak_i = p_{j}$;
If there are two such pairs with different amounts
(this is possible with $m\geq 4$),
select a pair with a largest amount. Break other ties arbitrarily.

Let $r_1,\ldots,r_{m-2}$ be the remaining issues whose budget remains fixed\;

Define magnitude step size: $\Delta \gets p_i / K$\;

\For{$t = 1$ to $K$}{
    $x \gets \;\text{round}(t\Delta)$
    Construct option $\other_1 \in \mathbb{R}^M$:
    \[
        \other_{1,i} \gets \peak_i + x, \quad
        \other_{1,j} \gets p_{j} - x, \quad
        \other_{1,r_k} \gets \peak_{r_k} \text{ for $k\in\{1,\ldots,m-2\}$}
    \]
    
    Construct option $\other_2 \in \mathbb{R}^M$:
    \[
        \other_{2,i} \gets \peak_i - x, \quad
        \other_{2,j} \gets p_{j} + x, \quad
        \other_{2,r_k} \gets \peak_{r_k} \text{ for $k\in\{1,\ldots,m-2\}$}
    \]

    \If{$\other_1$ and $\other_2$ are valid budget allocations}{
        Add pair $(\other_1,\other_2)$ to $S$\;
    }
}

\Return $S$\;
\end{algorithm}

\subsection*{Issue Asymmetry Results}
\label{issue_asymmetric_results}

\begin{table}[H]
\centering
\caption{Number of Users by Consistency Level
\label{asymmetric_results}}
\begin{tabular}{lc}
\toprule
\textbf{Consistency Level (\%)} & \textbf{\# of Users} \\
\midrule
50.0 & 2 (6.5\%) \\
60.0 & 4 (12.9\%) \\
70.0 & 5 (16.1\%) \\
80.0 & 4 (12.9\%) \\
90.0 & 3 (9.7\%) \\
100.0 & 13 (41.9\%) \\
\midrule
\textbf{Total} & 31 (100.0\%) \\
\bottomrule
\end{tabular}
\end{table}

\section{Appendix to \Cref{sub:consistency-in-asymmetry}: Consistency in Asymmetry}
\label[appendix]{app:consistency-in-asymmetry}

\begin{algorithm}[H]
\caption{Pair generation for testing consistency in issue-asymmetry.
\label{alg:consistency-in-issue-asymmetry}}

    \KwIn{Participant's ideal budget $\peak$; weights $\Lambda = \{0.2, 0.4\}$; vector length $m$}
    \KwOut{A set $S$ of questions with options generated by rotating difference vectors}
    
    Initialize empty set of questions $S \gets \emptyset$\; 
    Let $\min(\peak)$ denote the smallest component of $\peak$\;

    \For{each $\lambda \in \Lambda$}{ 
        Compute: $X_\lambda \gets \max(1, \;\text{round}(\lambda \cdot \min(\peak)))$         
        \CommentSty{//Define base difference vectors:}\; 
        $\diff_p \gets ((m-1)X_\lambda, -X_\lambda, \dots, -X_\lambda)$ \CommentSty{// Concentrated increase} \;
        $\diff_n \gets (-(m-1)X_\lambda, X_\lambda, \dots, X_\lambda)$
        \CommentSty{// Concentrated decrease} \; 
    
        \For{each $\diff \in \{\diff_p, \diff_n\}$}{ 
            \For{$j = 1$ \KwTo $m$}{ 
                Compute cyclic rotation $\Delta^{(j)} \gets cyclic\_shift(\diff, j)$\; 
            }
            Create a ranking-question with the $m$ options: $\other_1 \gets \peak + \Delta^{(1)}, \dots, \other_m \gets \peak + \Delta^{(m)}$\; 
            
            Add question to $S$\;
        }
    }    
    \Return $S$\;
\end{algorithm}

\section{Confidence Intervals for Reported Proportions }
\label{CI_appendix_asymmetry}

\begin{table}[H]
\centering
\small
\caption{Estimated proportions and Wilson 95\% confidence intervals for Constitency in Asymmetry}
\label{tab:confidence_intervals}
\begin{tabular}{lcc}
\hline
Property & Fraction & 95\% CI \\
\hline
Issue-asymmetric & $16/31$ & $[0.35,0.68]$ \\
\hline
\end{tabular}
\end{table}

\subsection*{Example of Generated Allocation Options}

To illustrate how alternative allocations were generated for a given participant, 
Table~\ref{tab:generated_options} presents the options produced for an ideal vector 
of $\peak = (85, 15, 5)$ under different values of $\lambda$. 
Each option is accompanied by its corresponding deviation vector $\Delta$, 
showing the directional adjustment applied to the original ideal allocation.

\begin{table}[H]
\centering
\small
\caption{Generated allocation options for $\peak=(80,15,5)$ under different $\lambda$ values, with corresponding deviation vectors.
\label{tab:generated_options}}
\begin{tabular}{c c c c c c c c}
\toprule
Question & $\lambda$ & Option 1 & $\diff_1$ & Option 2 & $\diff_2$ & Option 3 & $\diff_3$ \\
\midrule
1 & 0.2 
& $[82,14,4]$ & $[2,-1,-1]$ 
& $[79,17,4]$ & $[-1,2,-1]$ 
& $[79,14,7]$ & $[-1,-1,2]$ \\

2 & 0.4 
& $[84,13,3]$ & $[4,-2,-2]$ 
& $[78,19,3]$ & $[-2,4,-2]$ 
& $[78,13,9]$ & $[-2,-2,4]$ \\

3 & 0.2 
& $[78,16,6]$ & $[-2,1,1]$ 
& $[81,13,6]$ & $[1,-2,1]$ 
& $[81,16,3]$ & $[1,1,-2]$ \\

4 & 0.4 
& $[76,17,7]$ & $[-4,2,2]$ 
& $[82,11,7]$ & $[2,-4,2]$ 
& $[82,17,1]$ & $[2,2,-4]$ \\
\bottomrule
\end{tabular}
\end{table}

\subsection*{Participants’ Choice Consistency}
\begin{table}[H]
\centering
\begin{tabular}{lccccc}
\toprule
 & over $1/3$ & over $2/3$ & $3/3$ consistent \\
\midrule
Number of Participants & 27 & 16 & 7 \\
Percentage & 72.9\% & 43.2\% & 18.9\% \\
\bottomrule
\end{tabular}
\label{consistent_number}
\end{table}

\begin{algorithm}[H]
    \caption{Pair generation for testing consistency in sign asymmetry
\label{alg:consistency-in-sign-asymmetry}    
    }    
    \KwIn{Participant's ideal budget $\peak \in \mathbb{R}^M$ }
    \KwOut{A set $S$ of comparison pairs }
    
    Initialize empty set of questions $S \gets \emptyset$\; 
    Let $\min(\peak)$ denote the smallest component of $\peak$\; 
    
    Define base magnitude: $X_{\text{base}} \gets \max\left(1,\; \left\lfloor \tfrac{\min(\peak)}{10} \right\rfloor \right)$\; 
    
    Define magnitude levels: $\Lambda \gets \{X_{\text{base}},\; 2X_{\text{base}},\; 3X_{\text{base}},\; 4X_{\text{base}}\}$\;
    
    \For{each target category $i \in \{1,\dots,M\}$}{ 
        \For{each magnitude $X \in \Lambda$}{ 
            Define a \textbf{concentrated loss} vector $\diff_1 \in \mathbb{R}^M$: \;
            $d_{1,i} \gets -(M-1)X, \qquad d_{1,j} \gets X \;\; \text{for all } j \neq i$\; 
            
            Define a \textbf{concentrated gain} vector $\diff_2 \in \mathbb{R}^M$: \; 
            $d_{2,i} \gets (M-1)X, \qquad d_{2,j} \gets -X \;\; \text{for all } j \neq i$\;
            
            Construct options $\other_1 \gets \peak + \diff_1$ and $\other_2 \gets \peak + \diff_2$\; 
            \If{$\other_1$ and $\other_2$ are valid budget allocations (all entries in $[0,100]$)}{ 
                Add pair $(\other_1, \other_2)$ to $S$\; 
            }
        }
    }
    
    \Return $S$\; 
\end{algorithm}

\begin{algorithm}[H]
\caption{Fallback procedure for pair generation.
\label{alg:fallback}
}

\KwIn{Participant's ideal budget $p$; target category $i$; magnitude index $k \in \{1,2,3,4\}$; Pre-defined fixed vectors $F_k$}
\KwOut{A set $S$ of comparison pairs updated with fallback allocations}

\tcp{Triggered when the primary method fails for a given target category and magnitude}

Retrieve pre-defined difference vectors $(fd_1, fd_2) \gets F_k$\;
Rotate $fd_1$ and $fd_2$ based on target category $i$ to get $fd_1', fd_2'$\;
Construct fallback option A: $q_{f1} \gets p + fd_1'$\;
Construct fallback option B: $q_{f2} \gets p + fd_2'$\;

\If{$q_{f1}$ and $q_{f2}$ are valid budget allocations}{
    Add pair $(q_{f1}, q_{f2})$ to the poll set $S$\;
}
\end{algorithm}

\subsection*{Example of a generated comparison pair}
\begin{table}[H]
\centering
\small
\caption{Example of a generated comparison pair for a participant with an ideal budget of $p=(60,30,10)$, using magnitude level 2 ($X=2$). The concentrated change is applied to the first category, resulting in one option with a concentrated loss and another with a concentrated gain.
\label{tab:example_pair_magnitude2}}
\begin{tabular}{c c c}
\toprule
 & \textbf{Option A} & \textbf{Option B} \\
 & (Concentrated Loss) & (Concentrated Gain) \\
\midrule
\textbf{Deviation Vector} & $(-4, +2, +2)$ & $(+4, -2, -2)$ \\
\textbf{Resulting Allocation} & $(56, 32, 12)$ & $(64, 28, 8)$ \\
\bottomrule
\end{tabular}
\end{table}

\subsection*{Participant Preference Matrix}
Preference matrix for a participant where the rows correspond to topics, and the columns correspond to magnitude levels. Each cell is colored to indicate whether the participant preferred a distributed decrease (orange) or a concentrated decrease (blue).

\subsection*{Distribution of Participants by Concentrated and Distributed Levels}
\label{tab:concentrated_vs_distributed}

\begin{table}[H]
\centering
\caption{Number of Participants Across Different Levels of Budget Concentration and Distribution}
\begin{tabular}{lcc}
\toprule
\textbf{Percentage (\%)} & \textbf{Concentrated} & \textbf{Distributed} \\
\midrule
16.7 & 2  & 0  \\
25.0 & 6  & 1  \\
33.3 & 9  & 13  \\
41.7 & 11  & 6  \\
50.0 & 15  & 15  \\
58.3 & 6  & 11  \\
66.7 & 13  & 9  \\
75.0 & 1  & 6  \\
83.3 & 0  & 2  \\
\bottomrule
\end{tabular}
\end{table}

\section{Appendix to \Cref{sub:biennial}: Comparing Biennial Budgets}
\label[appendix]{app:biennial}

\begin{algorithm}[H]
    \caption{Pair generation for testing preferences among biennial budgets.
    \label{alg:biennial}}
    
    \KwIn{Participant's ideal budget $\peak$; number of repetitions per sub-poll $k=4$}
    \KwOut{A set $S$ of 12 questions comparing biennial budget allocations}
    
    $(x,y)$ denotes $(\text{year 1}, \text{year 2})$\;
    Initialize empty set of questions $S \gets \emptyset$\;
    
    \For{$i = 1$ \KwTo $k$}{
        Randomly generate a budget vector $r_i$\;
        
        \tcp{Sub-poll 1: Ideal year 1 vs. Ideal year 2}
        Add question to $S$: Option 1 is $(\peak, r_i)$, Option 2 is $(r_i, \peak)$\;
    
        \tcp{Sub-poll 2: Fixed year 1, Ideal year 2 vs. Balanced year 2}
        Fix year 1 budget to $r_i$\;
        Option 1: $(r_i, \peak)$\;
        Option 2: $(r_i, \other_i)$ such that $\tfrac{r_i + q_i}{2} = \peak$\;
        Add question to $S$\;
    
        \tcp{Sub-poll 3: Fixed year 2, Ideal year 1 vs. Balanced year 1}
        Fix year 2 budget to $r_i$\;
        Option 1: $(\peak, r_i)$\;
        Option 2: $(\other_i, r_i)$ such that $\tfrac{q_i + r_i}{2} = \peak$\;
        Add question to $S$\;
    }
    
    \Return $S$\;
\end{algorithm}

An example of questions with an ideal budget $\peak = (50, 30, 20)$ across two years:
\begin{table}[H]
\centering
\small
\begin{tabular}{|c|c|c|c|}
\hline
\textbf{Sub-poll} & \textbf{Year 1} & \textbf{Year 2}  & \textbf{Description} \\ \hline
1 & (50,30,20) & (40,25,35)  & Ideal in year 1, random in year 2 \\ \hline
1 & (40,25,35) & (50,30,20)  & Random in year 1, ideal in year 2 \\ \hline
2 & (40,25,35) & (50,30,20) & Ideal in year 2 \\ \hline
2 & (40,25,35) & (60,35,5)  & Average = ideal \\ \hline
3 & (50,30,20) & (40,25,35) & Ideal in year 1 \\ \hline
3 & (60,35,5) & (40,25,35) & Average = ideal \\ \hline
\end{tabular}
\end{table}

\subsection*{Biennial Poll Results}
\begin{table}[H]
\centering
\begin{tabular}{lcccc}
\toprule
\textbf{Sub-poll 1} \\
\midrule
\textbf{Consistency level} & \textbf{Number of users} & \textbf{Ideal Year 1} & \textbf{Random} \\
50\%  & 2  & 50.00\% & 50.00\% \\
75\%  & 12  & 66.7\% & 33.30\% \\
100\% & 25 & 96.00\% & 4.00\%  \\
\textbf{Total} & 39 & 84.60\% & 15.40\% \\
\midrule
\textbf{Sub-poll 2} \\
\midrule
\textbf{Consistency level} & \textbf{Number of users} & \textbf{Ideal Year 2} & \textbf{Balanced Year 2} \\
50\%  & 2  & 50.00\% & 50.00\% \\
75\%  & 10  & 55.00\% & 45.00\% \\
100\% & 27 & 100.00\% & 0.00\% \\
\textbf{Total} & 39 & 85.90\% & 14.10\% \\
\midrule
\textbf{Sub-poll 3} \\
\midrule
\textbf{Consistency level} & \textbf{Number of users} & \textbf{Ideal Year 1} & \textbf{Balanced Year 1} \\
50\%  & 3  & 50.00\% & 50.00\% \\
75\%  & 10  & 60.00\% & 40.00\% \\
100\% & 26 & 100.00\% & 0.00\% \\
\textbf{Total} & 39 & 85.90\% & 14.10\% \\
\bottomrule
\end{tabular}
\label{tab:subsurveys_results}
\end{table}

\subsection*{Biennial Poll Results (Cumulative)}
\begin{table}[H]
\centering
\begin{tabular}{lcccccc}
\toprule
\textbf{Sub-poll} & \textbf{over 50\%} & \textbf{over 75\%} & \textbf{100\%} & \textbf{Participants} \\
\midrule
Sub-poll 1 & 100.00\% (39) & 94.87\% (37)  & 64.10\% (25) & 39 \\
Sub-poll 2 & 100.00\% (39) & 94.87\% (37) & 69.23\% (27) & 39 \\
Sub-poll 3 & 100.00\% (39) & 92.31\% (36) & 66.67\% (26) & 39 \\
\bottomrule
\end{tabular}
\label{tab:cumulative_subsurveys_results}
\end{table}

\subsection*{Triangle Inequality}
\label{appendix:triangle_algorithm}

\begin{algorithm}[H]
    \caption{Pair generation for testing the triangle inequality.
    \label{alg:triangle}}
    
    \KwIn{Participant's ideal budget $\peak$; positive integer $k$ (number of base change vectors per rotation)}
    \KwOut{A set $S$ of comparisons between concentrated and distributed changes }
    
    Initialize empty set of questions $S \gets \emptyset$ \;
    
    \For{each of $k$ random base change vectors}{
        Sample $\other = [x_1,x_2,x_3]$ such that $\sum x_i = 0$, each $x_i$ is multiple of 5, and $\other \neq [0,0,0]$ \;
        
        Decompose $\other$ as $\other = \other_1 + \other_2$, where: \; 
        $\other_1 \gets [x_1,\, 0,\, -x_1], \quad \other_2 \gets [0,\, x_2,\, -x_2]$ \; 
        
        Verify that $\other_1 \neq [0,0,0]$ and $\other_2 \neq [0,0,0]$ \; 
        
        \If{all vectors $(\peak \pm \other, \peak \pm \other_1, \peak \pm \other_2)$ result in valid budgets in $[0,100]$ }{
            Add to $S$: $(\peak,\; \peak + \other)$ vs.\ $(\peak + \other_1,\; \peak + \other_2)$ \; 
            Add to $S$: $(\peak,\; \peak - \other)$ vs.\ $(\peak - \other_1,\; \peak - \other_2)$ \; 
            
            \tcp{Repeat for coordinate rotations $[x_2,x_3,x_1]$ and $[x_3,x_1,x_2]$}
            Repeat the same construction for the two coordinate rotations of $\other$, adding their comparisons to $S$ \; 
        }
    }
    
    \Return $S$ \; 
\end{algorithm}

\paragraph{Notes.}
\begin{itemize}
  \item Using $k=2$ base vectors per rotation yields $2 \times 3 \times 2 = 12$ experimental comparisons (plus 2 initial screening questions).
  \item Sampling constraints (multiples of 5, sum zero) preserve interpretability and ensure all resulting budgets are valid.
  \item Both positive and negative variants of each change vector are included to examine symmetry with respect to the direction of change.
\end{itemize}

\subsubsection*{Triangle Inequality Results}

\begin{table}[H]
\centering
\small
\caption{Distribution of Concentrated and Distributed Changes by Consistency Level
\label{tab:concentrated_vs_distributed_by_consistency}}
\begin{tabular}{lccc}
\toprule
\textbf{Consistency Level (\%)} & \textbf{\# of Users} & \textbf{Concentrated Change} & \textbf{Distributed Change} \\
\midrule
50.0  & 13 & 50.0\% & 50.0\% \\
58.3  & 20 & 45.9\% & 54.1\% \\
66.7  & 11 & 42.4\% & 57.6\% \\
75.0  & 16 & 34.4\% & 65.6\% \\
83.3  & 11 & 16.7\% & 83.3\% \\
91.7  & 7  & 8.3\%  & 91.7\% \\
100.0 & 10 & 40.0\% & 60.0\% \\
\midrule
\textbf{Total} & \textbf{88} & \textbf{36.7\%} & \textbf{63.3\%} \\
\bottomrule
\end{tabular}
\end{table}

\subsection{Confidence Intervals for Triangle Inequality}
\label{CI_appendix_triangle_inequality}

\begin{table}[H]
\centering
\small
\caption{Estimated proportions and Wilson 95\% confidence intervals.}
\label{tab:confidence_intervals_triangle_inequality}
\begin{tabular}{lcc}
\hline
Property & Fraction & 95\% CI \\
\hline
Triangle inequality & $17/88$ & $[0.12 , 0.29]$ \\
\hline
\end{tabular}
\end{table}

\section{Appendix to Section 6: Municipal vs.\ National Comparisons}
\label{municipal_vs_gov}

\subsection*{$\ell_1$ vs $\ell_2$ Rank Comparison}
\begin{table}[H]
\centering
\caption{Comparison of $\ell_1$ vs $\ell_2$ Rank Preferences Between Municipal and Government Budget Polls
\label{tab:l1_l2_rank_comparison}}
\begin{tabular}{llcccc}
\toprule
\textbf{Framing} & \textbf{Consistency Level} & \textbf{\# Users} & \textbf{$\ell_1$ (Rank)} & \textbf{$\ell_2$ (Rank)} & \textbf{Neutral} \\
\midrule
\multirow{7}{*}{Municipal Budget}
 & 50.0\%  & 12 & 0.0\%  & 0.0\%  & 100.0\% \\
 & 60.0\%  & 11 & 54.5\% & 45.5\% & 0.0\% \\
 & 70.0\%  & 4  & 25.0\% & 75.0\% & 0.0\% \\
 & 80.0\%  & 4  & 75.0\% & 25.0\% & 0.0\% \\
 & 90.0\%  & 1  & 100.0\%& 0.0\%  & 0.0\% \\
 & 100.0\% & 2  & 50.0\% & 50.0\% & 0.0\% \\
 & Total   & 34 & 35.3\% & 29.4\% & 35.3\% \\
\midrule
\multirow{6}{*}{Government Budget}
 & 50.0\%  & 4  & 0.0\%  & 0.0\%  & 100.0\% \\
 & 60.0\%  & 9  & 22.2\% & 77.8\% & 0.0\% \\
 & 70.0\%  & 8  & 50.0\% & 50.0\% & 0.0\% \\
 & 80.0\%  & 8  & 25.0\% & 75.0\% & 0.0\% \\
 & 90.0\%  & 2  & 50.0\% & 50.0\% & 0.0\% \\
 & Total   & 31 & 29.0\% & 58.1\% & 12.9\% \\
\bottomrule
\end{tabular}
\end{table}

\subsection*{Star-Shaped Preference}
\begin{table}[H]
\centering
\caption{Comparison of Star-Shaped Preference Metrics Between Municipal and Government Budget Polls
\label{tab:star_shaped_comparison}}
\begin{tabular}{lcc}
\toprule
\textbf{Framing} & \textbf{Random} & \textbf{Weighted Average} \\
\midrule
Municipal Budget  & 9.0\%  & 91.0\% \\
Government Budget & 11.2\% & 88.8\% \\
\bottomrule
\end{tabular}
\end{table}

\subsection*{Multi-Dimensional Single-Peaked}
\begin{table}[H]
\centering
\caption{Comparison of Multi-Dimensional Single-Peaked Test Results Between Municipal and Government Budget Polls
\label{tab:multi_dimensional_single_peaked_comparison}}
\begin{tabular}{llccc}
\toprule
\textbf{Framing} & \textbf{Consistency Level} & \textbf{\# Users} & \textbf{Far Vector} & \textbf{Near Vector} \\
\midrule
\multirow{4}{*}{Municipal Budget}
 & 80.0\%  & 4 (12.5\%) & 20.0\% & 80.0\%\\
 & 90.0\%  & 6 (18.7\%) & 10.0\% & 90.0\% \\
 & 100.0\% & 22 (68.7\%)& 0.0\%  & 100.0\% \\
 & Total   & 32 (100.0\%)& 4.4\%  & 95.6\% \\
\midrule
\multirow{3}{*}{Government Budget}
 & 90.0\%  & 4  & 10.0\% & 90.0\% \\
 & 100.0\% & 26 & 0.0\%  & 100.0\% \\
 & Total   & 34 & 2.4\%  & 97.6\% \\
\bottomrule
\end{tabular}
\end{table}

\subsection*{Peak Linear}
\begin{table}[H]
\centering
\caption{Comparison of Peak-Linear Consistency Metrics Between Municipal and Government Budget Polls
\label{tab:peak_linear_comparison}}
\begin{tabular}{lccc}
\toprule
\textbf{Framing} & \textbf{Overall Consistency} & \textbf{Transitivity Rate} & \textbf{Order Consistency} \\
\midrule
Municipal Budget  & 92.0\% & 98.3\% & 87.8\% \\
Government Budget & 78.3\% & 96.0\% & 70.1\% \\
\bottomrule
\end{tabular}
\end{table}

\subsection*{Issue Symmetry}
\begin{table}[H]
\centering
\caption{Comparison of Component-Symmetric Consistency Between Municipal and Government Budget Polls
\label{tab:component_symmetric_comparison}}
\begin{tabular}{lc}
\toprule
\textbf{Framing} & \textbf{Average Consistency Rate} \\
\midrule
Municipal Budget  & 30.6\% \\
Government Budget & 42.5\% \\
\bottomrule
\end{tabular}
\end{table}

\subsection*{Sign Symmetry}
\begin{table}[H]
\centering
\caption{Comparison of Sign-Symmetry Consistency Between Municipal and Government Budget Polls
\label{tab:sign_symmetry_comparison}}
\begin{tabular}{lc}
\toprule
\textbf{Framing} & \textbf{Average Consistency Rate} \\
\midrule
Municipal Budget  & 42.4\% \\
Government Budget & 61.8\% \\
\bottomrule
\end{tabular}
\end{table}

\subsection*{Identity Asymmetry}
\begin{table}[H]
\centering
\caption{Comparison of Identity Asymmetry Consistency Levels Between Municipal and Government Budget Polls
\label{tab:identity_asymmetry_comparison}}
\begin{tabular}{llc}
\toprule
\textbf{Framing} & \textbf{Consistency Level} & \textbf{\# Users} \\
\midrule
\multirow{7}{*}{Municipal Budget}
 & 50.0\%  & 1 (3.2\%) \\
 & 60.0\%  & 2 (6.5\%) \\
 & 70.0\%  & 2 (6.5\%) \\
 & 80.0\%  & 0 (0.0\%) \\
 & 90.0\%  & 6 (19.4\%) \\
 & 100.0\% & 20 (64.5\%) \\
 & Total   & 31 (100.0\%) \\
\midrule
\multirow{7}{*}{Government Budget}
 & 50.0\%  & 2 (6.5\%) \\
 & 60.0\%  & 4 (12.9\%) \\
 & 70.0\%  & 5 (16.1\%) \\
 & 80.0\%  & 4 (12.9\%) \\
 & 90.0\%  & 3 (9.7\%) \\
 & 100.0\% & 13 (41.9\%) \\
 & Total   & 31 (100.0\%) \\
\bottomrule
\end{tabular}
\end{table}

\subsection*{Asymmetric Loss Distribution}
\begin{table}[H]
\centering
\caption{Comparison of Asymmetric Loss Distribution Preferences Between Municipal and Government Budget Polls
\label{tab:asymmetric_loss_comparison}}
\begin{tabular}{lcc}
\toprule
\textbf{Framing} & \textbf{Concentrated (Target Decreases)} & \textbf{Distributed (Target Increases)} \\
\midrule
Municipal Budget  & 48.0\% & 52.0\% \\
Government Budget & 50.5\% & 49.5\% \\
\bottomrule
\end{tabular}
\end{table}

\subsection*{Preference Ranking}
\begin{table}[H]
\centering
\caption{Comparison of Preference Ranking Scores Between Municipal and Government Budget Polls
\label{tab:preference_ranking_comparison}}
\begin{tabular}{lc}
\toprule
\textbf{Framing} & \textbf{Final Score} \\
\midrule
Municipal Budget  & 45.3\% \\
Government Budget & 45.0\% \\
\bottomrule
\end{tabular}
\end{table}

\subsection*{Biennial Budget Preference}
\begin{table}[H]
\centering
\caption{Comparison of Biennial Budget Preferences Between Government and Municipal Budget Polls
\label{tab:biennial_budget_comparison}}
\begin{tabular}{llccc}
\toprule
\textbf{Framing} & \textbf{Consistency Level} & \textbf{Number of Users} & \textbf{Ideal Year 1} & \textbf{Random} \\
\midrule
\multirow{4}{*}{\shortstack{Government\\Budget}}
 & 50\%  & 2  & 50.0\% & 50.0\% \\
 & 75\%  & 9  & 63.9\% & 36.1\% \\
 & 100\% & 25 & 96.0\% & 4.0\% \\
 & Total & 36 & 85.4\% & 14.6\% \\
\midrule
\multirow{4}{*}{\shortstack{Municipal\\Budget}}
 & 50\%  & 2  & 50.0\% & 50.0\% \\
 & 75\%  & 12 & 66.7\% & 33.3\% \\
 & 100\% & 25 & 96.0\% & 4.0\% \\
 & Total & 39 & 84.6\% & 15.4\% \\
\bottomrule
\end{tabular}
\end{table}

\subsection*{Triangle Inequality}
\textbf{\begin{table}[H]
\centering
\small
\caption{Comparison of Triangle Inequality Test Results Between Municipal and Government Budget Polls
\label{tab:triangle_inequality_comparison}}
\begin{tabular}{llccc}
\toprule
\textbf{Framing} & \textbf{Consistency Level} & \textbf{Number of Users} & \textbf{Concentrated Change} & \textbf{Distributed Change} \\
\midrule
\multirow{8}{*}{\shortstack{Municipal\\Budget}}
 & 50.0\%  & 16 & 50.0\% & 50.0\% \\
 & 58.3\%  & 13 & 48.1\% & 51.9\% \\
 & 66.7\%  & 12 & 47.2\% & 52.8\% \\
 & 75.0\%  & 10 & 40.0\% & 60.0\% \\
 & 83.3\%  & 11 & 40.9\% & 59.1\% \\
 & 91.7\%  & 2  & 91.7\% & 8.3\% \\
 & 100.0\% & 3  & 100.0\% & 0.0\% \\
 & Total   & 67 & 49.6\% & 50.4\% \\
\midrule
\multirow{8}{*}{\shortstack{Government\\Budget}}
 & 50.0\%  & 10 & 50.0\% & 50.0\% \\
 & 58.3\%  & 13 & 44.3\% & 55.7\% \\
 & 66.7\%  & 5  & 40.0\% & 60.0\% \\
 & 75.0\%  & 8  & 43.8\% & 56.2\% \\
 & 83.3\%  & 8  & 16.7\% & 83.3\% \\
 & 91.7\%  & 4  & 8.3\%  & 91.7\% \\
 & 100.0\% & 8  & 25.0\% & 75.0\% \\
 & Total   & 56 & 35.6\% & 64.4\% \\
\bottomrule
\end{tabular}
\end{table}
}

\section{System Architecture and Reproducibility Guide}
\label{app:software_architecture}

This appendix provides a practical guide for researchers wishing to replicate this study or utilize the open-source polling framework for new experiments. For comprehensive documentation, including detailed API endpoints, troubleshooting guides, and full database schemas, please refer to the \texttt{README.md} file located in the root of the repository, at URL \url{https://github.com/ariel-research/budget-survey.git}.

The system is designed using a modular \textit{Strategy Pattern}, allowing researchers to inject new budget subjects (for example: municipal, national, or organizational budgets) and new comparison algorithms without modifying the frontend user interface. The system automatically adapts to the number of subjects ($m$) defined in the database, having been validated for $m \in \{3, 4, 5\}$.

\subsection{Setup and Configuration}
The system is containerized using Docker. The following steps outline the process from cloning the repository to configuring the environment.

\paragraph{Prerequisites:} Docker and Docker Compose.

\begin{enumerate}
    \item \textbf{Clone the repository:}
    \begin{verbatim}
git clone https://github.com/ariel-research/budget-survey.git
cd budget-survey
    \end{verbatim}

    \item \textbf{Environment Configuration:}
    Copy the example environment file and configure the critical application settings, including database credentials, the secret key, and the base URL.
    \begin{verbatim}
cp .env.example .env
# Edit .env to set:
# - SURVEY_BASE_URL (Your hosting domain or localhost:5001)
# - FLASK_SECRET_KEY (For session security)
# - MYSQL_PASSWORD (Database credentials)
    \end{verbatim}

    \item \textbf{External Provider Integration:}
    The system is designed to work with external panel providers. The configuration in \texttt{config.py} defines the redirection logic based on the participant's completion status.
    \begin{verbatim}
# config.py
EXTERNAL_PROVIDER_CONFIG = {
    "BASE_URL": "http://provider-url.com/status.php",
    "STATUS": {
        "COMPLETE": "finish",           # Successful completion
        "ATTENTION_FAILED": "filter",   # Failed attention checks (alertness tests)
        "FILTEROUT": "screenout",       # Failed pre-screening
    }
}
    \end{verbatim}
\end{enumerate}

\subsection{System Deployment}
To facilitate easy deployment, the repository includes a helper script (`deploy.sh`) that handles secret key generation and container orchestration.

\textbf{Launch the environment:}
Use the deployment script to start the application. The `dev` argument enables hot-reloading for code editing, while `prod` optimizes for data collection and security.
\begin{verbatim}
# For Development (Coding/Testing):
./scripts/deploy.sh dev

# For Production (Running Experiments):
./scripts/deploy.sh prod
\end{verbatim}

The survey interface will be available locally at \texttt{http://localhost:5001}.

\subsection{Defining Survey Content}
To add a new research topic, researchers insert a JSON-structured record into the \texttt{stories} table. The system allows multiple experimental conditions to run simultaneously on a single deployment.

\textbf{Step 1: Define the Story (Subjects).}
Insert the narrative context and subjects (e.g., Education, Sanitation, Culture).
\begin{verbatim}
INSERT INTO stories (code, title, description, subjects)
VALUES (
    'municipal_2025',
    JSON_OBJECT('en', 'City Budget', 'loc', '...'),
    JSON_OBJECT('en', 'Allocate funds...', 'loc', '...'),
    JSON_ARRAY(
        JSON_OBJECT('en', 'Education', 'loc', '...'),
        JSON_OBJECT('en', 'Sanitation', 'loc', '...'),
        JSON_OBJECT('en', 'Culture', 'loc', '...')
    )
);
\end{verbatim}

\textbf{Step 2: Configure the Algorithm (The Survey).}
Create a survey entry linking the story to a specific algorithm strategy.
\begin{verbatim}
INSERT INTO surveys (id, story_code, active, pair_generation_config)
VALUES (
    114, -- Internal ID used for routing
    'municipal_2025',
    TRUE,
    JSON_OBJECT(
        'strategy', 'l1_vs_l2_rank_comparison',
        'params', JSON_OBJECT('num_pairs', 10)
    )
);
\end{verbatim}

\textbf{Step 3: Distribution.}
Participants are directed to specific experimental conditions using URL parameters. A valid URL requires three components:
\begin{center}
\texttt{.../take-survey/?userID=[UID]\&surveyID=[SID]\&internalID=[IID]}
\end{center}

\begin{itemize}
    \item \textbf{userID (UID):} A unique identifier for the participant (passed dynamically by the panel provider) to ensure data linkage and prevent duplicate submissions.
    \item \textbf{surveyID (SID):} An identifier used by the external panel provider to track the specific survey instance and link participant data.
    \item \textbf{internalID (IID):} The specific experimental condition ID (e.g., \texttt{114} from Step 2). This parameter forces the system to load the specific Story and Algorithm configuration defined for that ID, enabling precise A/B testing.
\end{itemize}

\subsection{Implementing New Preference Algorithms}
The framework supports extending research logic via Python classes.

\subsubsection{Method A: Metric-Based Rankings}
This method compares two mathematical models (e.g., testing \textit{L1} vs. \textit{L2}). The researcher defines the utility formula, and the system handles the grid search and normalization.

\textbf{Step 1: Define the Utility Model.} 

Create a class in \texttt{application/services/algorithms/utility\_models.py}. The example below implements the $L_2$ (Euclidean) metric.

\begin{verbatim}
class L2UtilityModel(UtilityModel):
    @property
    def name(self) -> str:
        return "l2"

    def calculate(self, user_vec: tuple, cand_vec: tuple) -> float:
        # Returns negative distance (higher score = better match)
        dist = np.sqrt(np.sum((np.array(user_vec) - np.array(cand_vec))**2))
        return -float(dist)
\end{verbatim}

\textbf{Step 2: Create the Strategy Wrapper.} 

Inherit from \texttt{GenericRankStrategy} in \texttt{rank\_strategies.py}.

\begin{verbatim}
class L1VsL2RankStrategy(GenericRankStrategy):
    def __init__(self, grid_step=None):
        super().__init__(
            utility_model_a_class=L1UtilityModel,
            utility_model_b_class=L2UtilityModel,
            # grid_step: Defines the resolution of the discrete simplex.
            # e.g., step=5 generates vectors with multiples of 5 (0, 5, 10...).
            # Lower steps increase precision but increase computation cost.
            grid_step=grid_step,
            min_component=10 # Constraint: Min 10% per category
        )
\end{verbatim}

\textbf{Step 3: Registration.} Register the new class in \texttt{\_\_init\_\_.py}:
\begin{verbatim}
StrategyRegistry.register(L1VsL2RankStrategy)
\end{verbatim}

\subsubsection{Method B: Custom Logic}
For experiments requiring complex dynamic logic (e.g., temporal consistency or cyclic shifts), researchers can implement a fully custom strategy.

\textbf{Step 1: Inherit from Base Strategy.} 

Create a new file in \texttt{application/services/pair\_generation/} inheriting from \texttt{PairGenerationStrategy}.

\begin{verbatim}
class MyCustomLogicStrategy(PairGenerationStrategy):
    def get_strategy_name(self) -> str:
        return "my_custom_logic"

    def generate_pairs(self, user_vec, n, vec_size) -> list:
        pairs = []
        # Custom logic to generate 'n' pairs based on 'user_vec'
        # ...
        return pairs
\end{verbatim}

\textbf{Step 2: Registration.} Register the strategy in \texttt{\_\_init\_\_.py} to make it callable via the database configuration.
\begin{verbatim}
StrategyRegistry.register(MyCustomLogicStrategy)
\end{verbatim}

\subsection{Localization}
The system supports bilingual interfaces (e.g., English and a local language). While dynamic content (subjects, titles) is stored in the database as JSON objects, static UI labels (buttons, error messages) are managed in the application code (\texttt{application/translations.py}). Researchers adding new interface elements should add keys to the \texttt{TRANSLATIONS} dictionary.

\end{document}